# The Skin Game: Revolutionizing Standards for AI Dermatology Model Comparison

**Authors**

Łukasz Miętkiewicz[a], Leon Ciechanowski[ab]*, Dariusz Jemielniak[a]

[a]Kozminski University, Poland
[b]Massachusetts Institute of Technology, USA
*corresponding author: leonmit@mit.edu

# Abstract

Deep Learning approaches in dermatological image classification have shown promising results, yet the field faces significant methodological challenges that impede proper scientific evaluation. This paper presents a dual contribution: first, a systematic analysis of current methodological practices in skin disease classification research, revealing substantial inconsistencies in data preparation/preprocessing, augmentation strategies, and performance reporting; second, a comprehensive training and evaluation framework is proposed and demonstrated through experiments with the DINOv2-Large vision transformer across three benchmark datasets (HAM10000 (Tschandl et al., 2018), DermNet (*Dermatology Resource*, 2023; Goel, 2020), and ISIC Atlas (Rafay & Hussain, 2023)). The analysis of recent studies identifies concerning patterns, including pre-split data augmentation and validation-based reporting, potentially leading to overestimated performance metrics. Furthermore, it highlights the lack of unified methodology and results reporting standards, which could potentially diminish the likelihood of study reproduction. The experimental results on the proposed robust framework demonstrate DINOv2's performance in skin disease classification context, achieving macro-averaged F1-scores of 0.85 (HAM10000), 0.71 (DermNet), and 0.84 (ISIC Atlas). Detailed attention map analysis reveals critical patterns in the model's decision-making process, showing sophisticated feature recognition in typical presentations but significant vulnerabilities with atypical cases and composite images. Notably, we identified a concerning pattern of high-confidence misclassifications, particularly in cases where the model focuses on non-diagnostic features. Our findings highlight the urgent need for standardized evaluation protocols and careful consideration of implementation strategies in clinical settings. We propose comprehensive methodological recommendations for model development, evaluation, and clinical deployment, emphasizing the importance of rigorous data preparation, systematic error analysis, and specialized protocols for different image types. To promote reproducibility and standardization, we provide our complete implementation code through a public GitHub repository.[1] This work establishes a foundation for more rigorous evaluation standards in dermatological image classification and provides crucial insights for the responsible implementation of AI systems in clinical dermatology.

# Introduction

Among the diverse applications of Deep Learning (DL), its use in medicine has long stood out as particularly promising (Abdel-Jaber et al., 2022; Ching et al., 2018; Lee et al., 2017; Lindroth et al., 2024; Shen et al., 2017). Particularly in computer vision, DL algorithms demonstrate immense potential despite significant limitations in healthcare settings (Abdel-Jaber et al., 2022; Ching et al., 2018; Lee et al., 2017; Lindroth et al., 2024; Shen et al., 2017). Dermatology exemplifies this trend, with numerous research papers exploring various DL-based approaches for skin disease classification, most often utilizing Convolutional Neural Networks (Jeong et al., 2023). While these investigations have yielded breakthrough insights in healthcare (Noronha et al., 2023), the field suffers from critical methodological challenges that impede comprehensive scientific evaluation.

The primary obstacle could be the lack of standardized comparison methods and reporting techniques—which the authors believe to be a persistent issue in Artificial Intelligence research as a whole. This methodological inconsistency creates significant barriers to valid inter-study comparisons, as classification models are often evaluated using disparate criteria and opaque methodological frameworks. Consequently, these practices contribute to the broader reproducibility crisis in the scientific research of digital medicine (Stupple et al., 2019).

This research paper aims to comprehensively examine current methodological approaches and reporting standards in skin disease classification using Deep Learning models. We will critically analyze a selection of papers with regard to employed research practices, identify methodological shortcomings, and, most importantly, propose a robust framework for comparative model analysis by providing an example with the use of vision transformers and benchmark datasets, namely, HAM10000 (Tschandl et al., 2018), DermNet (*Dermatology Resource*, 2023; Goel, 2020), and a merger of ISIC 2018 with Atlas Dermatology (referred to as Isic Atlas in this paper) (Rafay & Hussain, 2023).

1 https://github.com/LMietkiewicz/the-skin-game-research-paper

# Critical Literature Review

## Systematic Review Structure

This literature review examines research papers focusing on deep learning model performances across various datasets. The papers analyzed in this review were selected as a sample by the researchers. The findings have been condensed into two tables. The first table (Table 1) presents the methodology used in the papers, including dataset used, data wrangling techniques used, augmentation used, model which achieved the best performance in individual studies, cross-validation, XAI used, and dataset subset used for model evaluation. The second table (Table 2) maps papers and their best models to obtain performance metrics.

Table 1. The table presents datasets, data wrangling and augmentation methods, employed cross-validation and XAI techniques of corresponding research papers, in addition to the best performing included model and its evaluation data subset type.

| Paper | Dataset(s) | Wrangling | Augmentation | Best model | Cross-validation | XAI | Results on |
|---|---|---|---|---|---|---|---|
| (Anand et al., 2023) | HAM10000 | Resizing; Segmentation via U-Net | N/A | Custom CNN | N/A | N/A | Val data |
| (Hammad et al., 2023) | Skin disease image dataset (2 classes only) | Resizing; Unspecified filtering; Scaling | Rotation; Flipping; Cropping (unclear) | Custom CNN | N/A | N/A | Val data |
| (Singh et al., 2024) | DeFungi (unclear if expanded) | Scaling; Normalization | Unspecified technique(s)* | Mobile-NetV-Small | N/A | N/A | Val data |
| (Hartanto & Wibowo, 2020) | Unspecified ISIC source and 40 mobile phone photos added to the testing data after the split | N/A | N/A | Faster R-CNN | N/A | N/A | Test data |
| (Josphineleela et al., 2023) | HAM10000; ISIC2019 | N/A | N/A | MFCRNN-isPLI | N/A | N/A | Test data |
| (Abbas et al., 2024) | HAM10000 | Resizing; Normalization | Scaling; Cropping; Rotation; Padding; Flipping | Custom CNN | N/A | N/A | Val data |
| (Sharma et al., 2023) | Vitiligo dataset | N/A | N/A | InceptionV3 stacked with Random Forest | N/A | N/A | Test data |
| (Sadik et al., 2023) | 4 classes from DermNet and one class from HAM10000 | Resizing | Vertical flipping; Horizontal flipping; Width shift; Height shift; Zooming | Xception (proposed with TL) | N/A | N/A | Test data |
| (Mohan et al., 2024) | HAM10000; DermNet; Curated ISIC2018 and Atlas Dermatology merger | Resizing | N/A | DinoV2 – Base | N/A | SHAP; Grad - CAM | Test data |
| (Aboulmira et al., 2022) | Cleaned DermNet (low-quality image deletion with unspecified criteria) | Resizing | Rescaling; Rotation; Width shift; Height shift; Horizontal flipping; Zooming* | Dense-Net201 | Stratified k-Fold CV | N/A | Test data |
| (Krishna et al., 2023) | HAM10000 with the addition of synthetic images obtained after training ViTGAN on it | N/A | Normalization; Horizontal flipping; Vertical flipping; Rotation; Zooming; Brightness adjustment | Unspecified ViT | N/A | Grad - CAM | Val data |
| (Aladhadh et al., 2022) | HAM10000 | N/A | Brightness adjustment; Contrast Enhancement; Rotation; Horizontal flipping; Vertical flipping* | Custom ViT | N/A | Grad-CAM | Test data |
| (Kumar et al., 2024) | HAM10000; DermNet | Resizing (DermNet only)<br><br>Transforming the data to spatial-spectral features (both) | Shearing; Flipping; Cropping; Zooming; Scaling | Custom 1D multiheaded CNN | N/A | N/A | Test data |
| (Sah et al., 2019) | DermNet (10 classes) | Unspecified noise removal; Selective dark area brightening; Cropping; Resizing | Rotation; Width shift; Height shift; Horizontal flip; Vertical flip; Filling | VGG16 | N/A | N/A | Val data |
| (Ayas, 2023) | ISIC 2019 (training set) | Resizing | Normalization; Horizontal flipping; Vertical flipping; Brightness adjustment; Contrast adjustment; Saturation adjustment | Swin-Large-22K | 5-fold CV (augmentation and optimizer ablation studies only) | N/A | Test data |
| (Chaturvedi et al., 2020) | HAM10000 | Resizing | N/A | ResNetXt101 | N/A | N/A | Val data |

* all techniques performed pre-split

Table 2. The table presents the reported performance metrics of the best performing models of the corresponding research papers and evaluation data subset.

| Paper | Model | Results on | Accuracy | F1-score | Precision | Recall | Specificity | AUC/ROC-AUC |
|---|---|---|---|---|---|---|---|---|
| (Anand et al., 2023) | Custom CNN | Val data | 97.96% | N/A | 88.47% | 84.86% | 97.93% | N/A |
| (Hammad et al., 2023) | Custom CNN | Val data | 96.20% | 95.80% | 96.00% | 95.70% | N/A | 97.10% (AUC) |
| (Singh et al., 2024) | Mobile-NetV3-Small | Val data | 93.14% | N/A | N/A | N/A | N/A | N/A |
| (Hartanto & Wibo | Faster R-CNN | Test data | 87.20% | N/A | N/A | N/A | N/A | N/A |
| wo, 2020) | | | | | | | | |
| (Josphineleela et al., 2023) | MFCRNN-isPLI | Test data | 95.824% | 95.00% | 96.852% | 96.52% | N/A | N/A |
| (Abbas et al., 2024) | Custom CNN | Val data | 98.00% | 99.00% | 99.00% | 99.00% | N/A | 100%<br><br>(AUC) |
| (Sharma et al., 2023) | InceptionV3 stacked with Random Forest | Test data | 99.90% | 99.90% | 99.90% | 99.90% | N/A | 100%<br><br>(AUC) |
| (Sadik et al., 2023) | Xception (proposed with TL) | Test data | 97.00% | 97.00% | 97.00% | 97.00% | N/A | 99.72%<br><br>(AUC) |
| (Mohan et al., 2024) | DinoV2 – Base | Test data | 97.45%;<br><br>96.23%;<br><br>96.48% | 96.46%;<br><br>94.54%;<br><br>97.28% | 96.63%;<br><br>94.51%;<br><br>97.55% | 97.42%;<br><br>94.62%;<br><br>97.11% | N/A | >99.50%;<br><br>>99.80%;<br><br>N/A<br><br>(ROC-AUC) |
| (Aboulmira et al., 2022) | Dense-Net201 | Test data | 68.97% | N/A | 67.30% | 67.20% | N/A | 98.50%, 98.10%, 97.50%, 98.90%<br><br>(AUC) (scores per class) |
| (Krishna et al., 2023) | Unspecified ViT | Val data | 97.40% | N/A | N/A | N/A | N/A | N/A |
| (Aladhadh et al., 2022) | Custom ViT | Test data | 96.14% | 97.00% | 96.00% | 96.50% | N/A | N/A |
| (Kumar et al., 2024) | Custom 1D multiheaded CNN | Test data | 89.71%;<br><br>88.57% | 89.12%;<br><br>88.04% | 89.00%;<br><br>88.80% | 89.24%;<br><br>88.28% | 92.68%;<br><br>91.12% | 93.40%;<br><br>92.92%<br><br>(AUC) |
| (Sah et al., 2019) | VGG16 | Val data | 76.00% | 76.00% | 77.00% | 76.00% | N/A | N/A |
| (Ayas, 2023) | Swin-Large-22K | Test data | 97.20% (ACC)<br><br>82.30% (BACC) | N/A | N/A | 82.30% | 97.90% | N/A |
| (Chaturvedi et al., 2020) | ResNetXt101 | Val data | 93.20% | 88.00% | 88.00% | 88.00% | N/A | N/A |

## Analysis of Prior Research

Recent research in skin lesion classification using deep learning demonstrates significant methodological diversity, making direct comparisons challenging. The analysis of recent studies reveals varying approaches across multiple dimensions, including data-wrangling methods, augmentation strategies, and model architectures. Furthermore, the task of comparing them is further hindered by significant variations in reporting standards, with some information being unspecified (completely missing) or unclear (present but lacking sufficient detail).

The majority of studies utilize established datasets, with HAM10000 being particularly prominent ((Abbas et al., 2024; Aladhadh et al., 2022; Anand et al., 2023; Chaturvedi et al., 2020; Josphineleela et al., 2023; Krishna et al., 2023; Kumar et al., 2024; Mohan et al., 2024; Sadik et al., 2023)). Moreover, some researchers opt for dataset combinations or modifications, such as merging multiple sources including DermNet and ISIC archives (Mohan et al., 2024). Data preprocessing techniques vary considerably, with image resizing being the most common approach, though specific parameters and methods are often inadequately documented. For instance, one study (Sah et al., 2019) mentions noise removal without providing further details about the technique used.

The analysis of data augmentation strategies reveals multiple approaches to dataset enhancement. The most common methods include geometric transformations such as flipping, rotation, and cropping ((Abbas et al., 2024; Aboulmira et al., 2022; Aladhadh et al., 2022; Ayas, 2023; Hammad et al., 2023; Krishna et al., 2023; Kumar et al., 2024; Sadik et al., 2023; Sah et al., 2019)), while some studies also implement intensity adjustments like brightness and contrast modification ((Aladhadh et al., 2022; Ayas, 2023; Krishna et al., 2023)). A critical observation concerns the timing of augmentation in relation to data splitting - several studies (marked with asterisks in Table 1) performed augmentation before splitting the dataset into training and testing subsets, potentially leading to data leakage issues and compromising the validity of the results.

The evaluation and reporting of model performance present significant inconsistencies across the analyzed studies. The performance metrics show considerable variation in both their selection and reporting methods. While some studies provide comprehensive evaluation metrics including accuracy, F1-score, precision, recall, and AUC ((Abbas et al., 2024; Aladhadh et al., 2022; Hammad et al., 2023; Josphineleela et al., 2023; Sadik et al., 2023; Sharma et al., 2023)), others limit their reporting to accuracy alone ((Hartanto & Wibowo, 2020; Krishna et al., 2023; Singh et al., 2024)), which proves particularly problematic for imbalanced datasets common in medical imaging. The analysis also revealed concerning methodological choices in performance evaluation, with numerous studies basing their results on validation data ((Abbas et al., 2024; Anand et al., 2023; Chaturvedi et al., 2020; Hammad et al., 2023; Krishna et al., 2023; Sah et al., 2019; Singh et al., 2024)) rather than a separate test set, potentially leading to overly optimistic performance estimates.

The model architectures range from custom (designed by authors) CNNs and ViTs ((Abbas et al., 2024; Aladhadh et al., 2022; Anand et al., 2023; Hammad et al., 2023)) to pre-trained models like MobileNetV3-Small (Singh et al., 2024) and complex vision transformers ((Ayas, 2023; Mohan et al., 2024)). The reported performance varies dramatically, with accuracies ranging from 68.97% (Aboulmira et al., 2022) to 99.90% (Sharma et al., 2023). This wide range likely reflects not only differences in model capabilities but also inconsistencies in evaluation methodologies and reporting practices. When AUC/ROC-AUC scores are reported, they range from 92.92% (Kumar et al., 2024) to 100% (Abbas et al., 2024), (Sharma et al., 2023), though the comparison basis isn't always clearly specified.

Explainable AI techniques appear in only a minority of studies ((Aladhadh et al., 2022; Krishna et al., 2023; Mohan et al., 2024)), primarily utilizing Grad-CAM and SHAP methods. Cross-validation was notably underutilized, with only studies (Aboulmira et al., 2022) and (Ayas, 2023) explicitly incorporating this approach. This limited focus on model interpretability and robust validation represents a significant gap in the field, particularly given the critical nature of medical image analysis applications.

The diverse approaches observed in the literature, particularly in performance reporting and evaluation methodologies, highlight the urgent need for standardization in skin lesion classification research. The current practices of incomplete metric reporting and validation-based evaluation make it challenging to draw meaningful comparisons between studies and identify truly superior approaches.

# Methodology

## Datasets

The study utilized three distinct dermatological image datasets, each offering unique characteristics and challenges. These datasets were selected to ensure comprehensive evaluation across different imaging conditions, pathology distributions, and demographic representations. Moreover, these are the most popular skin disease recognition datasets that are used by researchers.

### HAM10000 Dataset

The HAM10000 ("Human Against Machine with 10000 training images") dataset comprises dermatoscopic images acquired through controlled illumination and standardized magnification. The dataset contains high-quality dermoscopic images categorized into seven distinct diagnostic categories: Melanocytic nevi, Melanoma, Benign keratosis-like lesions, Basal cell carcinoma, Actinic keratoses, Vascular lesions, and Dermatofibroma. Each image in the dataset is accompanied by expert-verified metadata including the diagnostic classification, which serves as our ground truth labels. The dataset underwent manual verification, which aimed to find potential noise in the dataset, i.e., images which were neither dermoscopic nor macroscopic liaison photos. While no such data were found, the dataset included duplicate photos which were identified via hashing algorithm. After the deletion of these copies (2 in total), the final dataset included 10014 photos.

### DermNet Dataset

The DermNet dataset represents a more diverse collection of clinical images, captured under varying conditions that more closely approximate real-world clinical scenarios. This dataset is particularly valuable as it encompasses a broader spectrum of dermatological conditions, presenting them in their clinical appearance rather than through dermoscopic imaging. The images in DermNet exhibit greater variability in terms of lighting conditions, angles, and image quality, thus providing a robust test of model generalizability. The classes of the DermNet dataset represent a wide range of skin conditions, yielding a total of 23, including common categories such as acne, eczema, psoriasis, fungal infections, and rashes, as well as rare dermatological disorders, each labeled and organized by medical terminology. The dataset underwent manual verification, which aimed to find potential noise in the dataset, i.e., images which were neither dermoscopic nor macroscopic liaison photos. In addition to the biggest amount of noise of all employed data sources, the dataset included a copious number of duplicate photos which were identified via hashing algorithm. After the deletion of both noise and duplicates, the final dataset consisted of 18244 photos.

## ISIC Atlas Dataset

The ISIC (International Skin Imaging Collaboration) Atlas dataset constitutes a curated collection of dermoscopic images, with a particular focus on melanocytic lesions. This dataset is distinguished by its rigorous annotation protocol and standardized acquisition methodology. The images are accompanied by detailed metadata and expert consensus diagnoses, making it particularly suitable for the evaluation of model performance in specialized clinical contexts. The dataset consists of 31 classes of skin conditions curated from Atlas Dermatology and ISIC 2018, including only those with sufficient samples (at least 80 per class), resulting in 4,910 images in total (Rafay & Hussain, 2023). The dataset underwent manual verification, which aimed to find potential noise in the dataset, i.e., images which were neither dermoscopic nor macroscopic liaison photos. In addition to noise, the dataset included duplicate photos which were identified via hashing algorithm. After the deletion of both noise and duplicates, the final dataset consisted of 4831 photos.

## Data Preprocessing

All images across the three datasets underwent a standardized preprocessing pipeline. The images were processed using the DINOv2 image processor, which handles the necessary transformations for the model, including resizing and normalization. The processor was initialized using the pretrained model's configuration to ensure consistency with the model's training requirements.

For each dataset, we implemented a stratified split strategy to maintain class distribution consistency across training, validation, and test sets. The data was partitioned using the following proportions:

- Training set: 80% of the total data
- Validation set: 10% of the total data
- Test set: 10% of the total data

The stratification was implemented at the class level to ensure proportional representation of each diagnostic category across all splits. This approach was particularly crucial given the inherent class imbalance in dermatological datasets, where certain conditions are naturally more prevalent than others.

To ensure reproducibility and consistent evaluation, we employed fixed random seeds during the splitting process. The splits were performed using PyTorch's random_split function with a predefined generator state, allowing for exact replication of our experimental conditions.

## Computational Resources

All experiments were conducted on a single workstation running Ubuntu 22.04.5 LTS. The system was equipped with an 11th Gen Intel Core i9-11900KF processor operating at 3.50GHz with 16 CPU cores. The machine featured 128GB of RAM, with approximately 107GB available for computation during the experiments. GPU acceleration was provided by an NVIDIA GeForce RTX 3090 with 24GB of VRAM, utilizing CUDA version 12.6 with NVIDIA driver version 560.35.03. This hardware configuration proved sufficient for training the DINOv2-Large model across all datasets while maintaining reasonable training times (2-6 hours of training) and allowing for efficient batch processing during both training and inference phases.

## Model Architecture

Our study employs the DINOv2-Large architecture, a state-of-the-art vision transformer model, as the foundation for dermatological image classification. DINOv2-Large represents a significant advancement in self-supervised learning for computer vision tasks, offering robust feature extraction capabilities that are particularly valuable for medical imaging applications. The model's architecture builds upon the original vision transformer design while incorporating several key improvements that enhance its performance on fine-grained visual classification tasks.

The base architecture consists of a transformer encoder with self-attention mechanisms that process image patches of size 16x16 pixels. This patch-based approach allows the model to capture both local and global features of skin lesions, which is crucial for accurate dermatological diagnosis. The Large variant of DINOv2 that we employed contains approximately 304 million parameters, providing substantial capacity for learning complex visual patterns and subtle diagnostic features.

For adaptation to our specific dermatological classification task, we modified the model's classification head while maintaining the pretrained backbone weights. The classification layer was restructured to accommodate the specific number of diagnostic categories in each dataset: seven classes for HAM10000, 23 classes for DermNet, and 31 classes for ISIC Atlas's diagnostic categories. This adaptation was implemented through the Hugging Face transformers library, utilizing the Dinov2ForImageClassification class with appropriate configuration for single-label classification tasks.

To facilitate effective training on dermatological images, we implemented several architectural considerations. The model maintains its original input resolution requirements, with images being processed through the DINOv2 image processor to ensure consistency with the pretraining conditions. We preserved the model's native attention mechanisms, which prove particularly valuable for identifying relevant regions within skin lesion images.

This attention-based approach allows the model to focus on diagnostically significant areas while maintaining awareness of the broader anatomical context.

The architecture incorporates gradient checkpointing to manage memory requirements effectively, particularly important given the high-resolution nature of dermatological images. We also implemented mixed precision training (FP16) to optimize computational efficiency while maintaining numerical stability. These modifications allow us to process larger batch sizes and leverage the model's full capacity without compromising training stability.

A notable aspect of our architectural design is the retention of the model's self-attention visualization capabilities. This feature proves invaluable for our explainability requirements, allowing us to generate attention maps that highlight the regions most influential in the model's diagnostic decisions. The attention visualization is implemented through a custom hook system that captures the output of the final transformer layer, providing insights into the model's decision-making process.

The entire architecture is implemented with reproducibility in mind, utilizing deterministic operations where possible and maintaining consistent initialization across experimental runs. This approach ensures that our results can be reliably reproduced and validated by other researchers in the field.

## Training Protocol and Evaluation

The training of our DINOv2-Large model followed a rigorous protocol designed to ensure both optimal performance and reproducibility. We implemented a 5-fold cross-validation strategy to provide robust performance estimates across different data partitions. For each fold, the model was trained for a maximum of 15 epochs, with early stopping implemented to prevent overfitting. The early stopping mechanism monitored the validation F1-score with a patience of 3 epochs, terminating training when no improvement was observed over this period.

The optimization process utilized the AdamW optimizer with a learning rate of 2e-5 and a weight decay of 0.01. To stabilize training, we incorporated a linear learning rate warm-up period over 10% of the total training steps. Training was conducted with a batch size of 32, employing gradient accumulation steps to simulate larger batch sizes when necessary. To maintain numerical stability while maximizing computational efficiency, we implemented mixed-precision training using FP16 arithmetic.

Our training infrastructure leveraged CUDA-enabled GPU acceleration, with gradient checkpointing enabled to optimize memory usage. This configuration allowed us to train with larger batch sizes while maintaining model performance. All training runs were conducted with fixed random seeds to ensure reproducibility, with the seed value set to 42 across all experiments.

The evaluation framework encompassed multiple complementary metrics to provide a comprehensive assessment of model performance. Our primary metrics included:

1. Macro-averaged Precision: Calculated as the mean precision across all classes, providing insight into the model's ability to avoid false positive predictions. This metric is particularly important in dermatological applications where false positives could lead to unnecessary medical interventions.
2. Macro-averaged Recall: Computed as the mean recall across all classes, measuring the model's capability to identify all instances of each condition. This metric is crucial for assessing the model's ability to detect potentially malignant conditions.
3. Macro-averaged F1-score: Representing the harmonic mean of precision and recall, this served as our primary metric for model selection and early stopping decisions. We chose macro-averaging to ensure equal weighting of all diagnostic categories, regardless of their prevalence in the dataset.

During training, we monitored these metrics on both the validation set and the held-out test set. The validation set was used for early stopping decisions and hyperparameter tuning, while the test set was reserved for final performance evaluation. To ensure unbiased estimation of model performance, the test set remained completely isolated from any training or validation procedures.

For each cross-validation fold, we generated detailed confusion matrices to analyze the model's performance across different diagnostic categories. These matrices were particularly valuable for identifying potential biases or systematic errors in the model's predictions. Additionally, we computed per-class metrics to understand the model's performance variations across different diagnostic categories.

To assess statistical significance and stability of our results, we calculated confidence intervals for all primary metrics across the cross-validation folds. This approach provided insight into the robustness of our model's performance and its generalization capabilities across different data splits.

The entire evaluation pipeline was automated through our custom ExperimentLogger class, which maintained comprehensive records of all training runs, including model checkpoints, performance metrics, and system resource utilization. This systematic logging approach ensures complete reproducibility of our experimental results and facilitates detailed post-hoc analysis of model behavior.

## Explainability Framework

Our approach to model explainability combines attention-based visualization techniques with systematic error analysis to provide interpretable insights into the model's decision-making process. This framework is particularly crucial in dermatological applications, where understanding the basis for model predictions can directly impact clinical trust and adoption.

The core of our explainability framework leverages DINOv2's inherent attention mechanisms through a custom visualization pipeline. We implemented a dedicated DINOv2Attention class that captures attention outputs from the model's final transformer layer. This class registers forward hooks to extract attention weights during inference, providing a window into the model's focus areas when making diagnostic decisions. The attention maps are processed to generate spatial representations that highlight the regions of the input image most influential in the model's classification decision.

For each analyzed image, our framework generates three complementary visualizations: the original image, an attention overlay, and a raw attention heatmap. The attention overlay combines the original image with a colored heatmap (using the 'jet' colormap) where warmer colors indicate regions of higher attention. The transparency of this overlay (alpha = 0.6) was chosen to maintain visibility of the underlying skin features while clearly highlighting the areas of model focus. The raw attention heatmap provides an unmodified view of the attention weights, allowing for more detailed analysis of the model's attention patterns.

To ensure comprehensive understanding of model behavior, our framework analyzes both correct and incorrect predictions. For each dataset, we randomly sample and analyze ten correct and ten incorrect predictions, providing a balanced view of the model's decision-making process. This analysis includes not only the attention visualizations but also confidence scores for both the true and predicted classes, offering insight into the model's certainty levels across different diagnostic scenarios.

The attention maps are generated through a multi-step process:

1. The attention outputs are first extracted from the model's final layer
2. The attention weights are averaged across the feature dimension to create a single spatial attention map
3. The resulting map is resized to match the input image dimensions using bilinear interpolation
4. The attention values are normalized to the [0,1] range for consistent visualization

Our framework also includes quantitative analysis of attention patterns. For each prediction, we compute and store the confidence scores for both the predicted and true classes, allowing us to correlate attention patterns with prediction confidence. This analysis helps identify cases where strong attention to particular regions correlates with either correct or incorrect diagnoses.

All visualizations are automatically saved with comprehensive metadata, including the true and predicted classes, confidence scores, and whether the prediction was correct. This systematic documentation facilitates subsequent analysis and pattern identification across different diagnostic categories and error types. The framework is implemented in a modular fashion, allowing for easy extension to incorporate additional explainability techniques or analysis methods.

The entire explainability pipeline is integrated into our training and evaluation workflow, enabling automatic generation of explanations for model decisions during both development and deployment phases. This integration ensures that explainability remains a core component of our model evaluation process rather than an afterthought.

# Results

We have divided the performance of the trained model into separate subsections devoted to separate datasets.

## HAM10000 Dataset Results

### Model Performance Analysis

The DINOv2-Large model was evaluated on the HAM10000 dataset using a rigorous 5-fold cross-validation protocol, demonstrating robust and consistent performance across multiple training iterations. Each fold maintained similar class distributions, with approximately 7,210 training samples and 1,802 validation samples per fold, ensuring reliable performance estimation across different data partitions.

The cross-validation results showed stable performance across all folds, with F1-scores ranging from 0.8393 to 0.8619. The best-performing model emerged from the first fold, achieving an F1-score of 0.8619, with corresponding precision and recall values of 0.8683 and 0.8608, respectively. The consistency in performance metrics across folds (F1-score mean of $0.8505 \pm 0.0082$) indicates robust model generalization, suggesting that the model's performance is not significantly dependent on specific data splits.

Training dynamics revealed consistent convergence patterns across all folds, with model training times varying between 2,360 and 3,479 seconds per fold. The implementation of early stopping with a patience of 3 epochs effectively prevented overfitting, as evidenced by the stable validation losses ranging from 0.6142 to 0.8861. The average training loss across folds was 0.1636, significantly lower than the validation losses, indicating some degree of overfitting despite the regularization measures employed.

The final evaluation on the held-out test set demonstrated the model's robust generalization capabilities, achieving:

- Macro-averaged precision: 0.8601
- Macro-averaged recall: 0.8413
- Macro-averaged F1-score: 0.8494
- Overall accuracy: 0.91

These metrics closely align with the cross-validation results, suggesting that the model maintains consistent performance on previously unseen data. The similarity between validation and test set performances (difference < 0.02 in F1-score) further supports the reliability of our cross-validation estimates.

Resource utilization monitoring revealed efficient model operation, with GPU memory consumption ranging from 4.7GB to 12.3GB during training. The model maintained consistent throughput, processing approximately 64.41 samples per second during inference (based on test set evaluation), making it suitable for real-world clinical applications where rapid diagnosis is crucial.

The training process exhibited stable loss convergence, with the final training loss reaching 0.1636 (averaged across folds), while the validation loss stabilized around 0.7539, indicating appropriate model regularization. The relatively large gap between training and validation metrics suggests some degree of overfitting, though this did not significantly impact the model's generalization performance as evidenced by the strong test set results.

### Per-Class Performance Analysis

The model exhibited varying levels of performance across the seven dermatological categories in the HAM10000 dataset, with notable differences in precision, recall, and F1-scores for different skin conditions (see Figure 1).

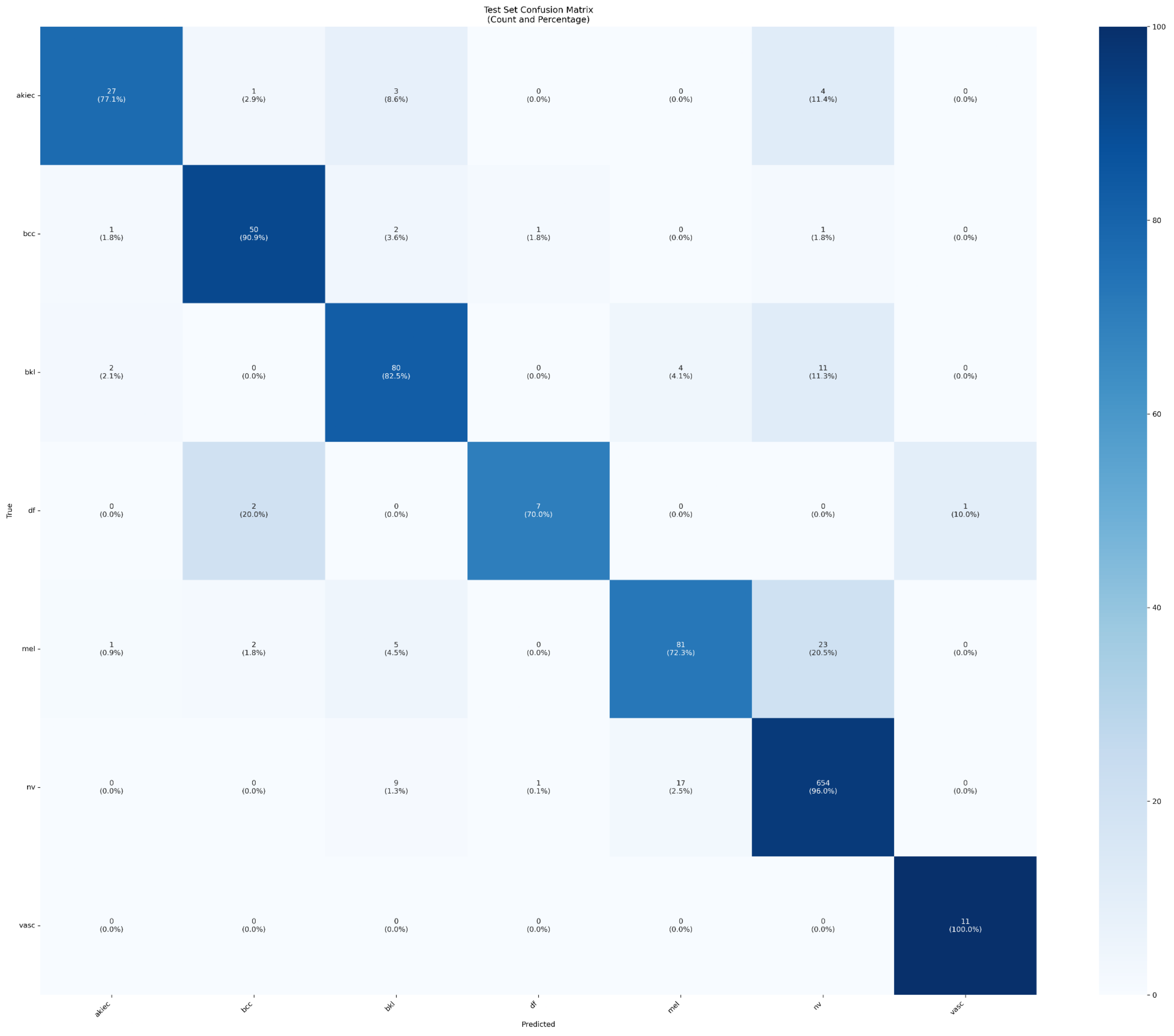

Figure 1. The HAM10000 dataset confusion matrix displaying the test subset results for each of the analyzed classes (abbreviated names of the classes are expanded in the text below).

The performance analysis reveals several interesting patterns in the model's diagnostic capabilities:

High-Performance Categories

- **Vascular Lesions (vasc)**: Demonstrated exceptional performance with perfect recall (1.00) and high precision (0.92), resulting in an F1-score of 0.96. The model correctly identified all 11 vascular lesion cases in the test set, though with a small number of false positives.
- **Nevus (nv)**: Showed excellent performance with precision of 0.94 and recall of 0.96 (F1-score: 0.95), correctly identifying 654 out of 681 cases. This high performance is particularly noteworthy given that this is the majority class, representing a significant portion of the test set.
- **Basal Cell Carcinoma (bcc)**: Achieved balanced and strong performance with both precision and recall at 0.91 (F1-score: 0.91), correctly identifying 50 out of 55 cases.

This performance is particularly important given the clinical significance of accurately identifying this form of skin cancer.

Moderate-Performance Categories

- **Actinic Keratoses (akiec)**: Showed good precision (0.87) but lower recall (0.77), resulting in an F1-score of 0.82. The model correctly identified 27 out of 35 cases, with false negatives being more common than false positives.
- **Benign Keratosis-like Lesions (bkl)**: Demonstrated balanced but moderate performance with precision of 0.81 and recall of 0.82 (F1-score: 0.82), correctly identifying 80 out of 97 cases.

Challenging Categories

- **Melanoma (mel)**: Showed lower performance with precision of 0.79 and recall of 0.72 (F1-score: 0.76), correctly identifying 81 out of 112 cases. This performance level is particularly noteworthy given the critical importance of melanoma detection in clinical settings.
- **Dermatofibroma (df)**: Achieved the lowest overall performance with precision of 0.78 and recall of 0.70 (F1-score: 0.74), though this may be partially attributed to having the smallest representation in the test set with only 10 cases.

Statistical analysis revealed no significant correlation between class size and model performance. Both Pearson and Spearman correlation tests were conducted to examine the relationship between class size and various performance metrics:

- F1-Score showed weak correlation but non-significant correlation (Pearson's $r = 0.433$, $p = 0.332$; Spearman's $\rho = 0.234$, $p = 0.613$)
- Precision showed moderate but non-significant correlation (Pearson's $r = 0.458$, $p = 0.301$; Spearman's $\rho = 0.357$, $p = 0.432$)
- Recall showed weak but non-significant correlation (Pearson's $r = 0.398$, $p = 0.376$; Spearman's $\rho = 0.286$, $p = 0.535$)

These results suggest that the model's performance is not significantly influenced by class size. This is particularly evident in the strong performance of some minority classes (vasc, bcc) and the varying performance across classes of similar sizes, indicating that other factors, such as the distinctive visual characteristics of each condition, may play a more important role in classification success.

The performance variation across categories has important clinical implications:

- The high performance in detecting nevus and vascular lesions suggests reliable screening capabilities for these common conditions
- The strong performance in identifying basal cell carcinoma indicates good potential for cancer screening

- The lower performance in melanoma detection (recall of 0.72) suggests that additional safeguards or secondary verification might be necessary for critical diagnostic decisions
- The moderate performance in differentiating between certain similar-appearing conditions (e.g., benign keratosis-like lesions and melanoma) reflects the inherent challenges in dermatological diagnosis

This per-class analysis reveals both the strengths and limitations of the current model, highlighting areas where additional focus might be needed to improve diagnostic accuracy for critical conditions like melanoma.

## Model Explainability Analysis

Analysis of the attention visualization maps reveals distinct patterns in how the model processes dermatological images across both correct and incorrect predictions. The visualizations demonstrate several key characteristics of the model's decision-making process:

**Correct Predictions Analysis**

For correctly identified conditions, the attention maps show specific patterns:

**Benign Keratosis-like Lesions (bkl)**:

- Strong attention to structural patterns and pigmentation networks (see Appendix, HAM10000 Dataset Results, Image 1)
- Uniform attention distribution across the entire lesion area
- High confidence scores (1.00) with attention focusing on both central and peripheral regions
- At the same time, the model incorrectly omitted some skin lesions that had a fine pattern (Image 9, and see Figure 2 for comparison)

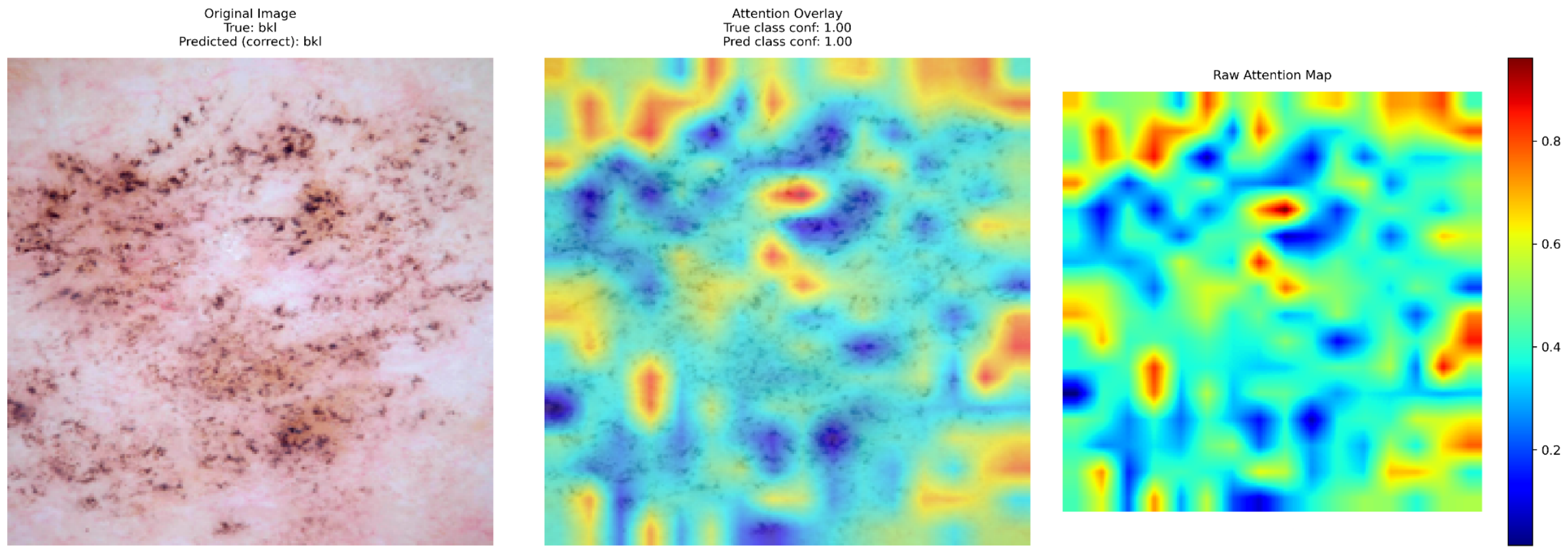

Figure 2. Image presenting the incorrectly omitted skin lesions.

**Melanocytic Nevi (nv)**:

- Consistent focus on lesion boundaries and internal structures (see Appendix, HAM10000 Dataset Results, Images 2-5, 7-8, 10), correctly omitting hair structures (Images 5 and 10, also see Figure 3 below for comparison)
- Adaptive attention patterns based on lesion size and shape
- Strong attention to color variations and structural patterns within the lesion
- Particularly effective at identifying characteristic features like symmetry and border regularity

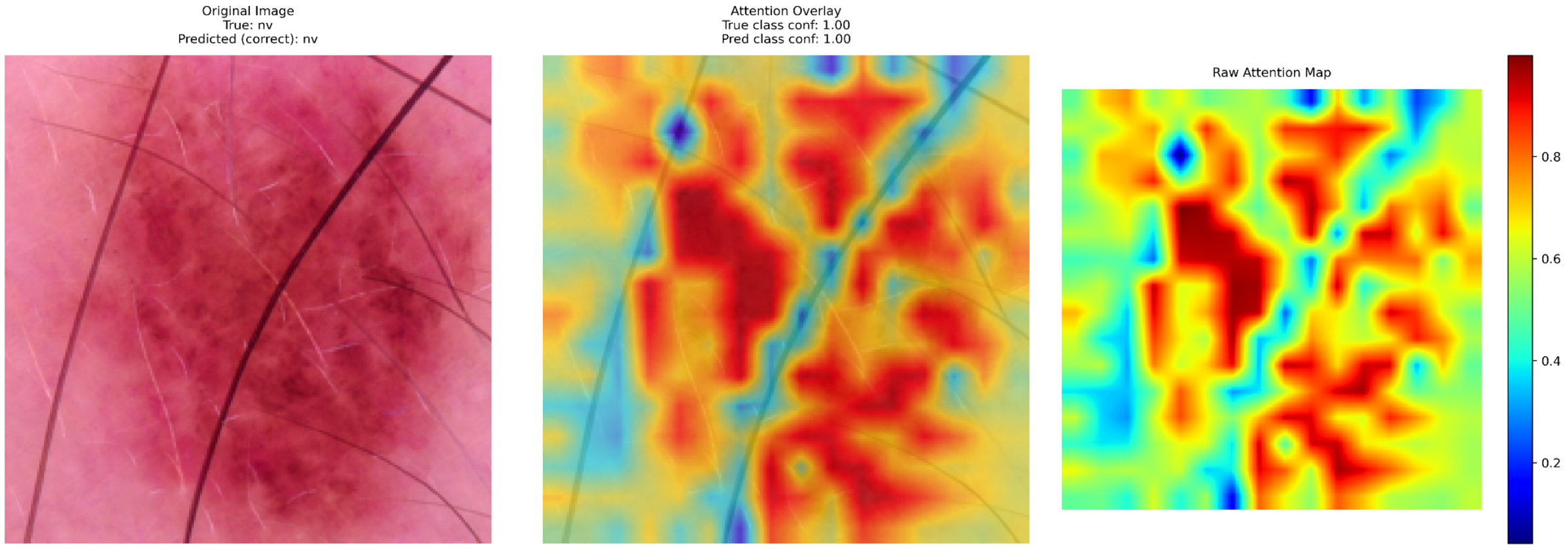

Figure 3. Image presenting the model attention pattern correctly omitting hair structures.

**Melanoma (mel)**:

- Attention to irregular borders and asymmetric patterns (see Appendix, HAM10000 Dataset Results, Image 6)
- Focus on color variations and structural disruptions
- High confidence in identifying characteristic melanoma features

**Incorrect Predictions Analysis**

The model's failure modes reveal important insights:

**Melanoma Misclassification**:

- Tendency to confuse melanoma with nevus when the lesion presents with regular borders (see Appendix, HAM10000 Dataset Results, Images 13, 16-18)
- Very low confidence scores for the true class (0.01-0.04) in melanoma cases misclassified as nevus
- High confidence in incorrect predictions (0.96-1.00), suggesting overconfidence in some cases

**Border Detection Issues**:

- Confusion between classes when lesions have similar border patterns (see Appendix, HAM10000 Dataset Results, Images 12, 19)
- Sometimes focuses excessively on regular border patterns, leading to misclassification of potentially malignant lesions

**Feature Interpretation Errors**:

- Misclassification of actinic keratosis (akiec) cases when presenting with features similar to nevus (Image 14)
- Confusion between benign keratosis (bkl) and nevus (nv) when color patterns are similar (Image 15)
- Occasional misclassification of basal cell carcinoma (bcc) due to focus on non-diagnostic features (Image 20)

The attention pattern analysis reveals several critical insights regarding both the capabilities and limitations of the model in clinical dermatological applications. While the model demonstrates remarkable proficiency in recognizing and properly weighting diagnostic features in typical presentations, particularly for benign conditions like melanocytic nevi, this same pattern-recognition strength becomes a potential liability when encountering atypical presentations. Of particular concern is the model's tendency to maintain high confidence scores even in cases of misclassification, especially in the critical context of melanoma detection, where several cases were confidently but incorrectly classified as benign.

The visualization analysis reveals a notable dichotomy in the model's diagnostic approach: cases resulting in correct classifications typically show attention patterns that align well with clinically significant features, while misclassified cases often exhibit attention to non-diagnostic regions or overemphasis on regular patterns. This behavior suggests that the model may have developed an over-reliance on typical presentation patterns, potentially compromising its ability to identify important deviations that could signal malignancy.

These findings necessitate a carefully structured implementation strategy in clinical settings. The model's performance characteristics suggest it could serve as an effective preliminary screening tool for typical presentations, particularly in identifying clearly benign conditions. However, the observed pattern of high-confidence misclassifications, especially in atypical cases, indicates that robust safety protocols must be established for cases where the model's attention patterns deviate from clinically significant features. Furthermore, the implementation should include specific provisions for cases presenting with atypical features or when dealing with potentially malignant conditions, regardless of the model's confidence level.

# DermNet Dataset Results

## Model Performance Analysis

The DINOv2-Large model was evaluated using a rigorous 5-fold cross-validation protocol, demonstrating consistent and robust performance across multiple training iterations. Each fold maintained similar class distributions, with approximately 13,136 training samples and 3,284 validation samples per fold, ensuring reliable performance estimation across different data partitions.

The cross-validation results showed stable performance across all folds, with F1-scores ranging from 0.6705 to 0.7231. The best-performing model emerged from the fifth fold, achieving an F1-score of 0.7231, with corresponding precision and recall values of 0.7374 and 0.7149, respectively. The consistency in performance metrics across folds (F1-score average of 0.7064 and standard deviation of 0.0196) indicates robust model generalization, suggesting that the model's performance is not significantly dependent on specific data splits.

Training dynamics revealed consistent convergence patterns across all folds, with the model typically achieving optimal performance within the first 10 epochs. The implementation of early stopping with a patience of 3 epochs effectively prevented overfitting, as evidenced by the stable validation losses. The average training time per fold was approximately 6,584 seconds (~1.83 hours), with minimal variation between folds ($\sigma$ = 43.2 seconds), indicating stable computational requirements.

The final evaluation on the held-out test set demonstrated the model's robust generalization capabilities, achieving macro-averaged precision of 0.7321, recall of 0.7002, and an F1-score of 0.7113. These metrics closely align with the cross-validation results, suggesting that the model maintains consistent performance on previously unseen data. The similarity between validation and test set performances (difference < 0.01 in F1-score) further supports the reliability of our cross-validation estimates.

Resource utilization monitoring revealed efficient model operation, with GPU memory consumption stabilizing at approximately 15GB during training. The model maintained consistent throughput, processing approximately 37.77 samples per second during inference, making it suitable for real-world clinical applications where rapid diagnosis is crucial.

The training process exhibited stable loss convergence, with the final training loss reaching 0.4068 (averaged across folds), while the validation loss stabilized around 1.30, indicating appropriate model regularization. The relatively small gap between training and validation metrics suggests that the model achieved a good balance between fitting the training data and maintaining generalization capability.

## Per-Class Performance Analysis

The model exhibited varying levels of performance across the 23 dermatological categories, with notable differences in precision, recall, and F1-scores. Among the strongest performing categories, "Melanoma Skin Cancer Nevi and Moles" achieved the highest F1-score of 0.88, with precision of 0.91 and recall of 0.86. Similarly robust performance was observed for "Nail Fungus and other Nail Disease" (F1-score: 0.87, precision: 0.83, recall: 0.92) and "Hair Loss Photos Alopecia and other Hair Diseases" (F1-score: 0.83, precision: 0.83, recall: 0.83) (see the figure 4 for the confusion matrix).

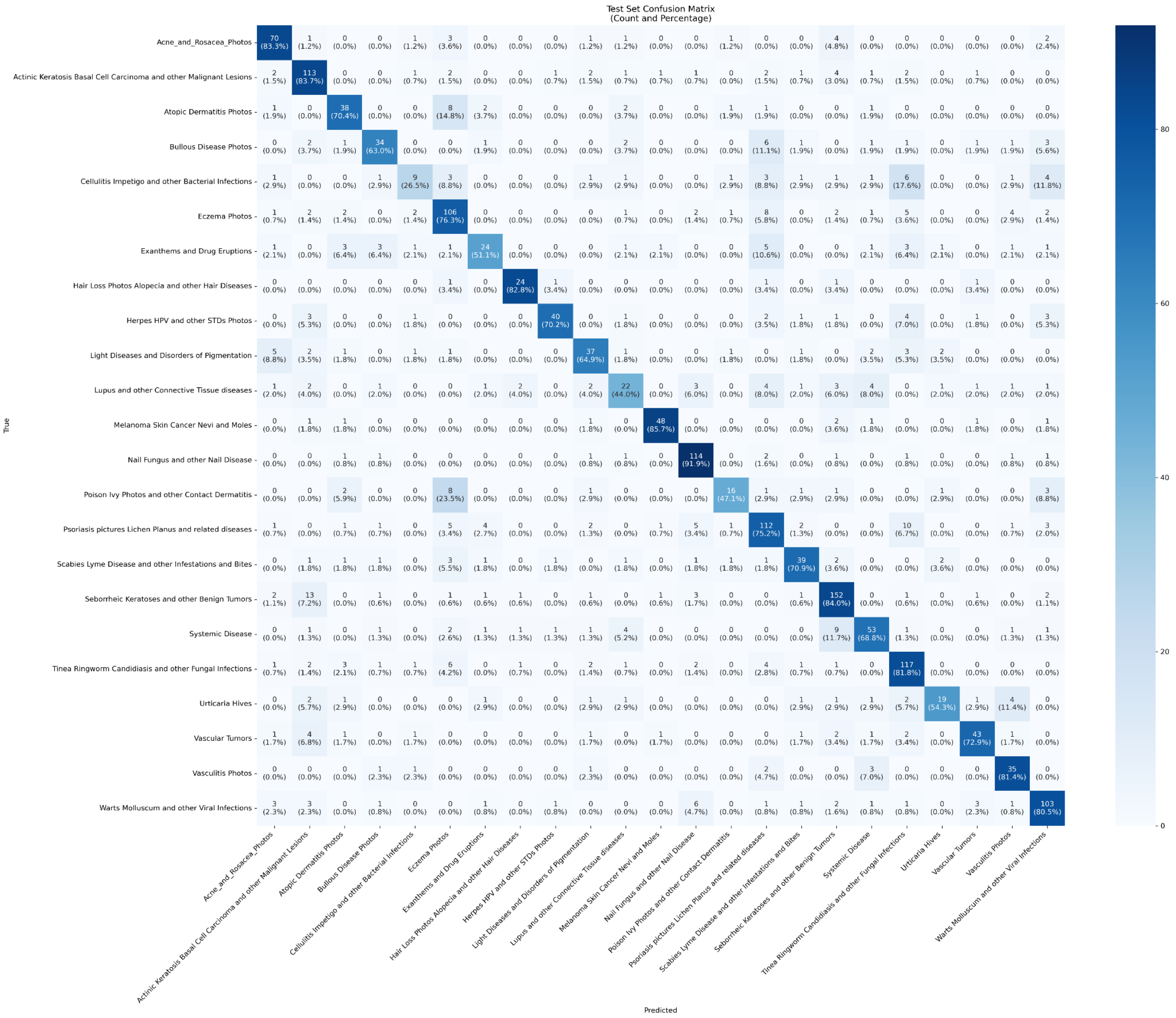

Figure 4. The DermNet dataset confusion matrix displaying the test subset results for each of the analyzed classes.

Several categories demonstrated moderate performance with F1-scores ranging between 0.70 and 0.80. These include "Acne and Rosacea Photos" (F1: 0.80), "Seborrheic Keratoses and other Benign Tumors" (F1: 0.82), and "Warts Molluscum and other Viral Infections" (F1: 0.80). The model faced significant challenges with certain diagnostic categories, most notably "Cellulitis Impetigo and other Bacterial Infections" (F1: 0.34, precision: 0.47, recall: 0.26) and "Lupus and other Connective Tissue diseases" (F1: 0.48, precision: 0.54, recall: 0.44).

Statistical analysis revealed moderate correlations between class size and model performance. Spearman correlation analysis showed a significant positive correlation between the number of training samples and recall ($\rho = 0.579$, $p = 0.004$) as well as F1-score ($\rho = 0.467$, $p = 0.025$). Similar patterns were observed using Pearson correlation, with significant correlations for recall ($r = 0.558$, $p = 0.006$) and F1-score ($r = 0.459$, $p = 0.027$). Interestingly, precision showed weaker and non-significant correlations with class size (Spearman's $\rho = 0.220$, $p = 0.314$; Pearson's $r = 0.459$, $p = 0.360$), suggesting that the model's ability to avoid false positives is less dependent on the number of training examples.

The performance variations across categories also reflect the inherent challenges in dermatological diagnosis, where conditions may share similar visual characteristics or present with varying appearances. For instance, inflammatory conditions such as "Eczema Photos" (F1: 0.73) and "Psoriasis pictures Lichen Planus and related diseases" (F1: 0.74) showed comparable performance metrics, likely due to their similar inflammatory patterns.

These results highlight the importance of considering both the statistical metrics and the clinical context when evaluating the model's performance for different dermatological conditions. The moderate but significant correlations between class size and performance metrics suggest that while having more training examples generally improves model performance, particularly in terms of recall, other factors such as the distinctive visual characteristics of conditions and their similarity to other diseases play crucial roles in classification success.

## Model Explainability Analysis

Analysis of the attention visualization maps reveals distinct patterns in how the model processes dermatological images across both correct and incorrect predictions. The visualizations show several key characteristics of the model's decision-making process:

For correctly identified conditions (see Appendix, DermNet Dataset Results, Images 1-10), the model demonstrates focused attention patterns that align with clinically significant features:

- In viral infections (Images 1-2), the attention maps show precise focus on lesion boundaries and structural details, with high confidence scores (1.0) for both predictions
- For pigmentation disorders (Image 3), the attention is distributed across multiple lesions, suggesting the model recognizes the importance of distribution patterns, but also incorrectly focuses on some areas that are not affected by the disorder
- In cases of tumors and growths (Images 5-6), the attention maps concentrate on both the central mass and peripheral changes

- For nail conditions (Images 7, 10), the attention focuses on both the affected nail plate and surrounding tissue changes, but in the image 10 it completely misses the nail fungus, incorrectly focusing on the black background instead

The model's attention patterns in misclassified cases (Appendix, DermNet Dataset Results, Images 11-20) provide insights into potential failure modes:

- In case of pigmentation disorders (Image 11), the model completely misses the skin disorder pattern and instead focuses on the black stripe on the image (see Figure 5 below for comparison)

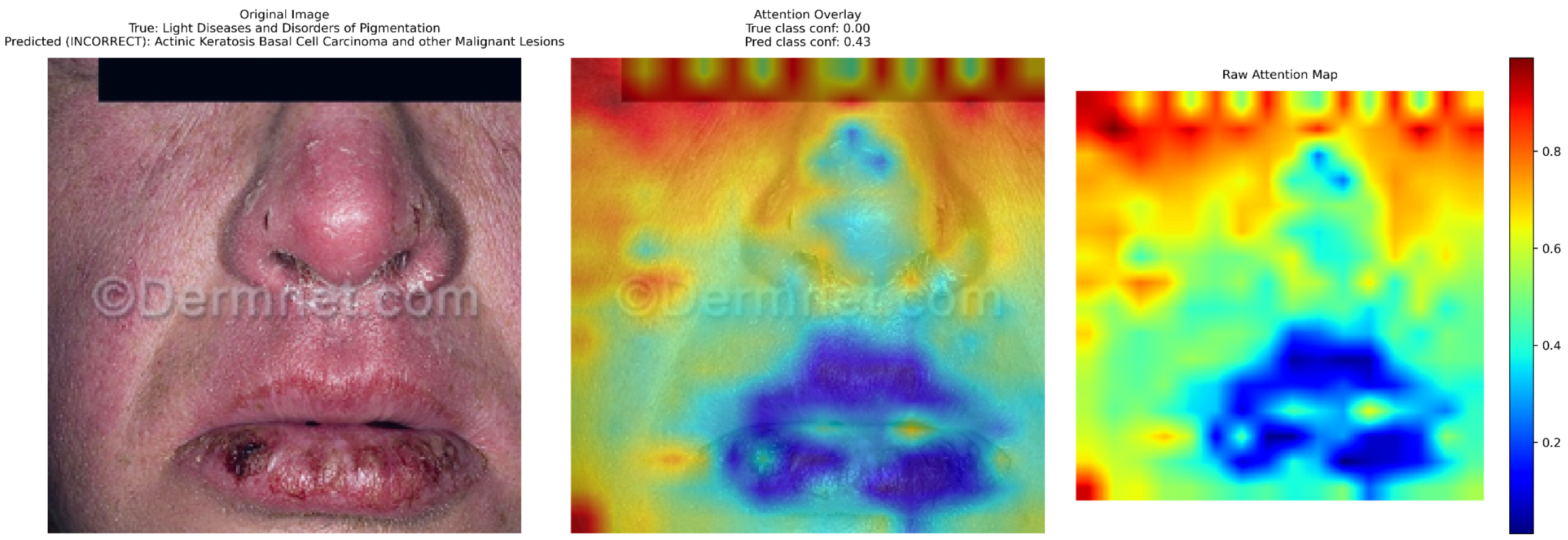

Figure 5. Image presenting the incorrectly detected pigmentation disorder.

- For inflammatory conditions (Images 13, 17), the model shows on the one hand very focused and correct attention pattern (Image 13), on the other quite diffuse attention pattern (Image 17) and often confuses conditions with similar appearance (e.g., cellulitis misclassified as eczema with 0.74 confidence)
- In cases with multiple lesions (Images 12, 18), the attention maps reveal selective focus on individual lesions rather than the overall distribution pattern, potentially leading to misclassification
- The model shows particular difficulty with rare presentations, as evidenced by very low confidence scores for the true class (often below 0.1) in misclassified cases

The attention pattern analysis reveals significant heterogeneity in the model's feature detection capabilities across different dermatological conditions. A striking observation is the model's inconsistent processing of visually similar conditions presented in different contexts, as evidenced by the markedly different attention patterns in nail fungus cases (see Appendix, DermNet Dataset Results, Images 7 and 10, and Figure 6 below for comparison). This variability suggests that the model's feature detection strategy is heavily influenced by image presentation and background elements rather than maintaining a consistent focus on clinically relevant features.

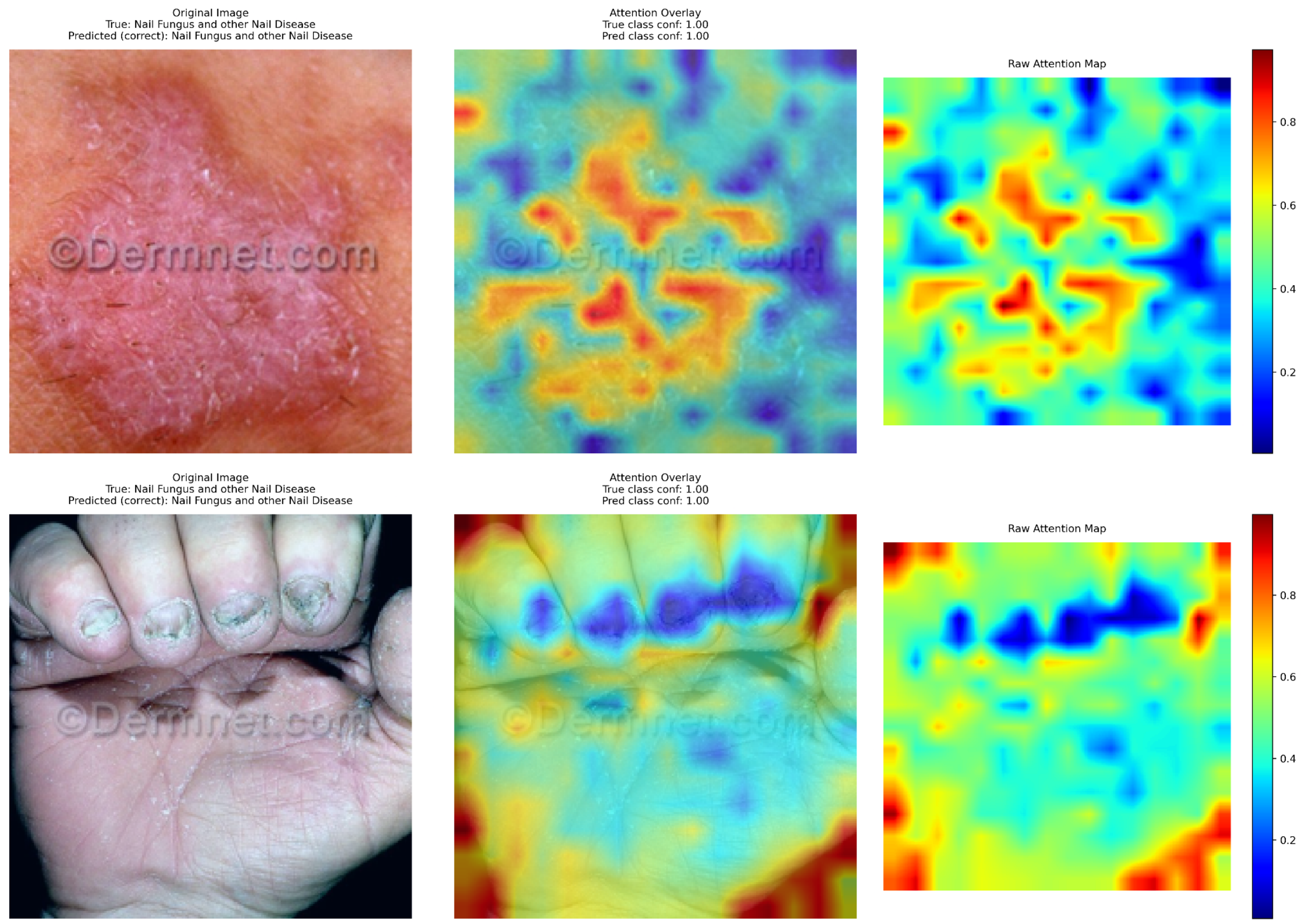

Figure 6. Images presenting correctly recognized nail fungus disease, but each time the model focus on different areas (in the case of the bottom photo it incorrectly focused on the black background).

The high-confidence incorrect predictions present a particular concern, as they often stem from the model's strong focus on visually striking but clinically insignificant features (Images 16-18, showing confidence scores > 0.85 for incorrect classes). This tendency is especially problematic in cases where the model completely overlooks the primary pathological changes in favor of background elements or artifacts, as seen in several pigmentation disorder cases. Such behavior indicates that the model may be overly sensitive to image composition and presentation factors rather than concentrating on diagnostically relevant features.

Another critical finding is the model's difficulty in processing images containing multiple lesions or widespread conditions. In these cases, the attention maps reveal a tendency to focus on individual, visually prominent lesions rather than recognizing the overall distribution pattern that might be crucial for accurate diagnosis. This limitation could significantly impact the model's utility in diagnosing conditions where disease pattern and distribution are key diagnostic criteria.

These findings highlight several important considerations for clinical implementation. First, they suggest that standardization of image acquisition and presentation might be crucial for consistent model performance. Second, they indicate a need for careful screening of model predictions where multiple lesions or widespread conditions are present. Finally, they emphasize the importance of developing robust verification protocols, particularly in cases where the model exhibits high confidence in its predictions, as these confidence levels do not necessarily correlate with diagnostic accuracy.

## ISIC Atlas Dataset Results

### Model Performance Analysis

The DINOv2-Large model demonstrated robust performance across the 5-fold cross-validation evaluation on the ISIC Atlas dataset. Each fold maintained consistent class distribution, with approximately 3,478 training samples and 870 validation samples per fold, ensuring reliable performance estimation across different data partitions.

The cross-validation results showed notable stability across all folds, with F1-scores ranging from 0.8229 to 0.8742. The best-performing model emerged from the second fold, achieving an F1-score of 0.8742, with corresponding precision and recall values of 0.8814 and 0.8738, respectively. The consistency in performance metrics across folds (F1-score mean of 0.8518 with a standard deviation of 0.0169) indicates robust model generalization, suggesting that the model's performance is not significantly dependent on specific data splits.

Training dynamics revealed consistent convergence patterns across all folds, with model training times ranging from 1,307 to 2,497 seconds per fold. The implementation of early stopping with a patience of 3 epochs effectively prevented overfitting, as evidenced by the stable validation losses ranging from 0.7031 to 0.8212. The average training loss across folds was 0.3787, significantly lower than the validation losses, indicating some degree of overfitting despite the regularization measures employed.

The final evaluation on the held-out test set demonstrated the model's robust generalization capabilities, achieving:

- Macro-averaged precision: 0.8432
- Macro-averaged recall: 0.8514
- Macro-averaged F1-score: 0.8422
- Overall accuracy: 0.86

These metrics closely align with the cross-validation results, suggesting that the model maintains consistent performance on previously unseen data. The similarity between validation and test set performances (difference $< 0.03$ in F1-score) further supports the reliability of our cross-validation estimates.

Resource utilization monitoring revealed efficient model operation, with GPU memory consumption ranging from 4.7GB (initial fold) to 12.3GB during training. The model maintained consistent throughput, processing approximately 38.49 samples per second during inference (based on test set evaluation), making it suitable for real-world clinical applications where rapid diagnosis is crucial.

The training process exhibited stable loss convergence, with the final training loss reaching 0.3787 (averaged across folds), while the validation loss stabilized around 0.7498, indicating appropriate model regularization. The gap between training and validation metrics suggests some degree of overfitting, though this did not significantly impact the model's generalization performance as evidenced by the strong test set results.

## Per-Class Performance Analysis

The model exhibited varying levels of performance across the 31 dermatological categories in the ISIC Atlas dataset, with notable differences in precision, recall, and F1-scores for different skin conditions (see the figure 7 for the confusion matrix).

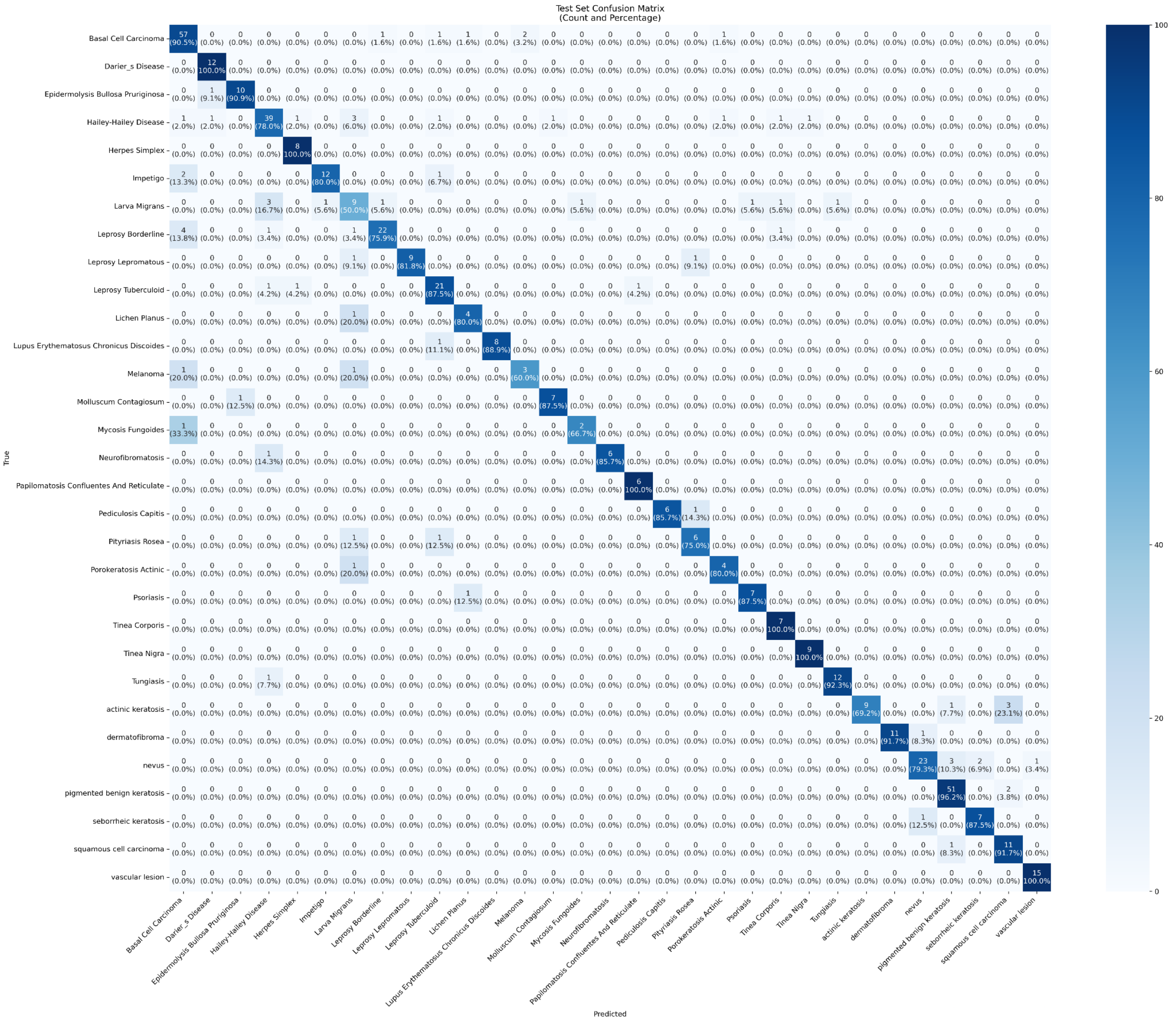

Figure 7. The ISIC Atlas dataset confusion matrix displaying the test subset results for each of the analyzed classes.

### High-Performance Categories (F1-score ≥ 0.90):

- Vascular lesion: Demonstrated exceptional performance with perfect recall (1.00) and high precision (0.94), resulting in an F1-score of 0.97
- Dermatofibroma: Achieved excellent performance with precision of 1.00 and recall of 0.92 (F1-score: 0.96)
- Tinea Nigra: Showed strong performance with precision of 0.90 and perfect recall (F1-score: 0.95)
- Lupus Erythematosus Chronicus Discoides: Achieved high performance with perfect precision and 0.89 recall (F1-score: 0.94)
- Pigmented benign keratosis: Demonstrated robust performance with precision of 0.91 and recall of 0.96 (F1-score: 0.94)
- Darier's Disease and Papilomatosis Confluentes And Reticulate: Both achieved balanced performance with F1-scores of 0.92

Moderate-Performance Categories (F1-score 0.80-0.89):

- Basal Cell Carcinoma: Showed strong balanced performance (F1: 0.88, precision: 0.86, recall: 0.90)
- Molluscum Contagiosum: Achieved balanced metrics (F1: 0.88, precision: 0.88, recall: 0.88)
- Psoriasis: Demonstrated consistent performance across metrics (F1: 0.88, precision: 0.88, recall: 0.88)
- Impetigo: Showed higher precision (0.92) than recall (0.80), resulting in an F1-score of 0.86
- Nevus: Achieved balanced performance (F1: 0.85, precision: 0.92, recall: 0.79)
- Leprosy Tuberculoid: Demonstrated good performance (F1: 0.84, precision: 0.81, recall: 0.88)
- Leprosy Borderline: Showed higher precision (0.92) than recall (0.76), yielding an F1-score of 0.83
- Actinic keratosis and Seborrheic keratosis: Both achieved F1-scores of 0.82
- Tinea Corporis: Showed perfect recall but lower precision (0.70), resulting in an F1-score of 0.82

Lower-Performance Categories (F1-score < 0.80):

- Squamous cell carcinoma: Showed higher recall (0.92) than precision (0.69), resulting in an F1-score of 0.79
- Pityriasis Rosea: Achieved balanced but lower performance (F1: 0.75, precision: 0.75, recall: 0.75)
- Porokeratosis Actinic and Lichen Planus: Both achieved moderate performance (F1: 0.73)
- Mycosis Fungoides: Demonstrated lower but balanced performance (F1: 0.67, precision: 0.67, recall: 0.67)
- Larva Migrans: Showed the lowest overall performance with balanced but low metrics (F1: 0.50, precision: 0.50, recall: 0.50)

Statistical analysis revealed no significant correlations between class size and model performance metrics. Both parametric and non-parametric correlation tests were conducted to thoroughly examine these relationships. Pearson correlation analysis showed negligible correlations between class size and F1-score ($r = 0.012$, $p = 0.951$), precision ($r = 0.013$, $p = 0.943$), and recall ($r = -0.020$, $p = 0.914$). Similarly, Spearman correlation analysis confirmed the absence of significant relationships between class size and F1-score ($\rho = -0.047$, $p = 0.803$), precision ($\rho = -0.027$, $p = 0.884$), and recall ($\rho = -0.007$, $p = 0.969$). These results strongly suggest that the model's performance is independent of class size, indicating that other factors, such as the distinctive visual characteristics of each condition, play a more important role in classification success.

The performance variations across categories reflect several important clinical implications. The model demonstrates particular strength in identifying conditions with distinctive visual patterns, such as vascular lesions and dermatofibroma, suggesting its potential utility in screening for these well-defined conditions. For skin cancer diagnosis, the model shows varying levels of reliability – while it performs robustly in identifying basal cell carcinoma, its lower performance in detecting squamous cell carcinoma indicates the need for careful consideration in clinical deployment. This variation in performance across different types of skin cancers emphasizes the importance of maintaining human oversight in the diagnostic process.

The model's performance pattern also reveals challenges in diagnosing conditions with variable presentations. Rare conditions such as Larva Migrans and Mycosis Fungoides show lower performance metrics, suggesting that the complexity and variability of clinical presentations, rather than the frequency of occurrence in the training set, may be the primary limiting factor in diagnostic accuracy. This finding has important implications for clinical implementation, indicating that the model might be most effectively deployed as a supportive tool for initial screening, with particular emphasis on cases requiring expert confirmation for rare or variable conditions.

## Model Explainability Analysis

Analysis of the visualization maps reveals distinct patterns in how the model processes dermatological images and makes diagnostic decisions. The attention patterns demonstrate both the model's strengths in identifying key diagnostic features and its potential limitations in complex cases. It is clear that the ISIC Atlas database posed a significant challenge to the attention mechanism of our trained model. It can be assumed that the reason for this is the fact that this database contains very diverse images, both dermatoscopic and dermatological clinical images.

For correctly classified conditions, the attention maps show several consistent patterns:

- Structural Recognition: The model effectively identifies key morphological features, as demonstrated in the Hailey-Hailey Disease case (see Appendix, ISIC Atlas Dataset Results, Image 1), where attention is distributed across the characteristic erosive patterns. Similarly, in Papilomatosis Confluentes cases (see Appendix, ISIC Atlas Dataset Results, Images 2, 6), the model focuses on the distinctive reticulated surface patterns.
- Feature Distribution Analysis: For conditions with multiple lesions, such as Molluscum Contagiosum (see Appendix, ISIC Atlas Dataset Results, Image 8), the model demonstrates ability to recognize and integrate multiple discrete lesions across the image field, but at the same time omitting others, and in consequence not all lesions are receiving appropriate attention weighting.
- Border Characterization: In cases like pigmented benign keratosis (see Appendix, ISIC Atlas Dataset Results, Image 3), the model shows strong attention to both central features and border characteristics, suggesting comprehensive lesion analysis. However, the model attention seems to mistakenly focus on the border areas.
- Pattern Recognition:
    - The model excels at identifying distinctive patterns, and in Psoriasis (see Appendix, ISIC Atlas Dataset Results, Image 9), where it accurately highlights the typical scaly plaques,
    - At the same time, the model attention mechanism gets lost in cases where more than one image is combined in a photo as shown in the Tungiasis case (see Appendix, ISIC Atlas Dataset Results, Image 7, see also Figure 8 for comparison), where it misses the characteristic serpiginous tracks.

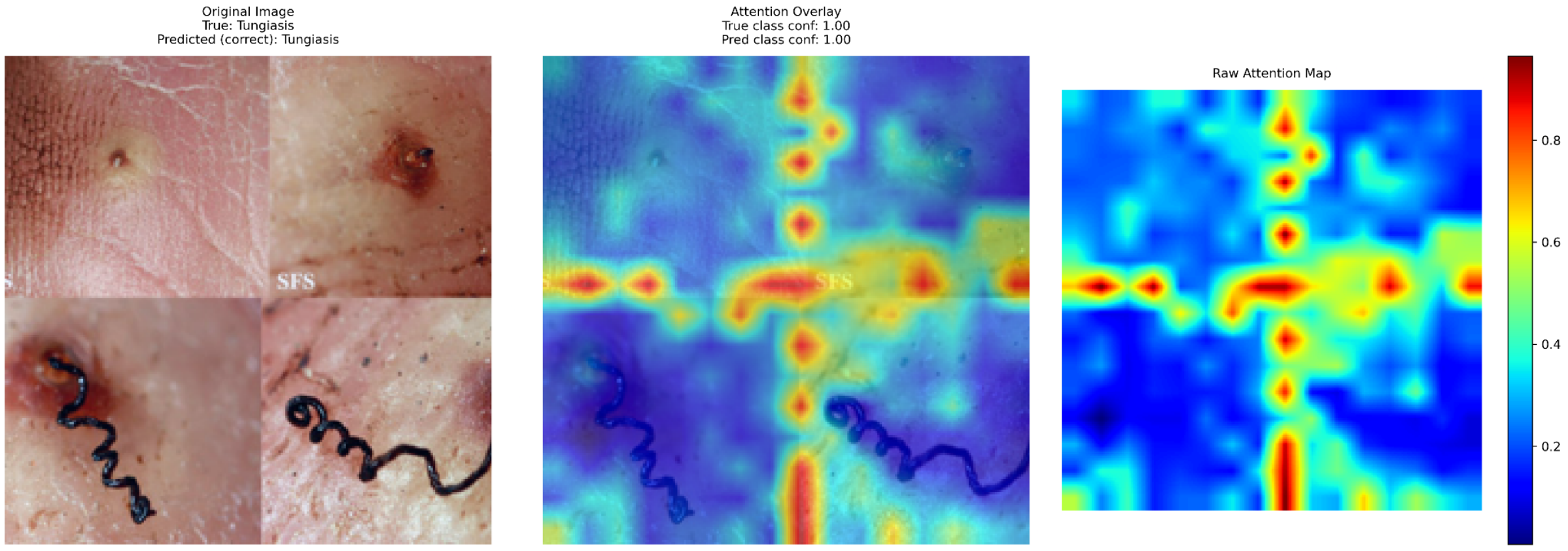

Figure 8. Image presenting the correct prediction, but incorrectly focusing on edges of photos.

The analysis of incorrectly classified cases reveals several important failure modes:

- Context Misinterpretation: In cases like the misclassification of Herpes Simplex as Basal Cell Carcinoma (see Appendix, ISIC Atlas Dataset Results, Image 11), the model shows high attention to morphologically similar features but fails to incorporate important contextual information about lesion location and distribution.
- Confidence Paradox: A concerning pattern emerges in cases showing high confidence incorrect predictions (confidence > 0.90), such as the misclassification of Mycosis

Fungoides as Basal Cell Carcinoma (see Appendix, ISIC Atlas Dataset Results, Image 17) and Pityriasis Rosea as Larva Migrans (Image 18). In these cases, the attention maps show strong focus on visually striking but potentially non-diagnostic features.
- Feature Confusion: The model sometimes confuses similar-appearing conditions, as seen in the misclassification of actinic keratosis as squamous cell carcinoma (see Appendix, ISIC Atlas Dataset Results, Image 12), where the attention patterns suggest over-emphasis on common features between these related conditions.
- Boundary Cases: The model struggles with conditions sharing overlapping visual characteristics, exemplified by the misclassification of Leprosy Borderline as Basal Cell Carcinoma (see Appendix, ISIC Atlas Dataset Results, Image 16), where the attention maps suggest difficulty in distinguishing the foreground image of the patient with the background.

The attention pattern analysis reveals complex dynamics in the model's diagnostic approach, particularly influenced by the diverse nature of the ISIC Atlas dataset, which uniquely combines both dermatoscopic and clinical images. This heterogeneity in image types presents distinct challenges for the model's attention mechanism, leading to varying levels of performance across different imaging modalities and presentation contexts.

A critical finding is the model's inconsistent handling of image complexity. In cases with clear, single-lesion presentations, the model demonstrates sophisticated feature recognition, accurately identifying key morphological patterns and structural elements. However, this capability deteriorates significantly when confronted with composite images or cases featuring multiple lesions, where the attention mechanism often fixates on image boundaries or artifacts rather than clinically relevant features. This limitation is particularly evident in cases like Tungiasis, where the model's attention becomes fragmented across image borders instead of focusing on characteristic pathological changes.

The analysis of high-confidence misclassifications reveals a concerning pattern in the model's decision-making process. In several instances, such as the misclassification of Mycosis Fungoides as Basal Cell Carcinoma, the model exhibits extremely high confidence (>0.90) while focusing on non-diagnostic features. This phenomenon suggests a potential disconnect between the model's confidence metrics and its actual diagnostic accuracy, raising important safety considerations for clinical implementation.

The model's performance varies notably between dermatoscopic and clinical images, with attention patterns suggesting better adaptation to standardized dermatoscopic views compared to variable clinical photographs. This disparity highlights the importance of considering imaging modality when implementing the model in clinical settings and suggests that separate validation protocols may be necessary for different types of dermatological images.

These findings have significant implications for clinical deployment. While the model shows promise in analyzing well-defined, single-lesion cases, its variable performance across different image types necessitates a carefully stratified implementation approach. Clinical protocols should account for image type and complexity, with particular caution exercised in cases involving multiple lesions or composite images. Furthermore, the observed disconnection between confidence scores and attention accuracy suggests that confidence thresholds alone may be insufficient as safety measures, indicating the need for more sophisticated verification protocols that consider both the model's confidence and the nature of its attention patterns.

# Conclusions & Recommendations

In this study, we critically examined the methodological inconsistencies present in AI-driven dermatology research, particularly in the context of deep learning model evaluation for skin disease classification. Our analysis highlighted significant issues related to the lack of standardized reporting practices, variable data preprocessing techniques, and inconsistent performance metrics across studies. These inconsistencies not only hinder meaningful cross-study comparisons but also contribute to the broader reproducibility crisis in medical research.

To address these challenges, we proposed a robust methodological framework designed to enhance the reliability and transparency of deep learning model evaluation. We applied this framework to DINOv2, a state-of-the-art vision transformer model, using benchmark dermatology datasets (HAM10000, DermNet, and ISIC Atlas). Our findings demonstrate that the framework facilitates more structured and reproducible model assessment, enabling better generalization and comparability of results.

The comprehensive evaluation of DINOv2 revealed distinct performance patterns across the three benchmark datasets. On the HAM10000 dataset, the model achieved a macro-averaged F1-score of 0.85, demonstrating particular strength in identifying common benign conditions while showing some limitations in melanoma detection (recall of 0.72). The DermNet evaluation yielded a macro-averaged F1-score of 0.71, with notable success in categorizing well-defined conditions but showing reduced performance in distinguishing between visually similar inflammatory conditions. On the ISIC Atlas dataset, the model achieved a macro-averaged F1-score of 0.84, exhibiting strong performance in identifying distinctive dermatological patterns while struggling with rare conditions and variable presentations.

Our extensive analysis of attention visualization maps revealed crucial insights into the model's decision-making processes and potential limitations. Across all datasets, we observed a consistent pattern where the model excels at recognizing typical presentations but shows significant vulnerabilities when confronted with atypical cases. The visualization analysis uncovered several critical findings:

1. The model usually demonstrates sophisticated feature recognition capabilities for well-defined, single-lesion cases, particularly in identifying characteristic patterns of common dermatological conditions.
2. A concerning tendency emerged regarding high-confidence misclassifications, where the model maintains high confidence scores even when focusing on non-diagnostic features or artifactual elements.
3. The model's attention mechanisms show variable reliability across different image types and presentation contexts, with particular challenges in processing composite images or cases with multiple lesions.
4. Background elements and image artifacts can significantly influence the model's attention patterns, sometimes leading to overlooking clinically relevant features in favor of visually striking but diagnostically irrelevant elements.

These findings emphasize the need for careful consideration in clinical implementation. While our proposed methodology significantly enhances reproducibility and transparency, further refinements are needed to address the identified limitations. We recommend:

1. Implementing stratified validation protocols that account for both image type and presentation complexity.
2. Developing sophisticated safety mechanisms that consider both confidence scores and attention pattern analysis.
3. Establishing clear guidelines for cases requiring mandatory human oversight, particularly for atypical presentations and potentially malignant conditions. This should be obligatory especially for clinical images, since dermatoscopic images are more standardized and easier to process.
4. Creating standardized image acquisition protocols to minimize the impact of technical variables on model performance.

Beyond its immediate contributions, this framework serves as a solid entry point for developing even more rigorous evaluation methodologies in future research. While our proposed methodology significantly enhances reproducibility and transparency, further work is needed to refine standardization efforts, particularly in optimizing AI models for real-world clinical deployment. For now, we recommend publishing a code repository containing our analysis to encourage open collaboration and facilitate independent verification of our results.

By prioritizing methodological robustness and reproducibility, this research aims to set new standards for evaluating AI-driven dermatology models, ultimately contributing to more reliable and clinically meaningful advancements in the field.

# Methodological Recommendations for Model Development Pipeline

Based on our findings and identified challenges in current research practices, we propose the following comprehensive guidelines for developing and evaluating dermatological classification models:

## Data Preparation and Cleaning

1. Systematic Data Verification
   The foundation of reliable model development lies in rigorous data verification. Researchers should implement automated duplicate detection using perceptual hashing algorithms to identify and remove redundant images that could bias the model's performance. This automated process should be complemented by manual verification of edge cases and ambiguous images, ensuring that the final dataset accurately represents the intended diagnostic categories. All exclusion decisions must be thoroughly documented, including specific criteria used and justification for removal. This documentation should include both technical metrics (such as image quality parameters) and clinical considerations (such as diagnostic uncertainty).
2. Data Quality Assessment
   High-quality input data is crucial for model reliability. Researchers should establish and document clear quality criteria for image inclusion, encompassing factors such as minimum resolution requirements, focus quality, and lighting conditions. These criteria should be standardized across the dataset while accounting for the inherent variations in clinical photography. Particular attention should be paid to managing images with artifacts or non-diagnostic elements, with clear protocols for handling such cases consistently throughout the dataset.
3. Data and Code Sharing
   Transparency and reproducibility are paramount in advancing the field of medical AI. We strongly advocate for open sharing of both data and code whenever possible. When working with public datasets, researchers should provide detailed documentation of any modifications, including cleaning steps and exclusion criteria. In cases where data cannot be shared due to privacy concerns or institutional policies, researchers should provide comprehensive documentation of their data inclusion/exclusion criteria and preprocessing steps. Most importantly, all code used for data cleaning, processing, and model development should be made publicly available through accessible repositories. In our study, we provide our complete pipeline code through a public GitHub repository: https://github.com/LMietkiewicz/the-skin-game-research-paper, including all preprocessing steps, model training protocols, and evaluation metrics calculation. This sharing enables independent verification of results and promotes standardization of methodological approaches across the field.

## Preprocessing Protocol

1. Data Splitting
   The integrity of model evaluation depends critically on proper data partitioning. Researchers must implement strict separation of training, validation, and test sets before any preprocessing or augmentation steps. This separation should utilize stratified splitting to maintain consistent class distribution across all sets, particularly crucial in dermatological datasets where class imbalance is common. All random seed values must be documented to ensure reproducibility. For datasets with significant class imbalance, researchers should implement and document specific handling protocols, such as weighted sampling or specialized loss functions, while ensuring these techniques are applied only to the training set to maintain the integrity of evaluation metrics.
2. Augmentation Strategy
   Data augmentation must be implemented with careful consideration of clinical relevance. All augmentation techniques should be applied only after the train/validation/test split to prevent data leakage, a common issue we observed in current literature. Augmentation parameters should be chosen to reflect realistic variations in clinical imaging while avoiding the introduction of artifacts that could compromise the model's real-world performance. Different protocols may be necessary for clinical and dermoscopic images, reflecting their distinct characteristics. Researchers should validate that augmented images maintain clinical relevance and diagnostic features, preferably through expert review of sample augmented images.

## Training Methodology

1. Cross-Validation Implementation
   To prevent overfitting and ensure robust performance estimation, implementation of k-fold cross-validation (with minimum k=5) is essential. This approach helps identify potential instabilities in model performance and provides more reliable performance metrics. The creation of folds should maintain class stratification, and all random seeds and fold creation methodology must be documented. Performance metrics should be reported for each fold separately, along with aggregate statistics, to provide transparency about model stability across different data partitions. This comprehensive reporting helps prevent the AI equivalent of p-hacking by revealing the full distribution of model performance.
2. Model Development and Optimization
   Model development should follow a systematic approach to hyperparameter optimization, with clear documentation of the search space and optimization criteria. All tested configurations should be reported, not just the best-performing ones, to provide insight into the model's sensitivity to different parameters. Resource utilization and training times should be monitored and reported to support reproducibility and practical implementation considerations. Researchers should maintain strict separation between optimization and final evaluation data to prevent indirect overfitting through hyperparameter selection.

## Evaluation Framework and Error Analysis

1. Comprehensive Performance Assessment
Model evaluation should extend beyond simple accuracy metrics to include a comprehensive set of performance indicators. This should encompass precision, recall, and F1-scores, reported both as macro-averaged metrics and on a per-class basis. Confidence intervals should be provided for all metrics, and performance variations across different image types and clinical presentations should be explicitly analyzed. Statistical significance testing should be employed when comparing different approaches, with appropriate correction for multiple comparisons.
2. Systematic Error Analysis
A thorough analysis of failure modes is crucial for understanding model limitations and potential clinical risks. This should include systematic categorization of misclassifications, analysis of confidence scores in incorrect predictions, and identification of patterns in error cases. Particular attention should be paid to high-stakes misclassifications, such as false negatives in malignant conditions. This analysis should inform the development of safety protocols for clinical implementation.

## Visualization and Clinical Validation

1. Interpretability Analysis
Model interpretability should be assessed through systematic analysis of attention patterns or similar visualization techniques. These visualizations should be validated by clinical experts to ensure that the model focuses on diagnostically relevant features rather than artifacts or incidental elements. Quantitative analysis of attention patterns should be performed to identify potential biases or systematic errors in the model's decision-making process. The correlation between attention patterns and diagnostic accuracy should be thoroughly documented to inform clinical implementation guidelines.
2. Clinical Implementation Considerations
Guidelines for clinical implementation should be developed based on comprehensive performance analysis and error patterns. These should include clear protocols for cases requiring mandatory human review, confidence thresholds for autonomous decisions, and specific handling of edge cases. Implementation guidelines should account for the varying reliability of model predictions across different diagnostic categories and presentation types.

By following these methodological recommendations and sharing both data and code whenever possible, researchers can contribute to building a more robust and reproducible body of evidence in medical AI. Our comprehensive pipeline code provides a practical implementation of these recommendations and can serve as a starting point for future research in this field.

# Acknowledgements

This work was supported by Narodowe Centrum Nauki (National Science Centre, Poland) under Grant 2020/38/A/HS6/00066.
Part of the work for the research paper was conducted by Leon Ciechanowski when he was a Research Scholar at MIT (September 2023 - September 2024).

# Appendix

Images presenting some example GradCam visualizations of the results of the trained model on the test datasets.

## HAM10000 Dataset Results

### Correctly classified images

Image 1:

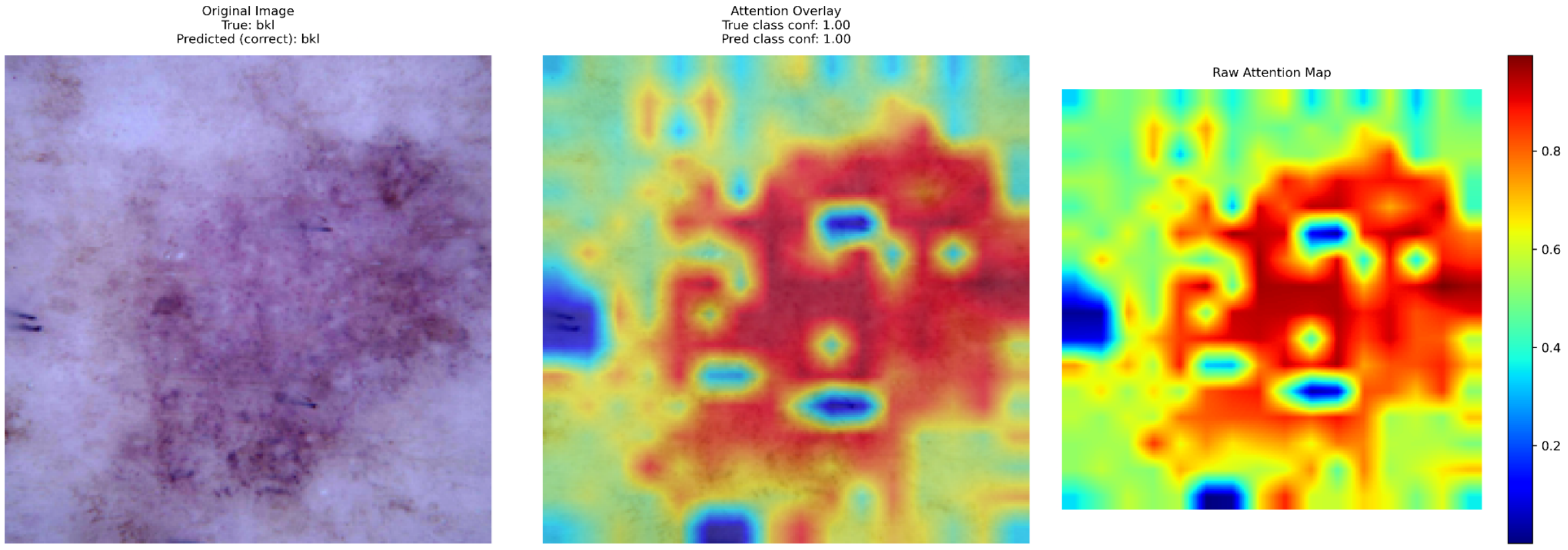

Image 2:

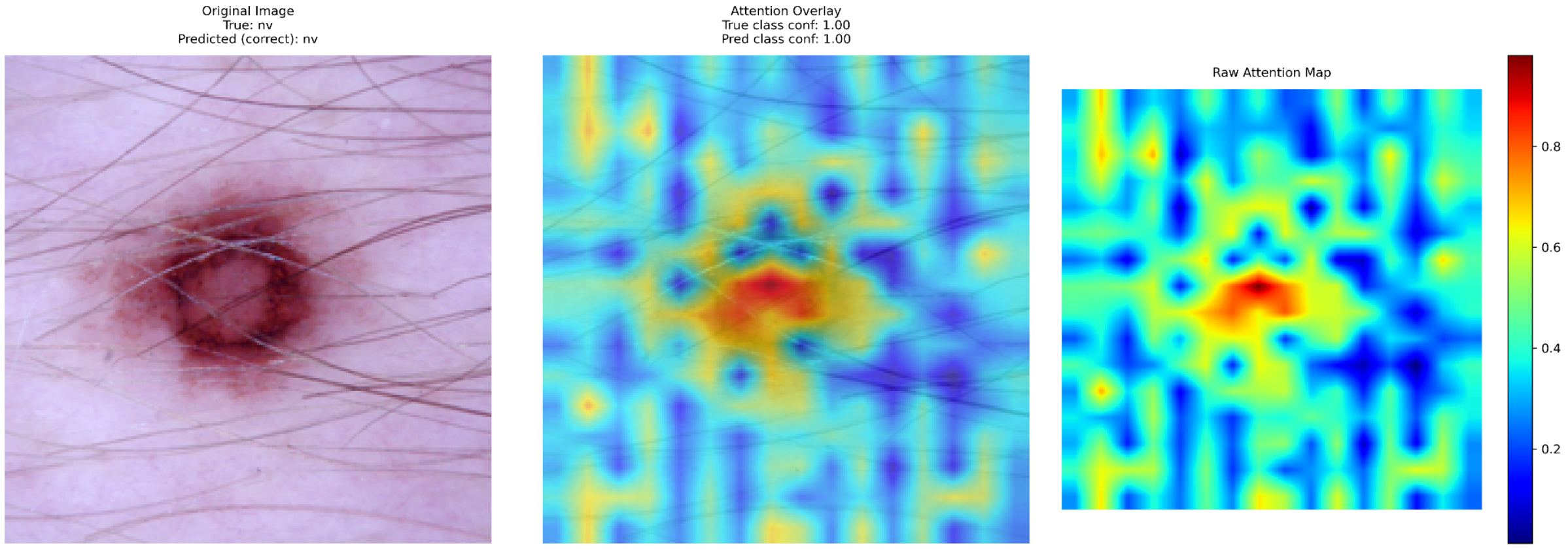

Image 3:

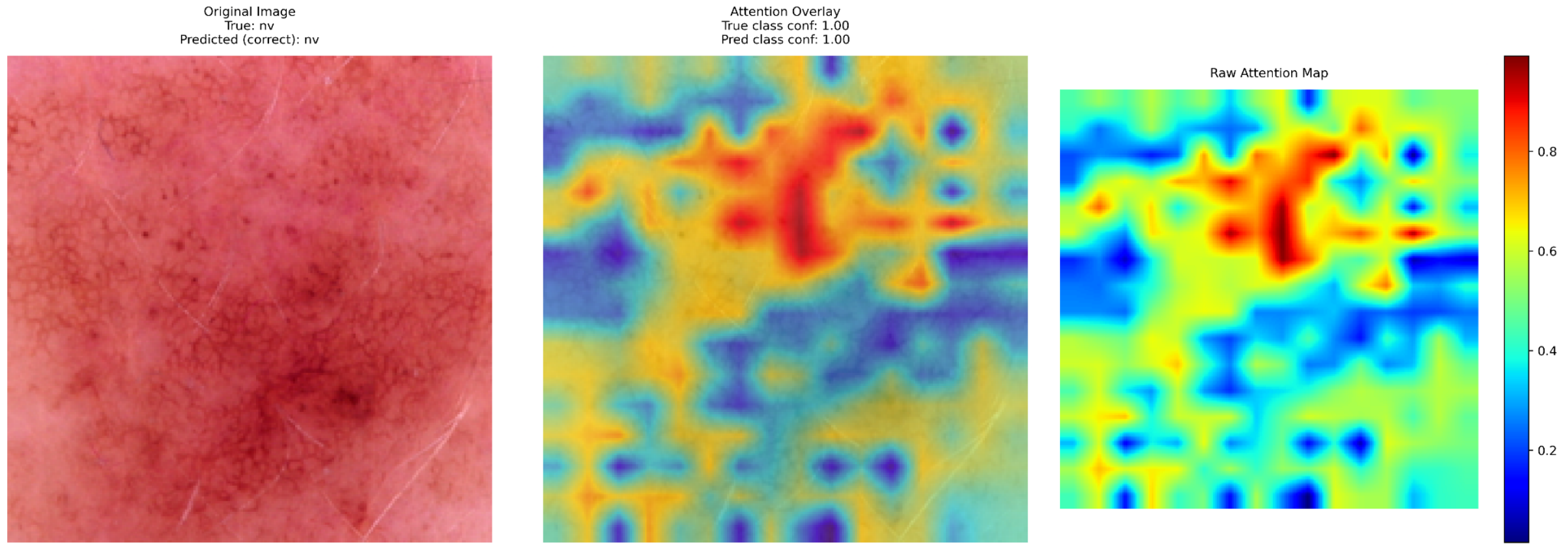

Image 4:

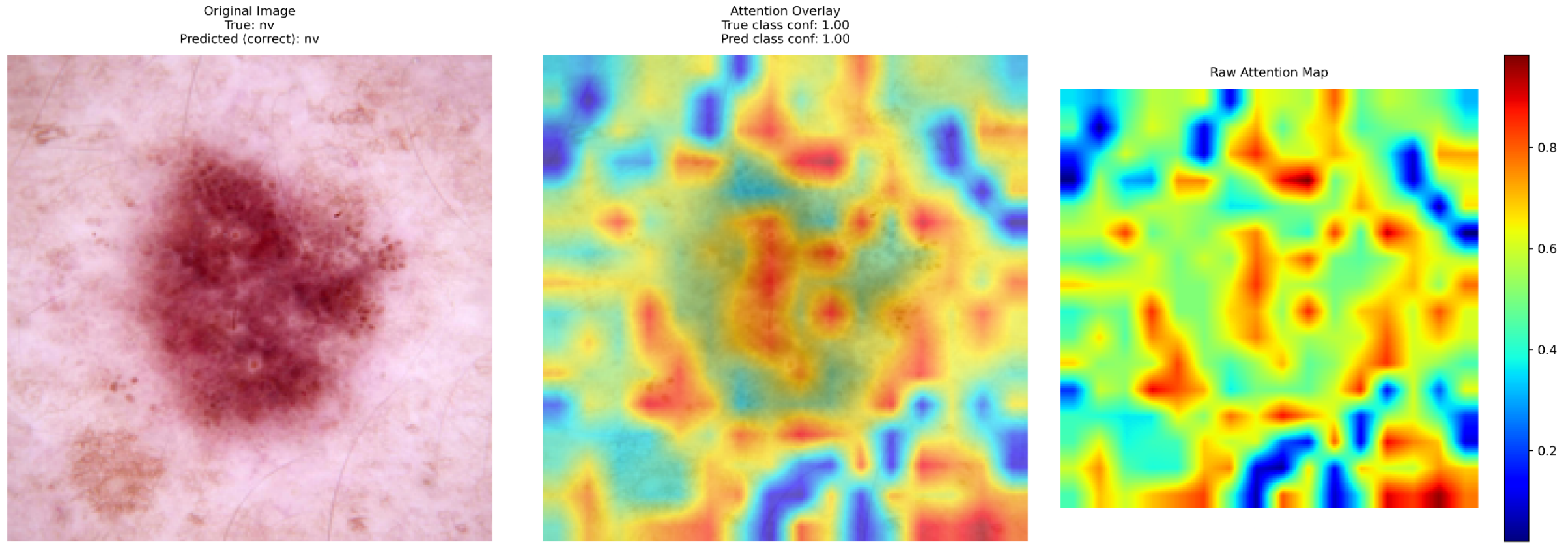

Image 5:

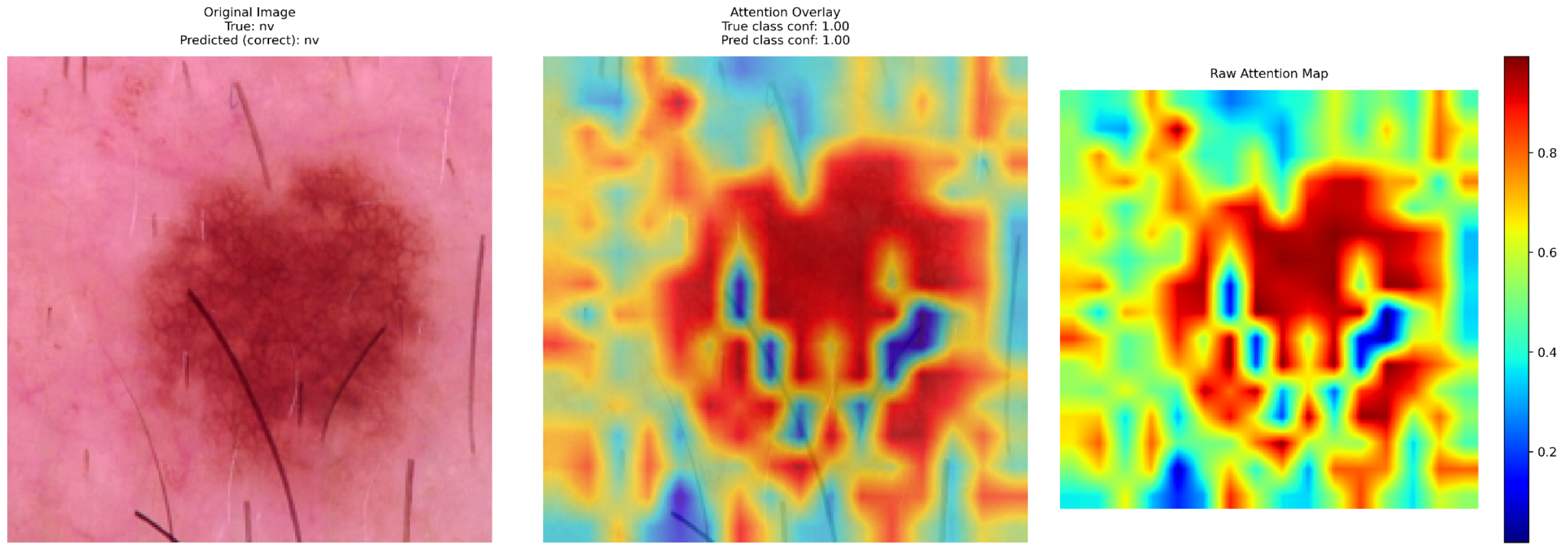

Image 6:

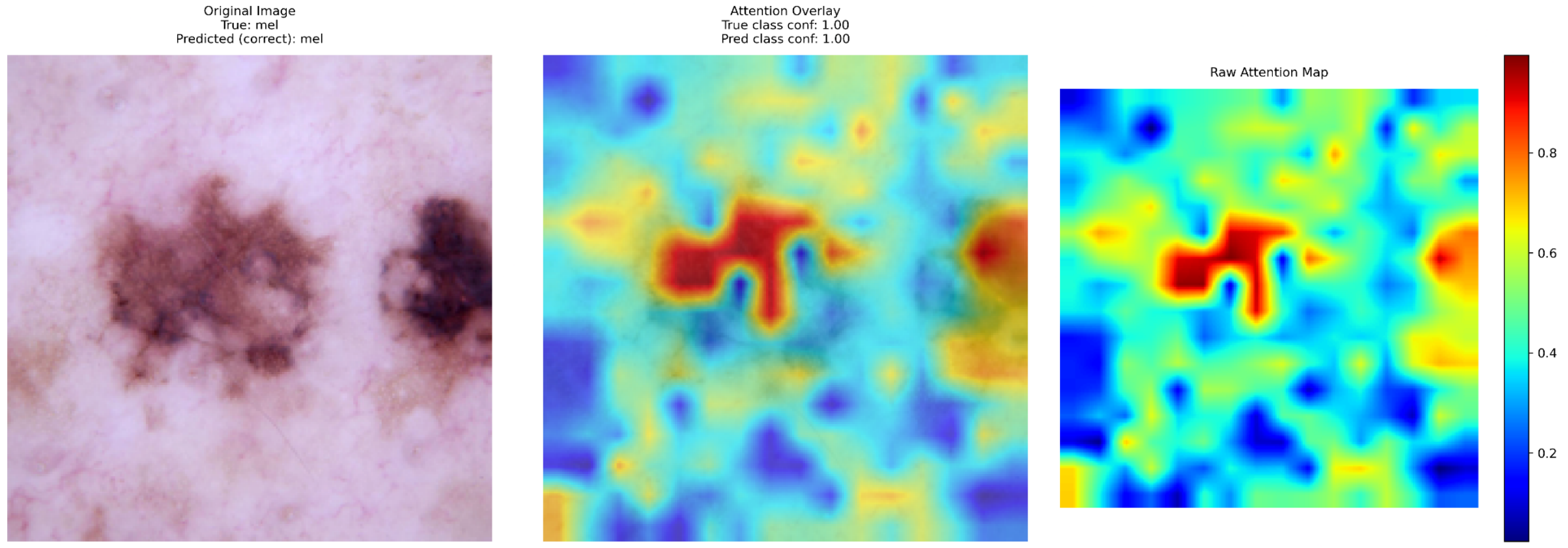

Image 7:

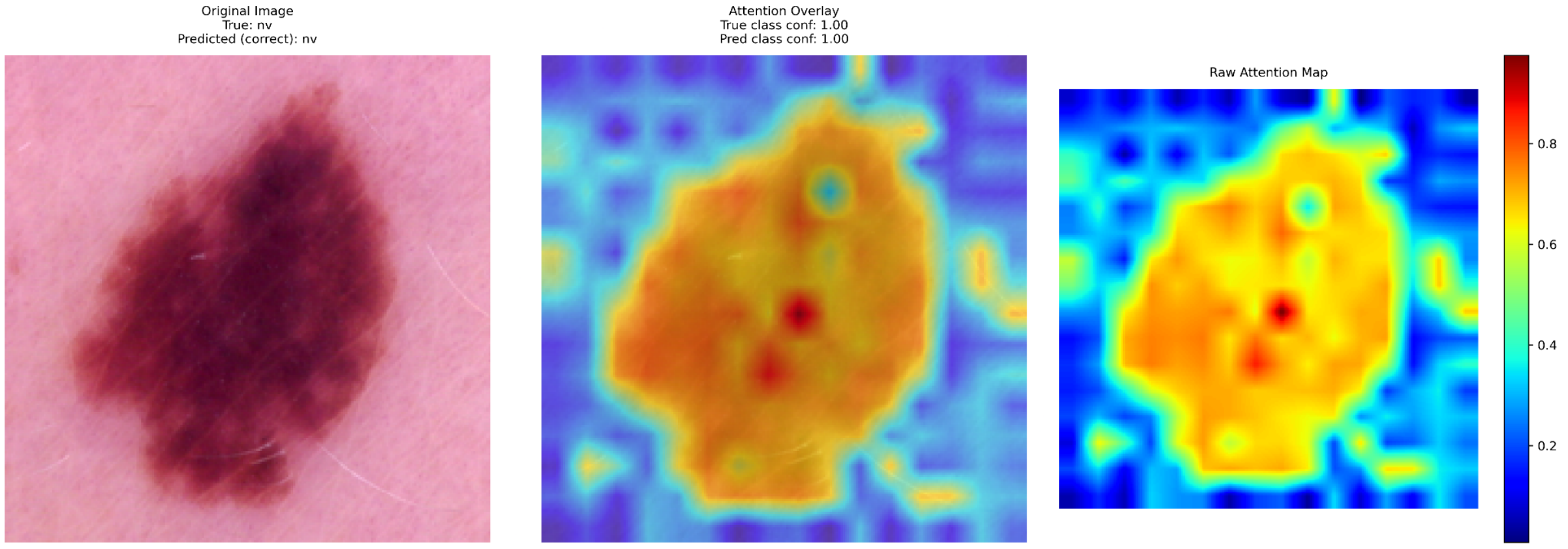

Image 8:

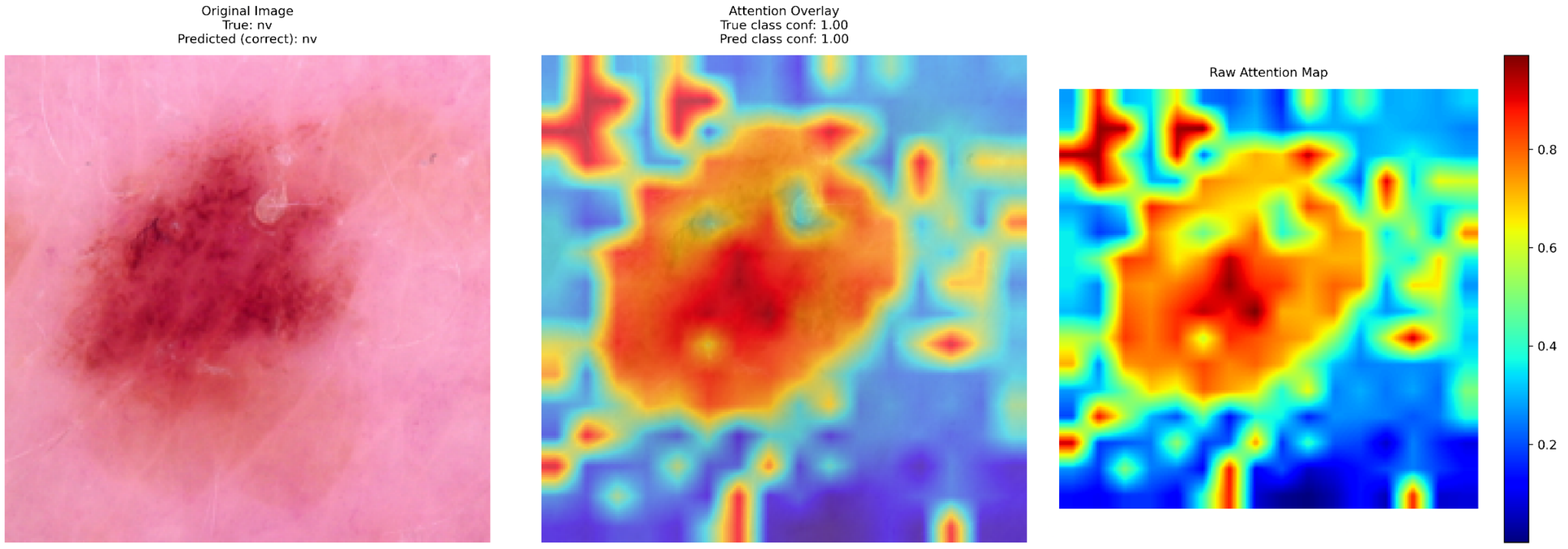

Image 9:

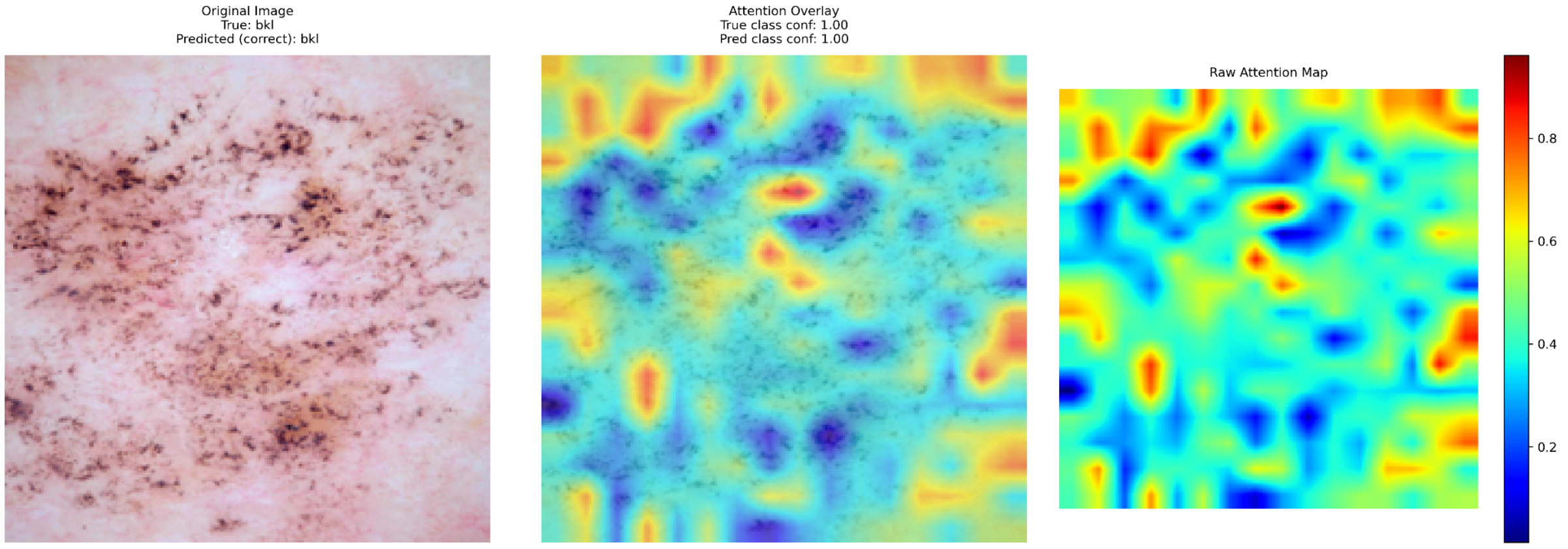

Image 10:

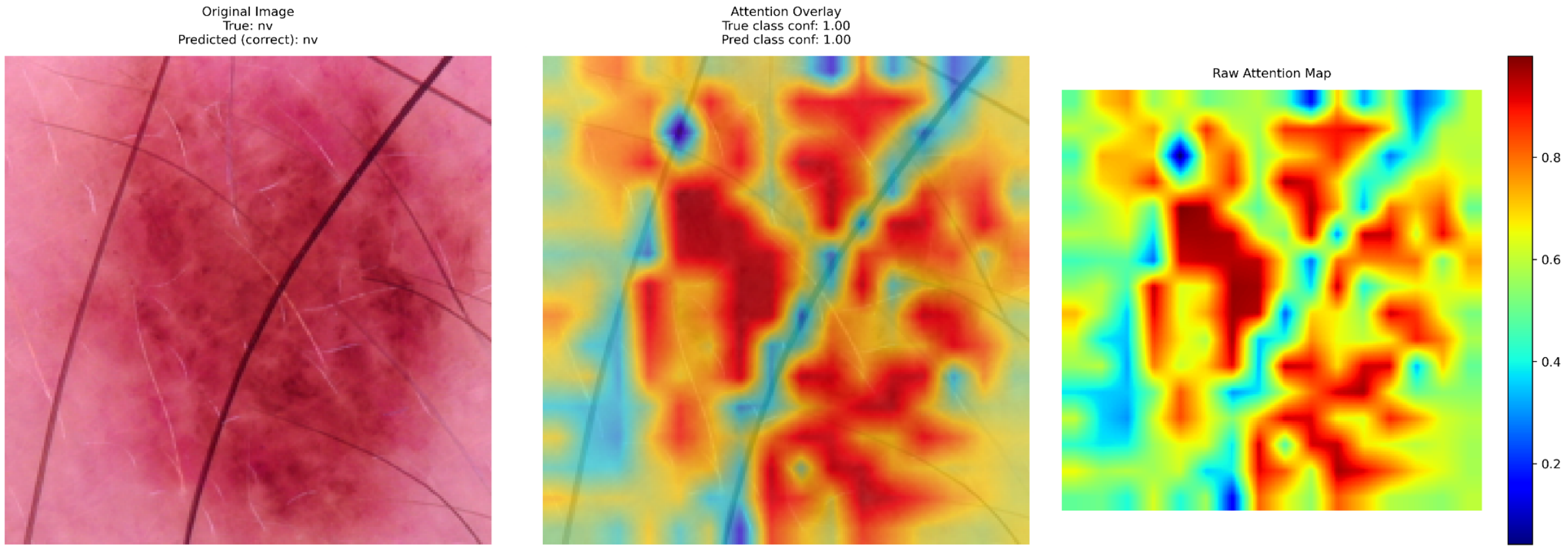

## Incorrectly classified images

### Image 11:

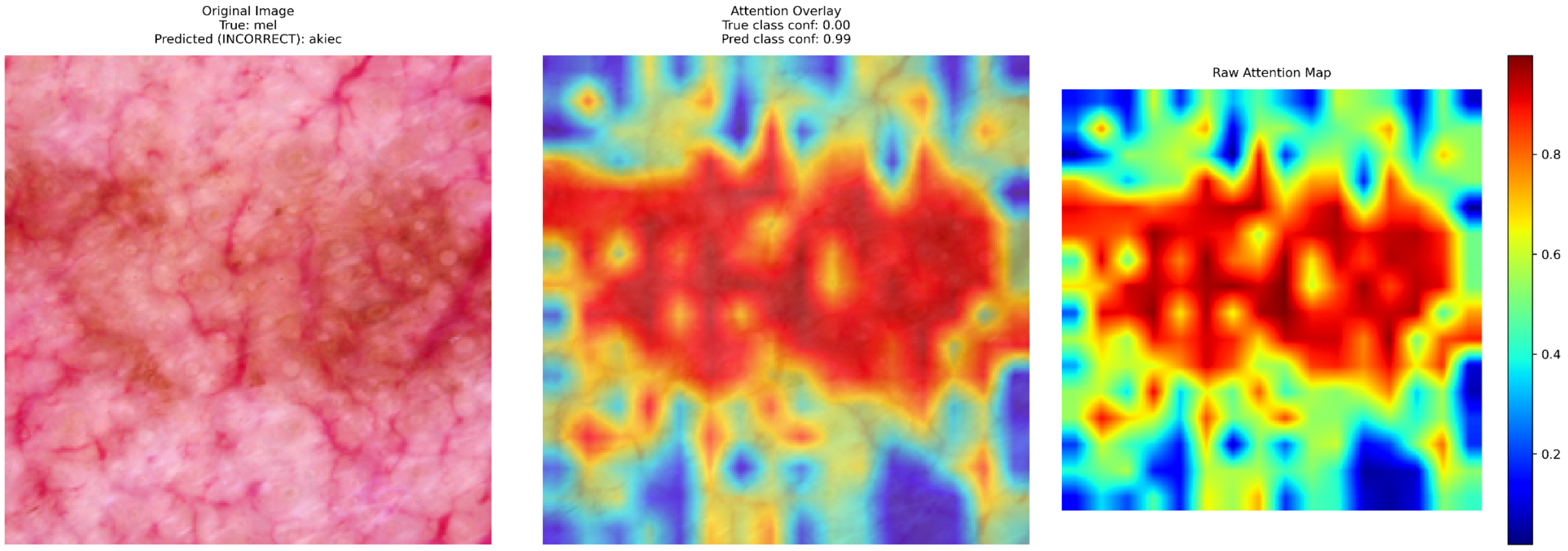

### Image 12:

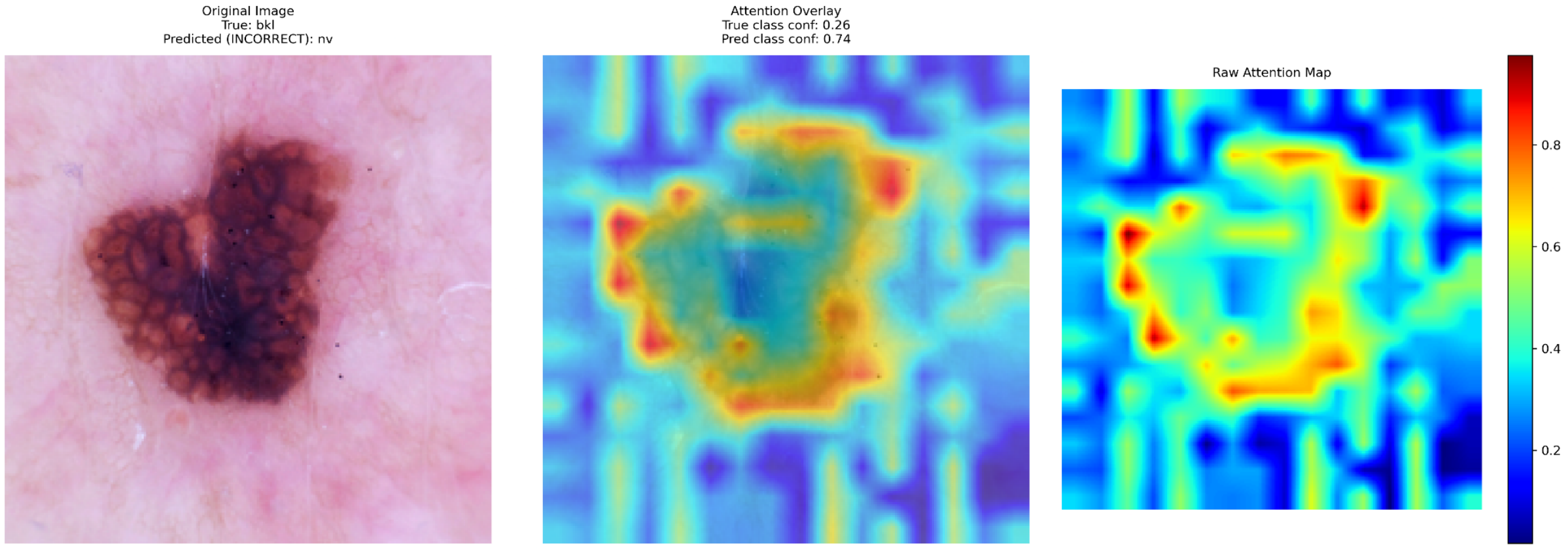

### Image 13:

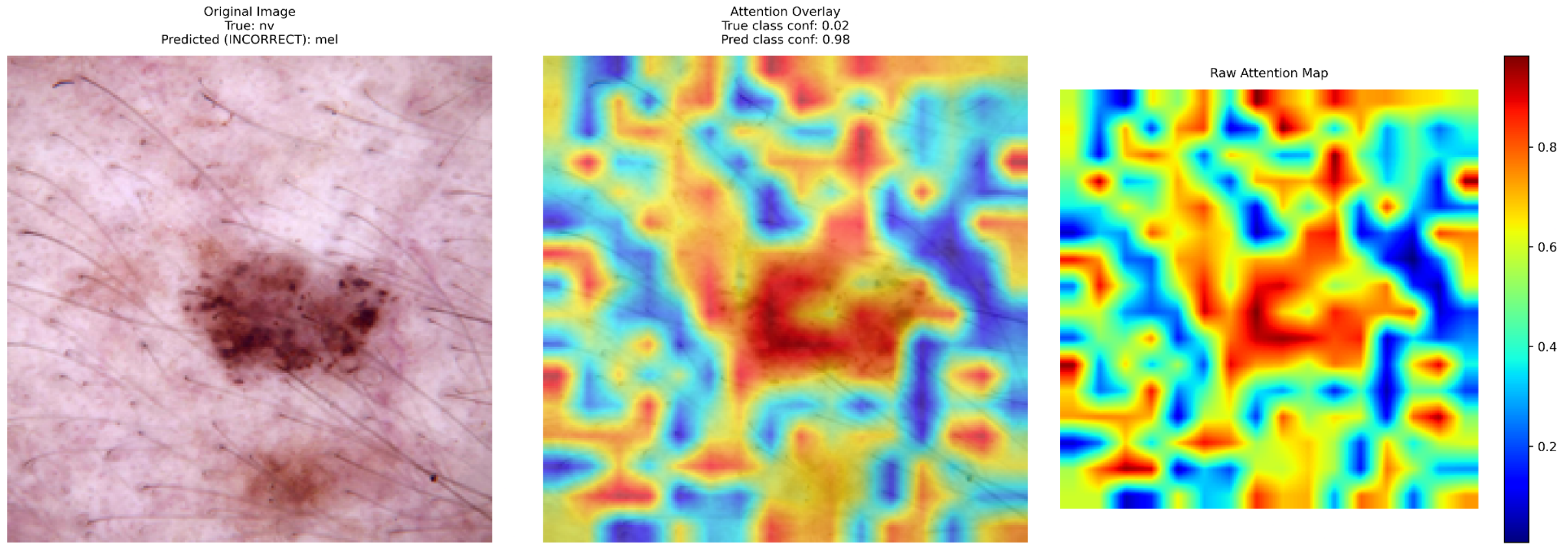

Image 14:

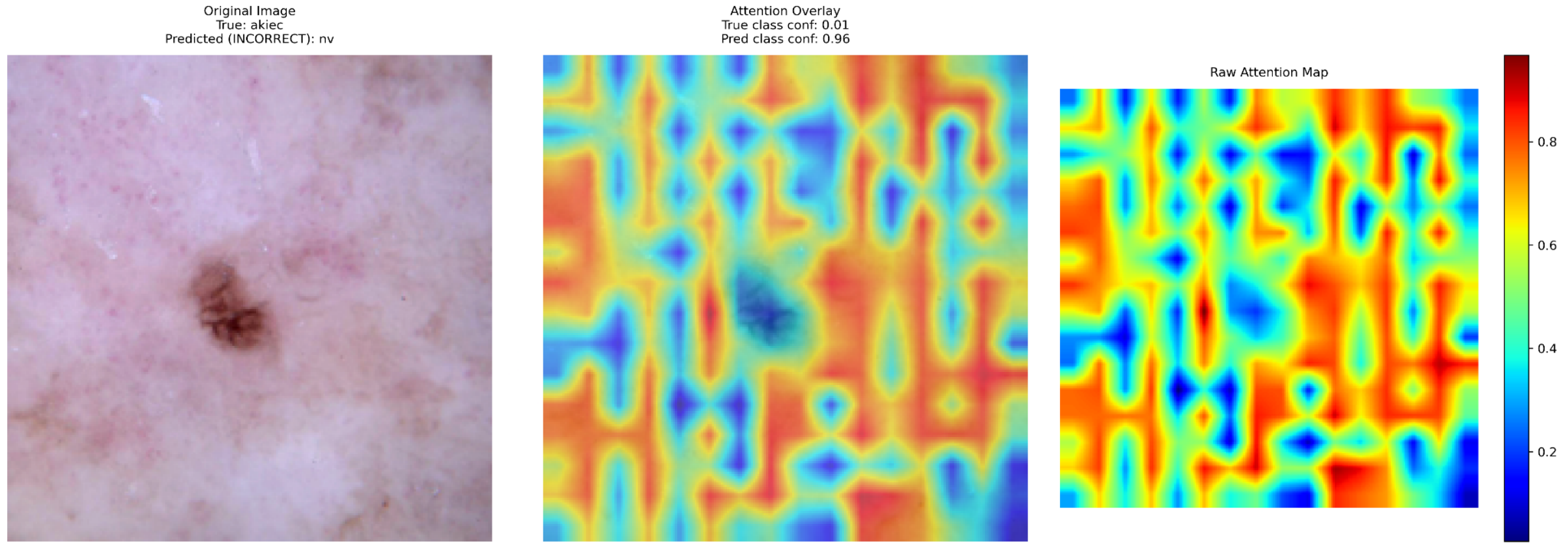

Image 15:

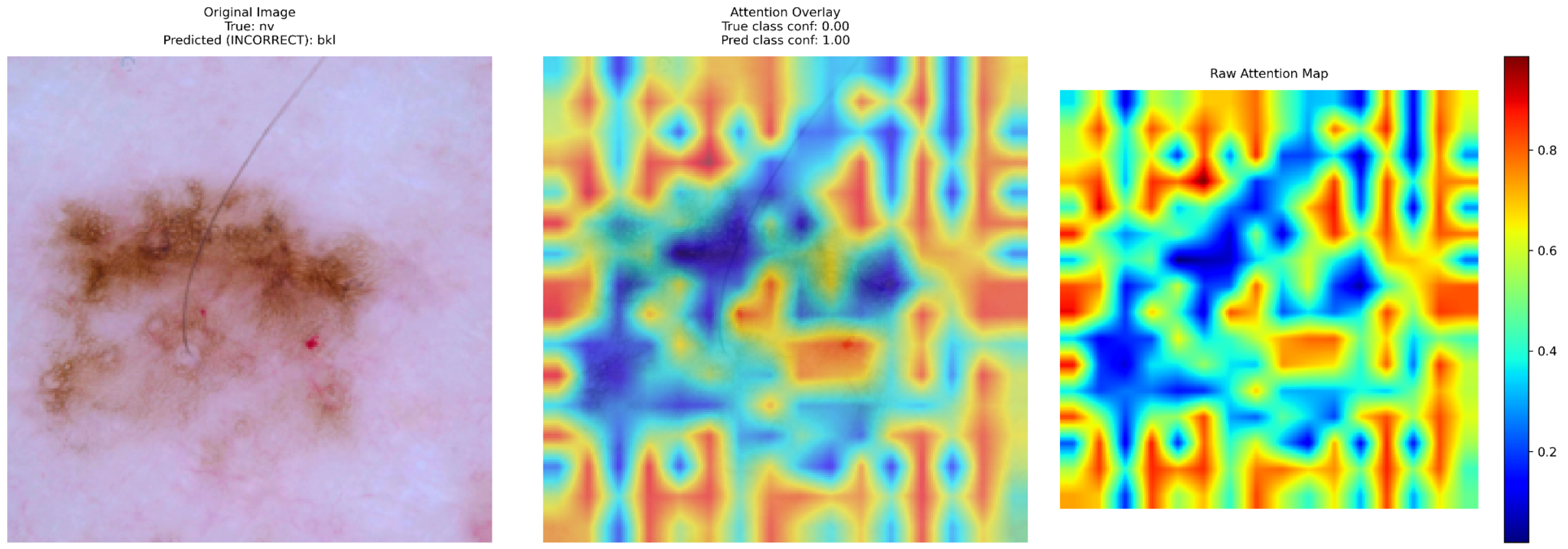

Image 16:

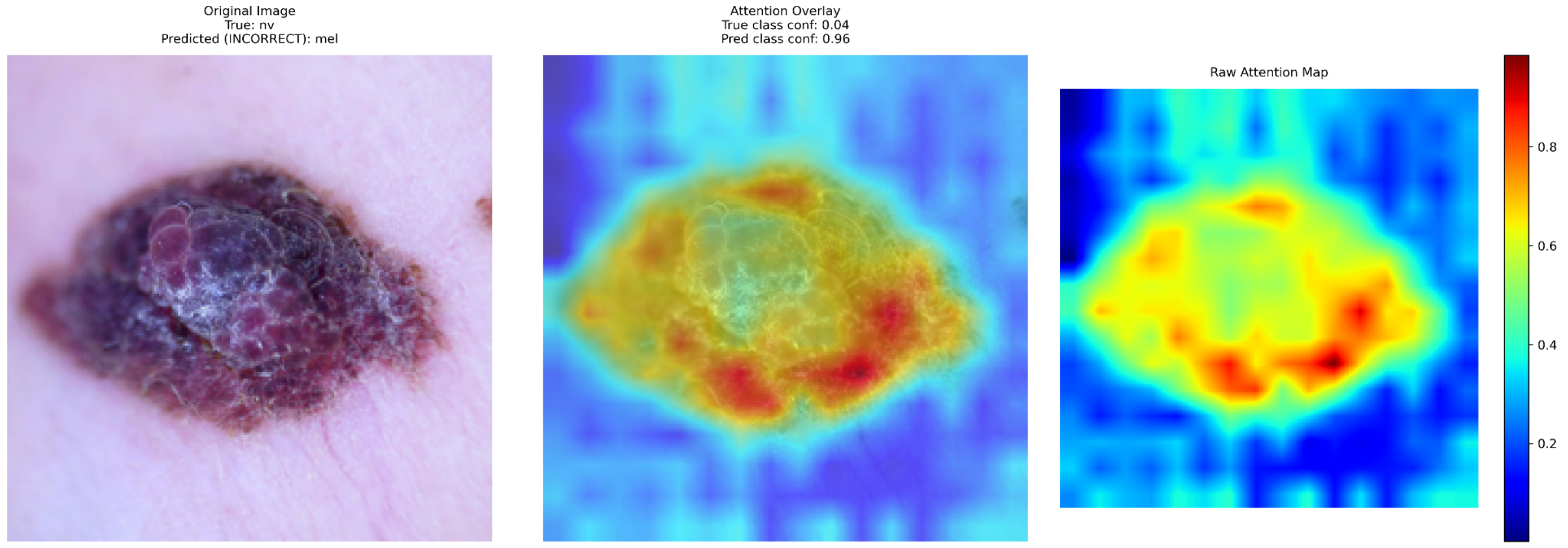

Image 17:

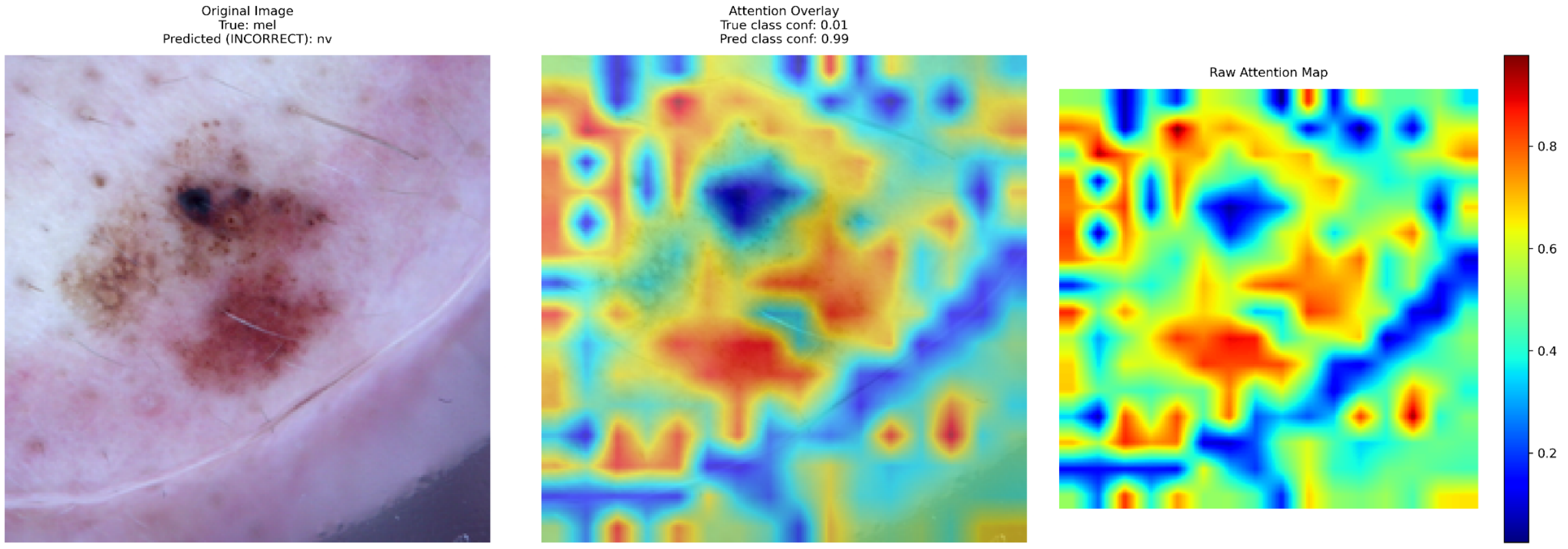

Image 18:

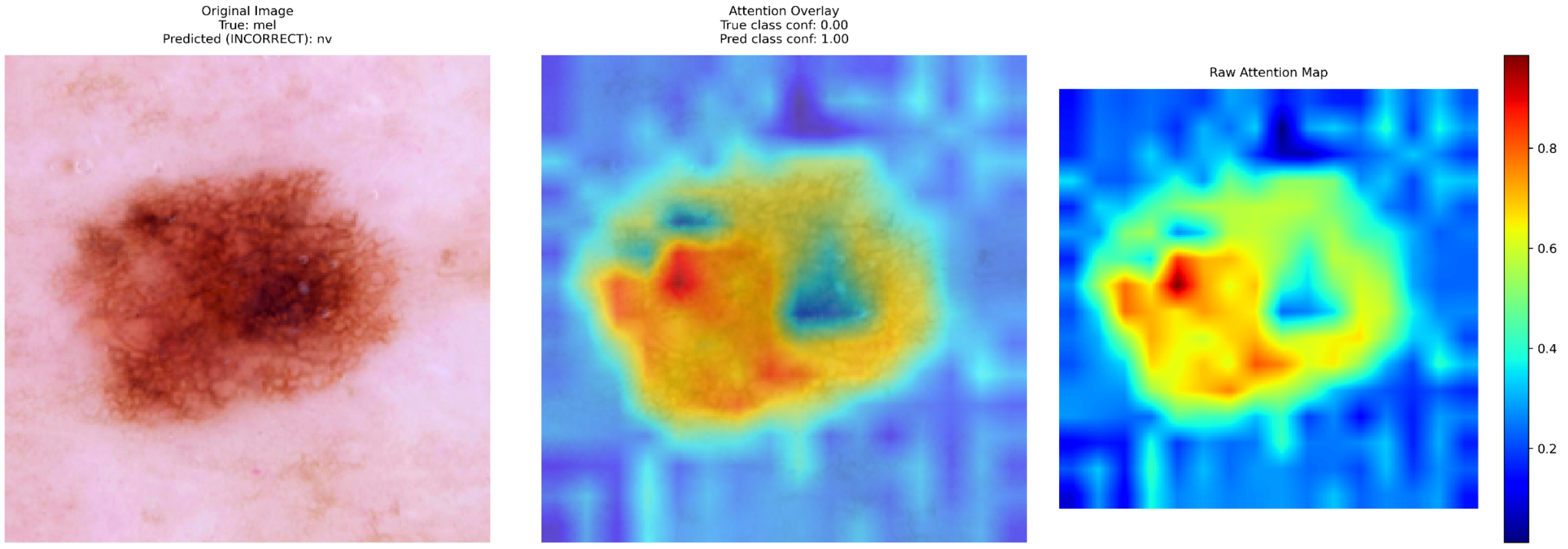

Image 19:

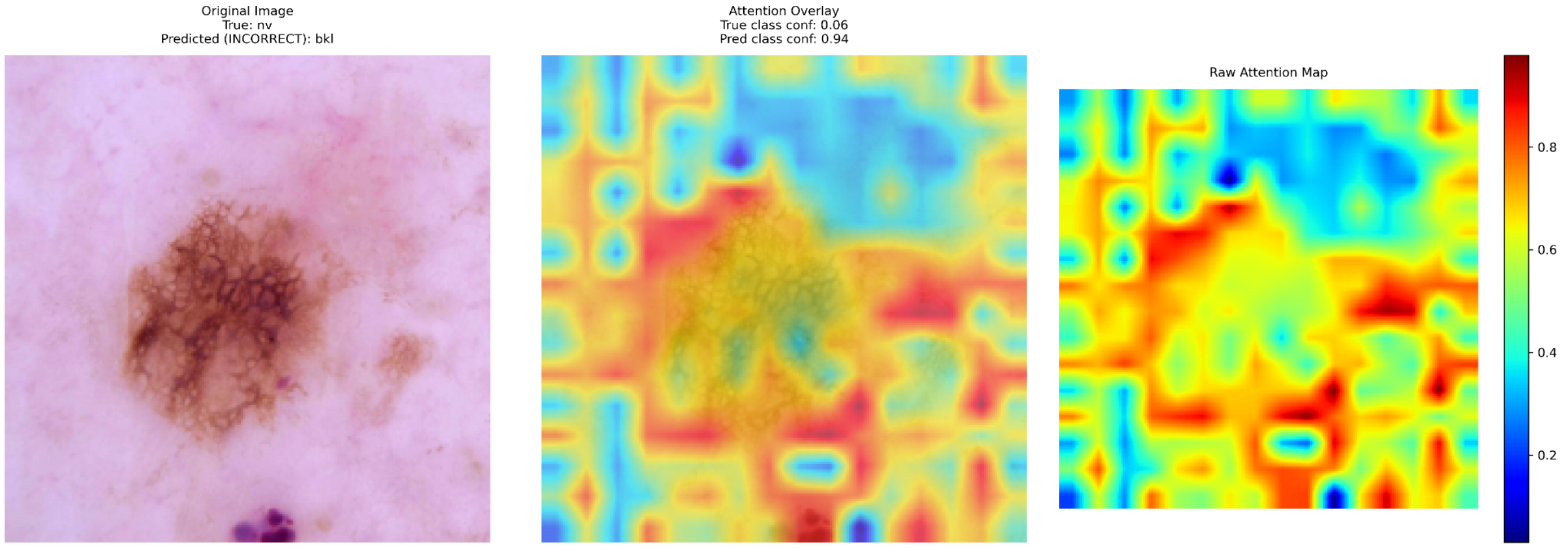

Image 20:

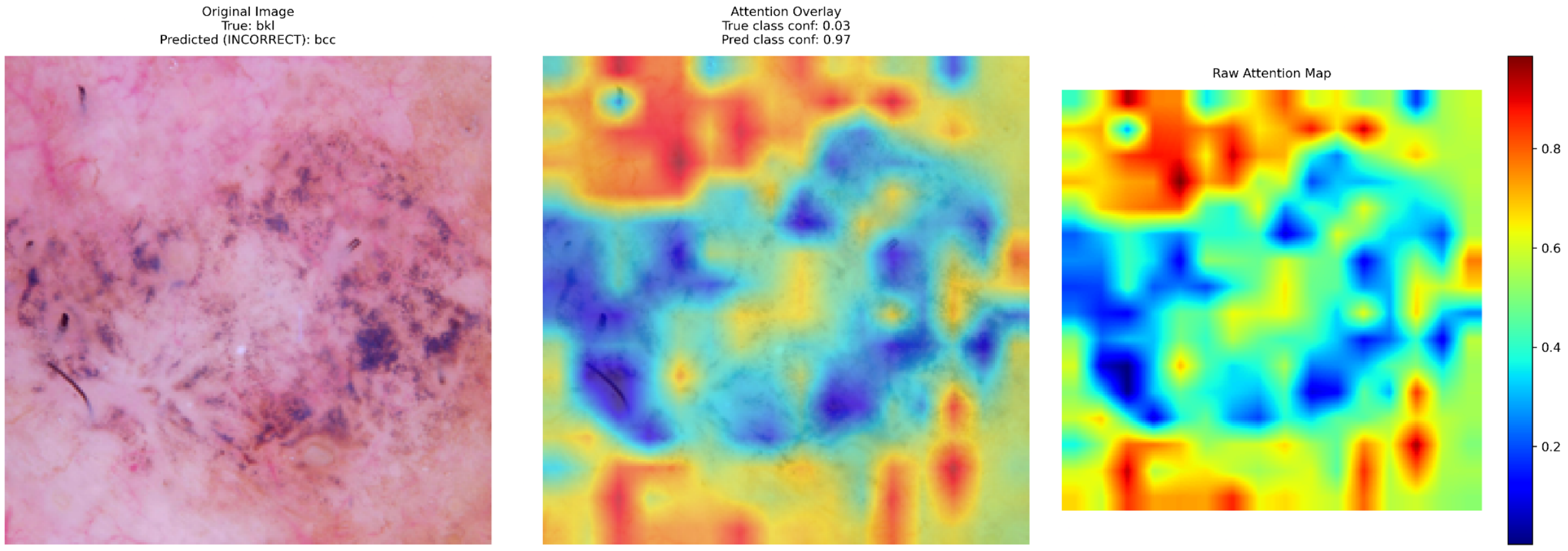

# DermNet Dataset Results

## Correctly classified images

### Image 1:

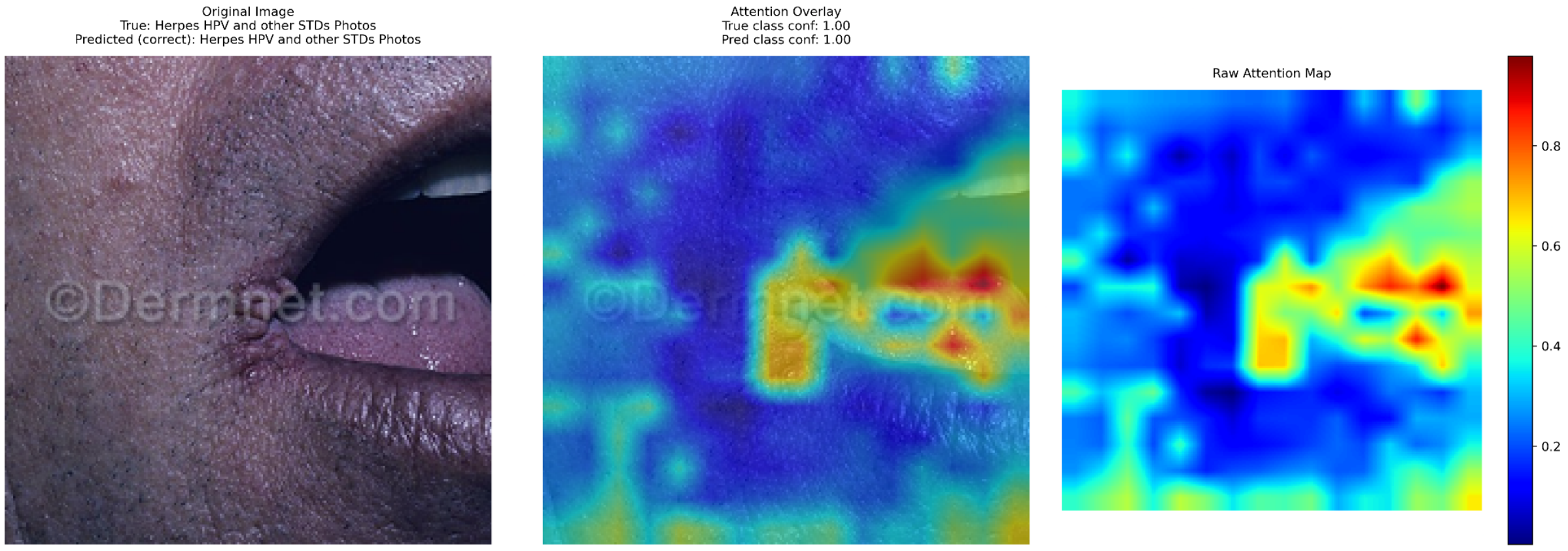

### Image 2:

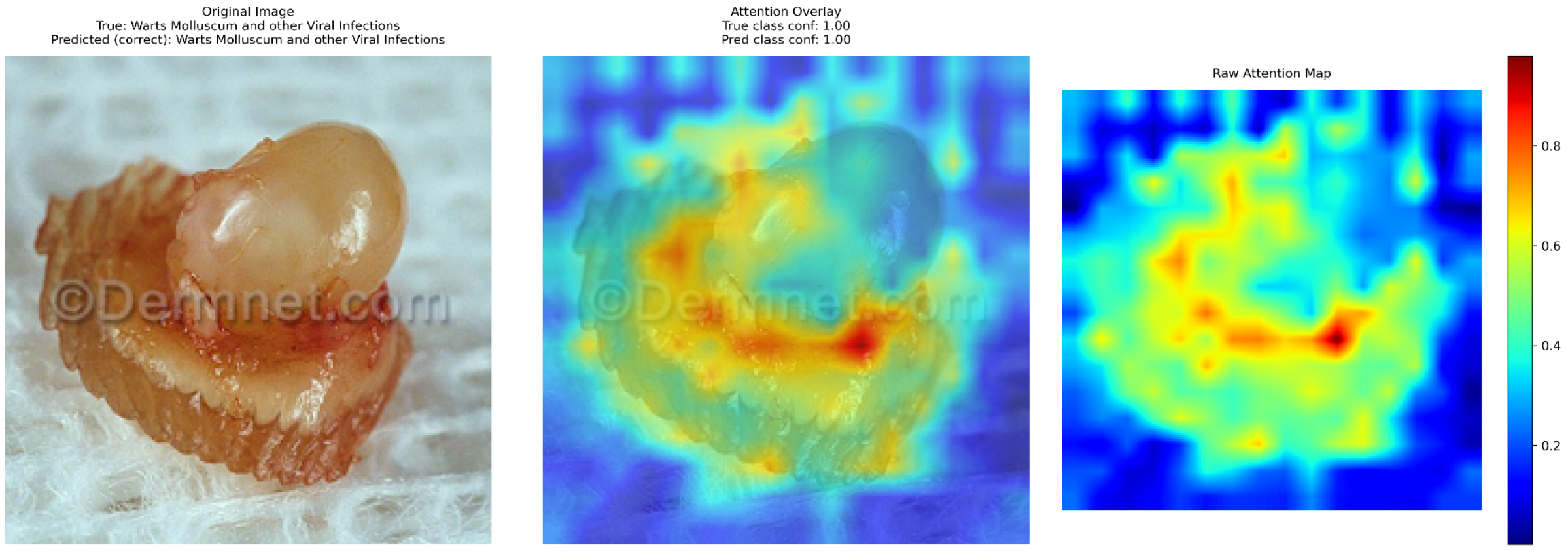

### Image 3:

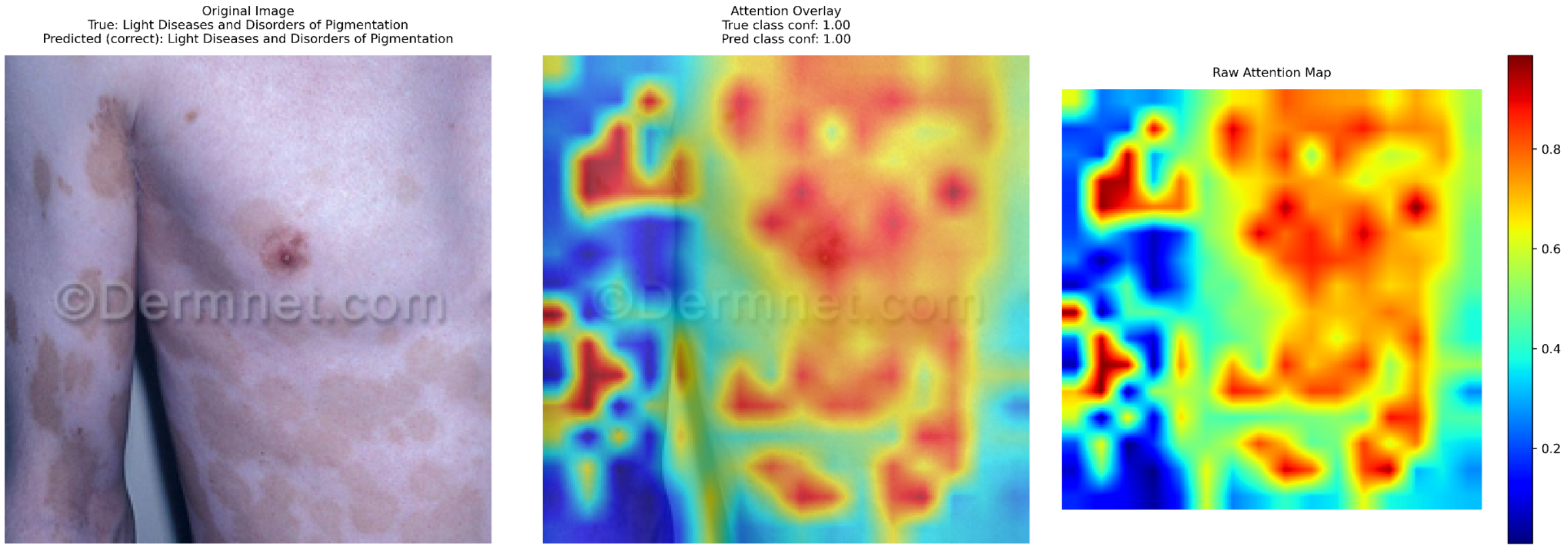

Image 4:

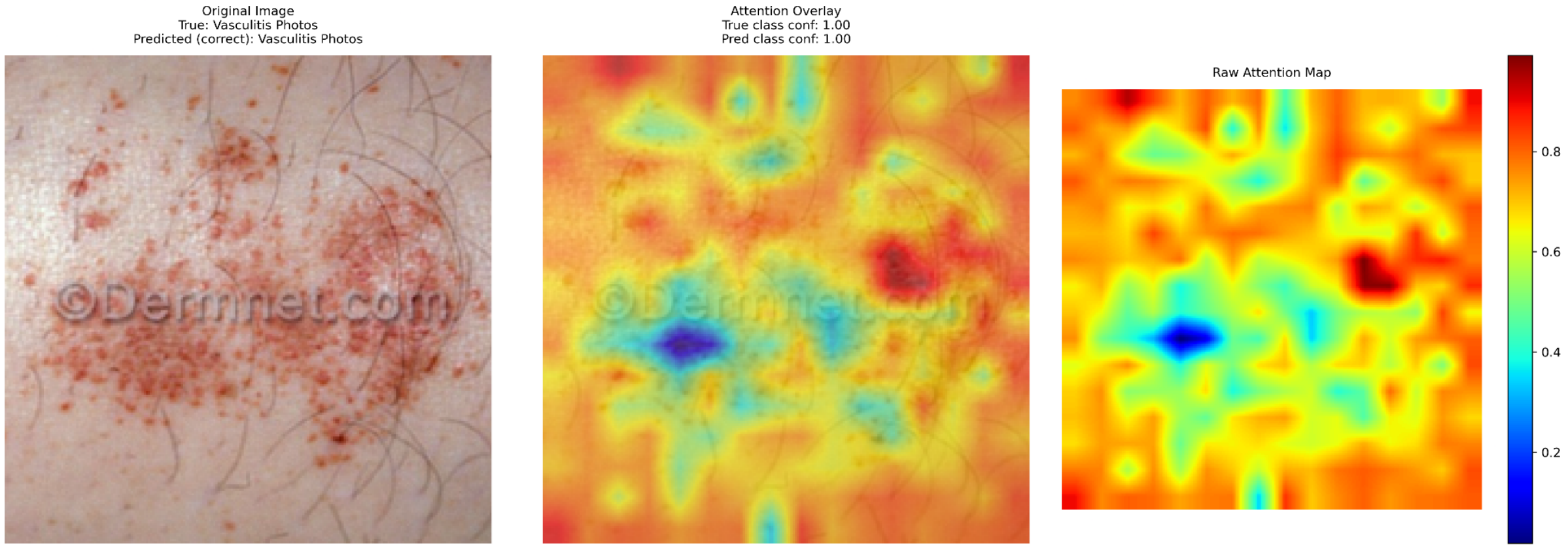

Image 5:

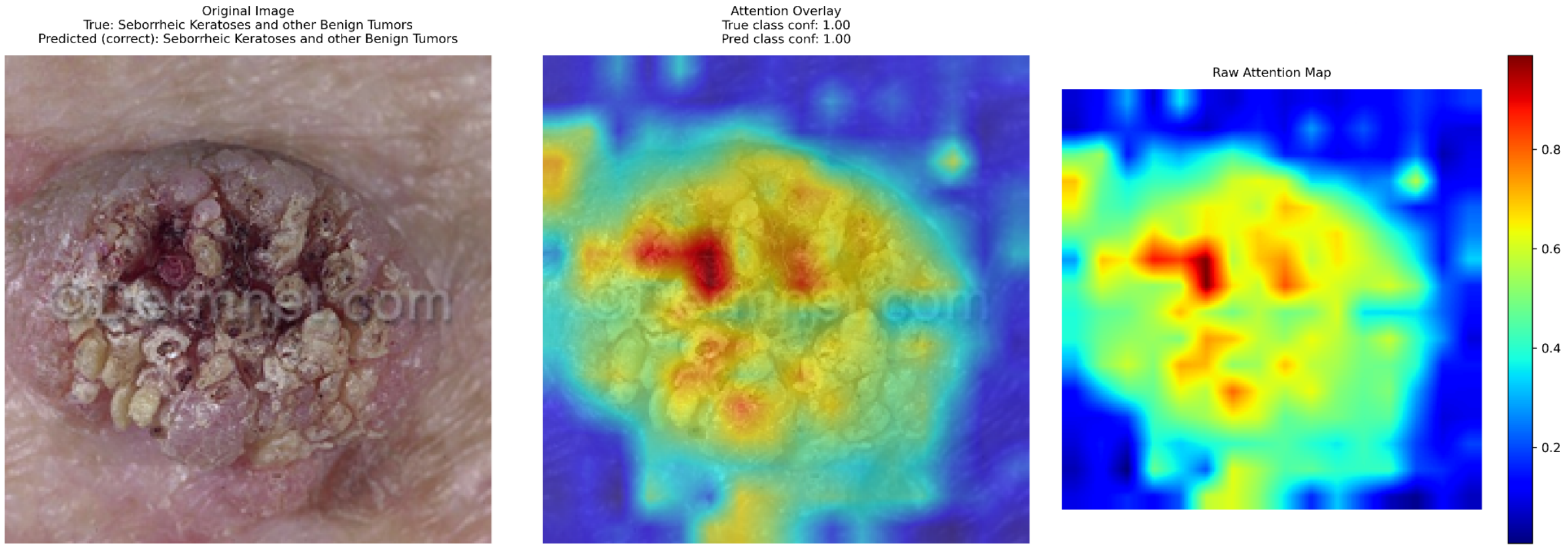

Image 6:

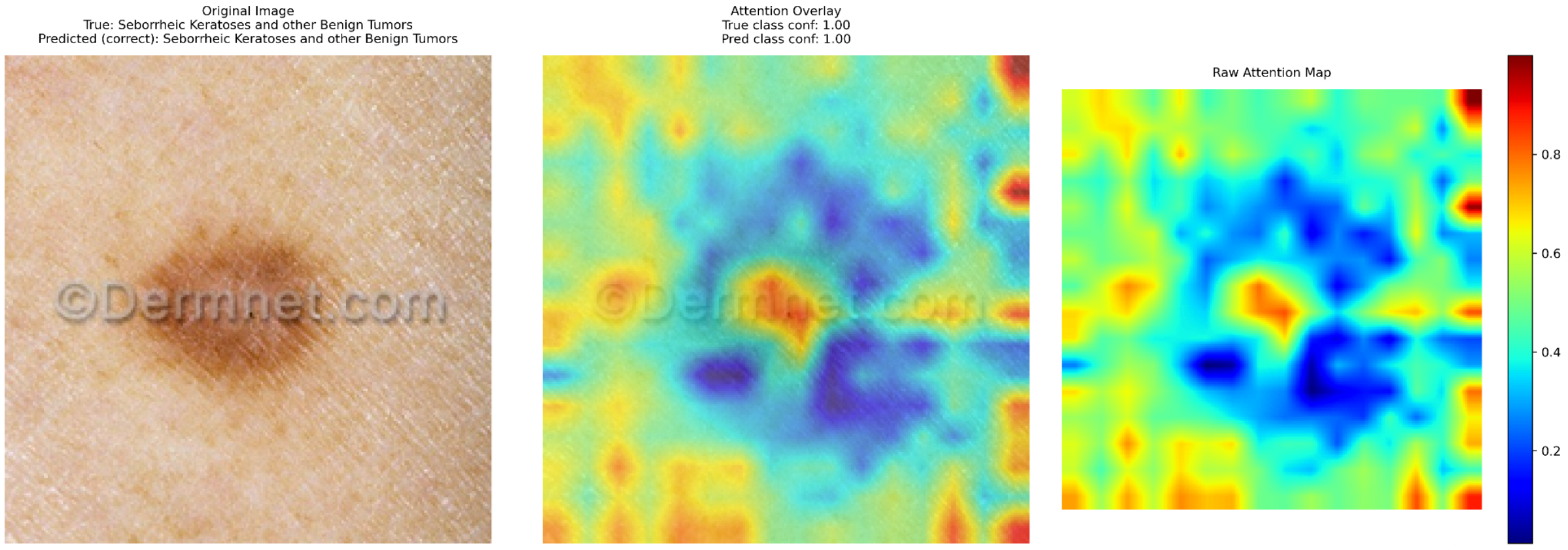

Image 7:

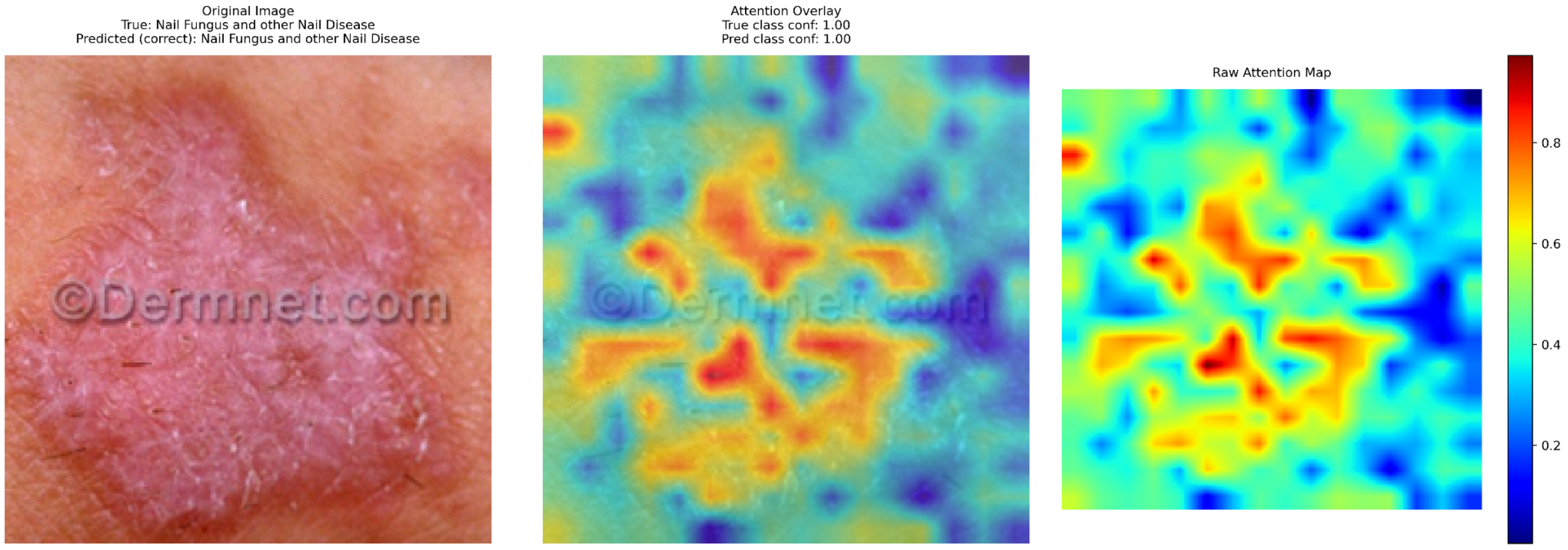

Image 8:

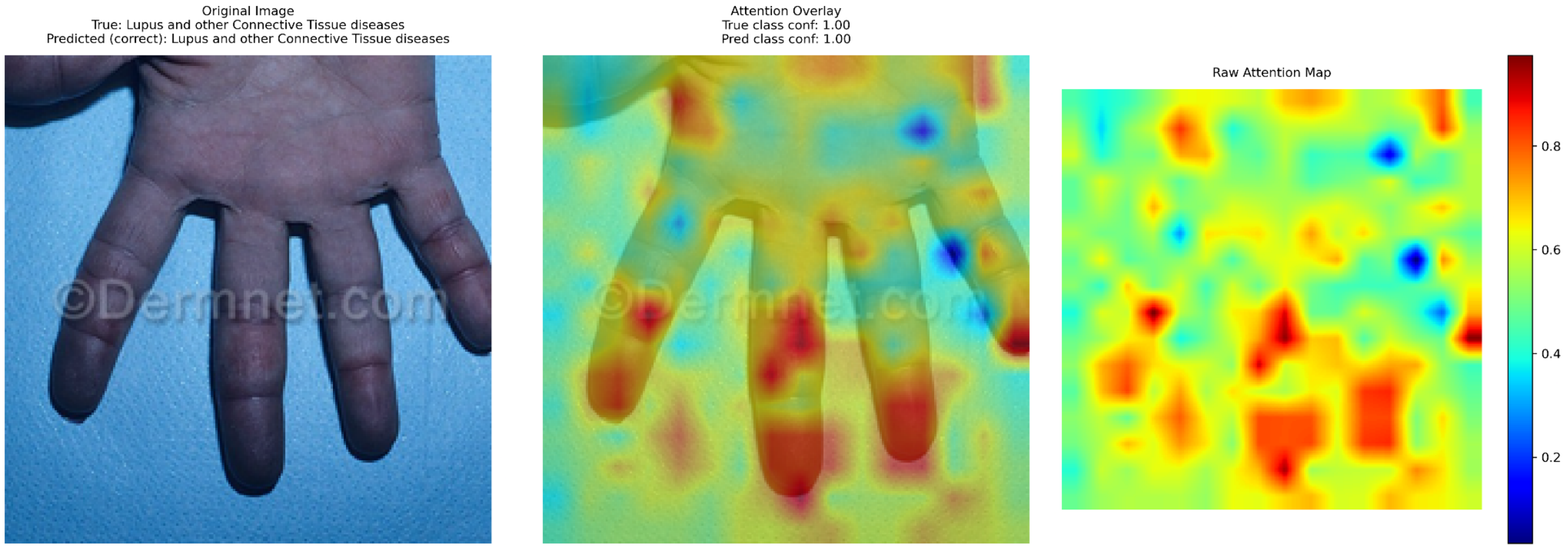

Image 9:

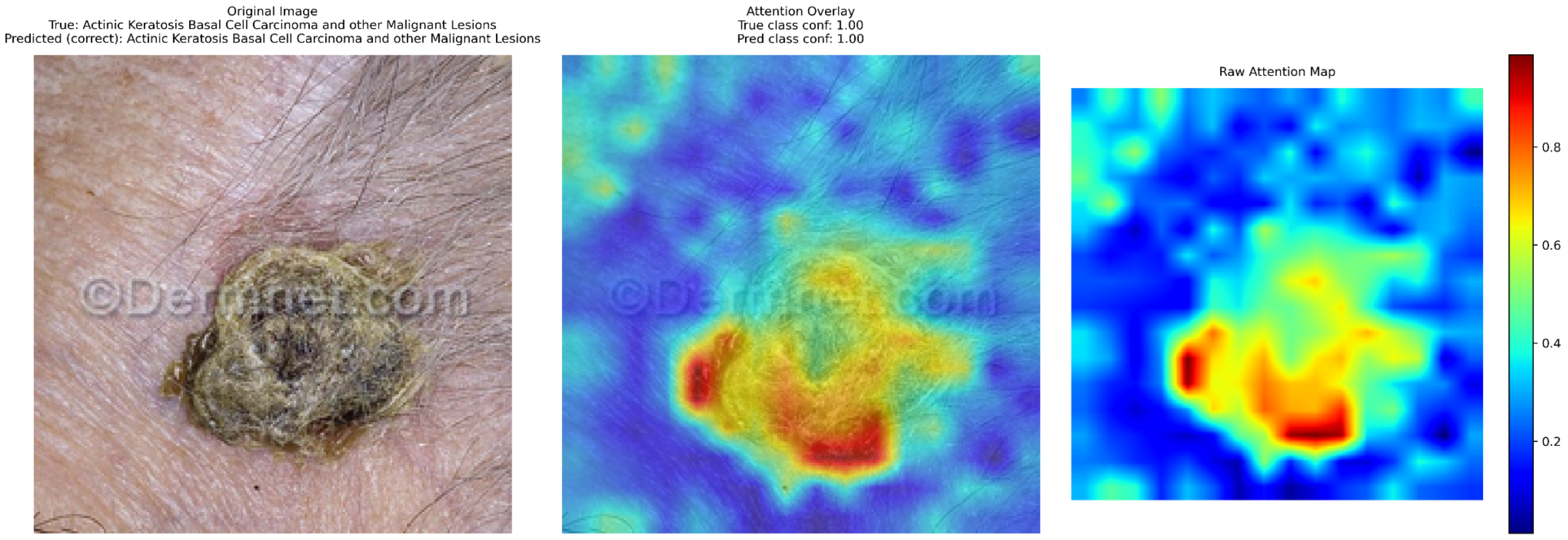

Image 10:

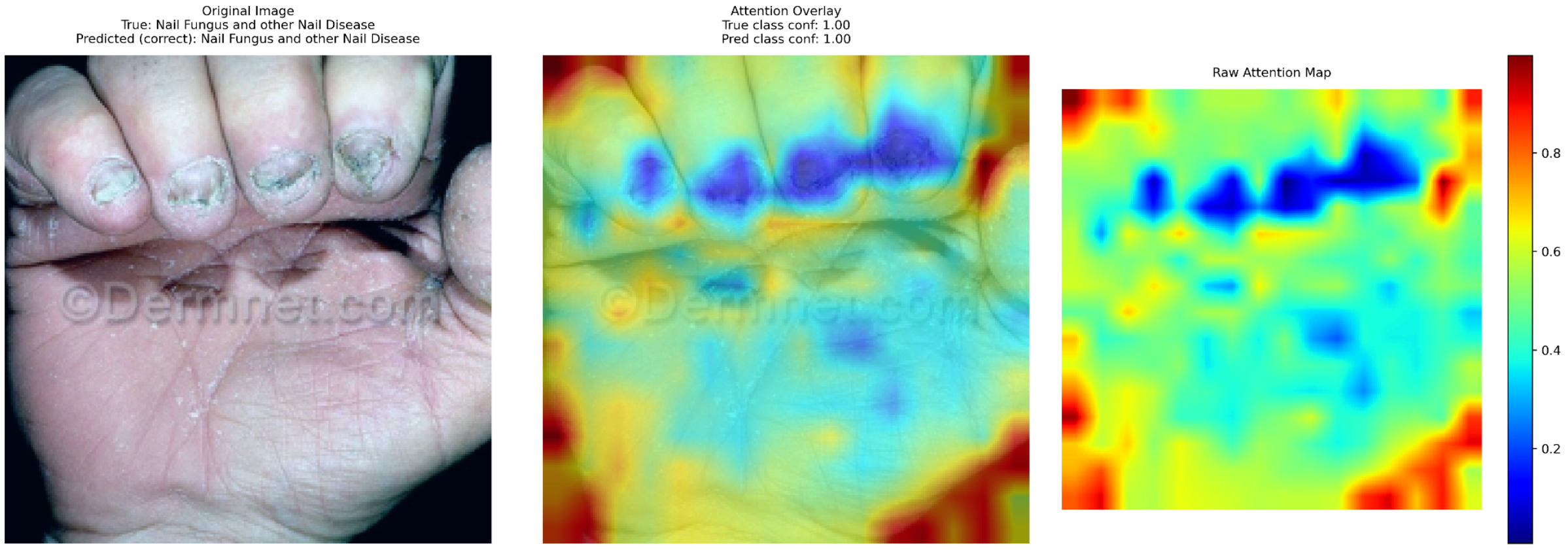

## Incorrectly classified images

### Image 11:

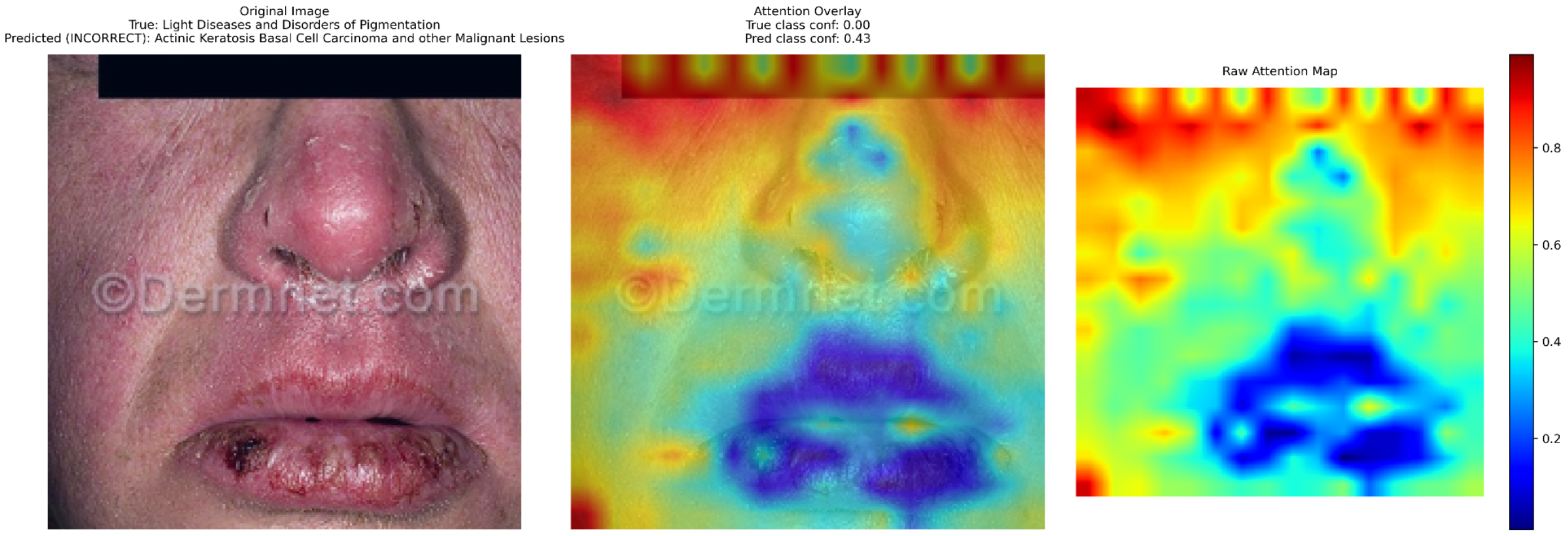

### Image 12:

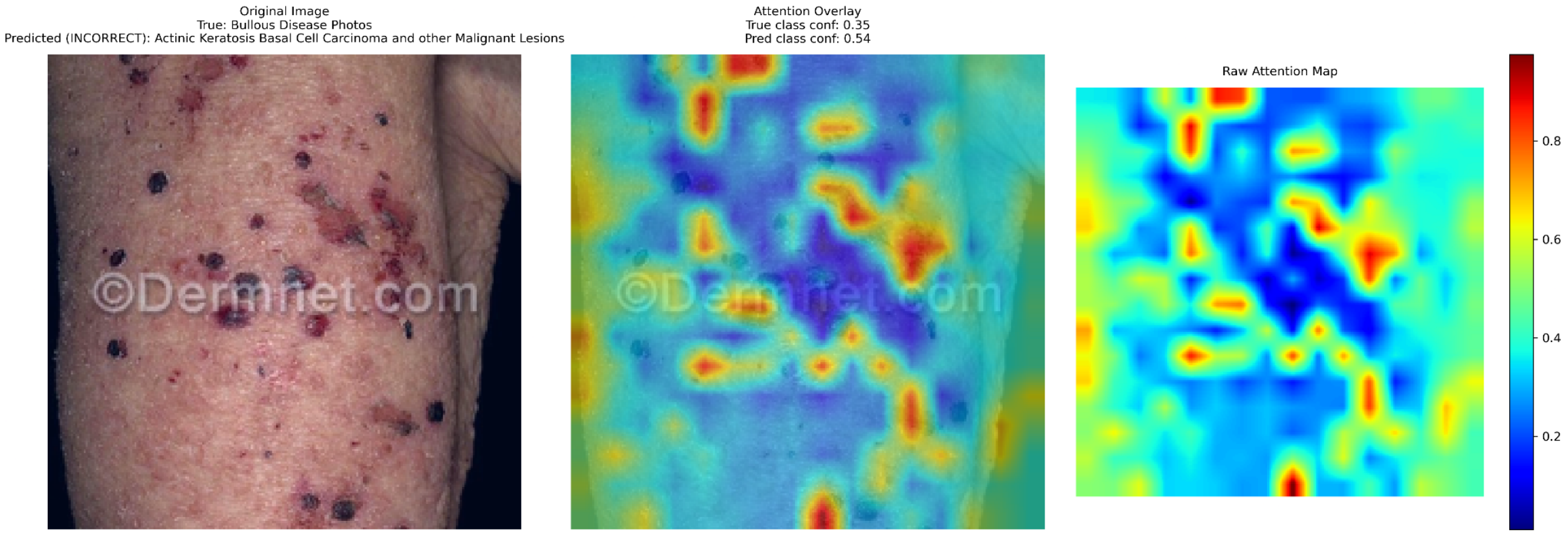

### Image 13:

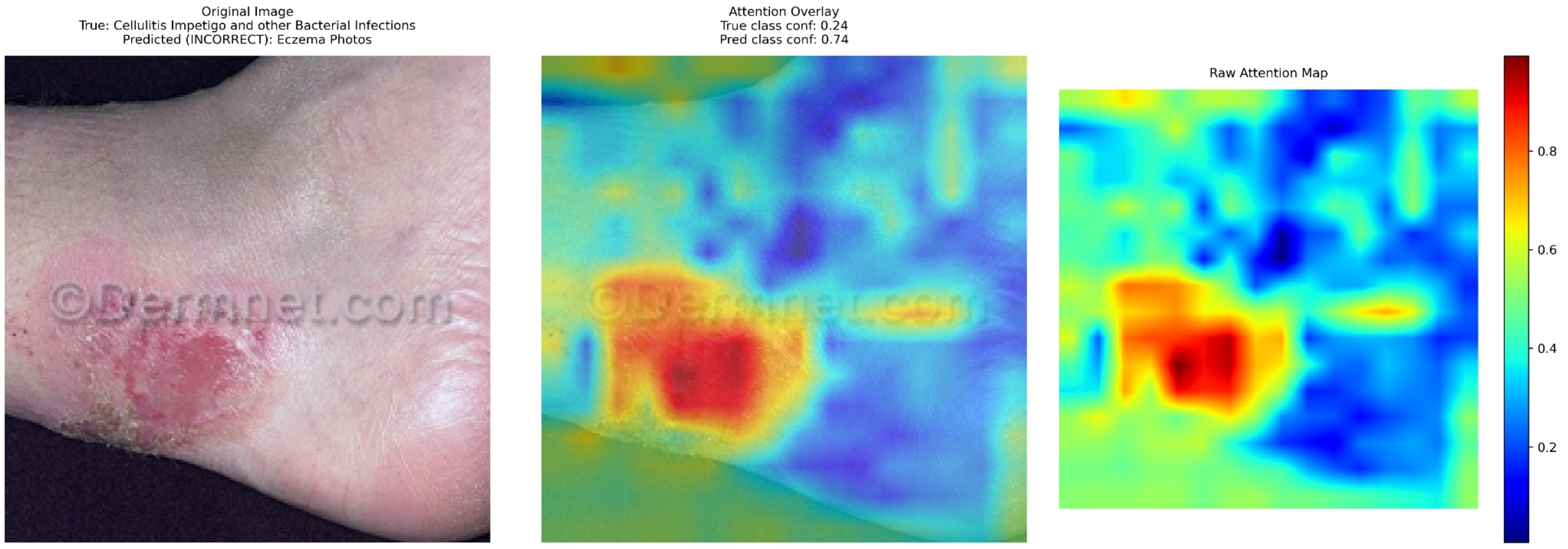

Image 14:

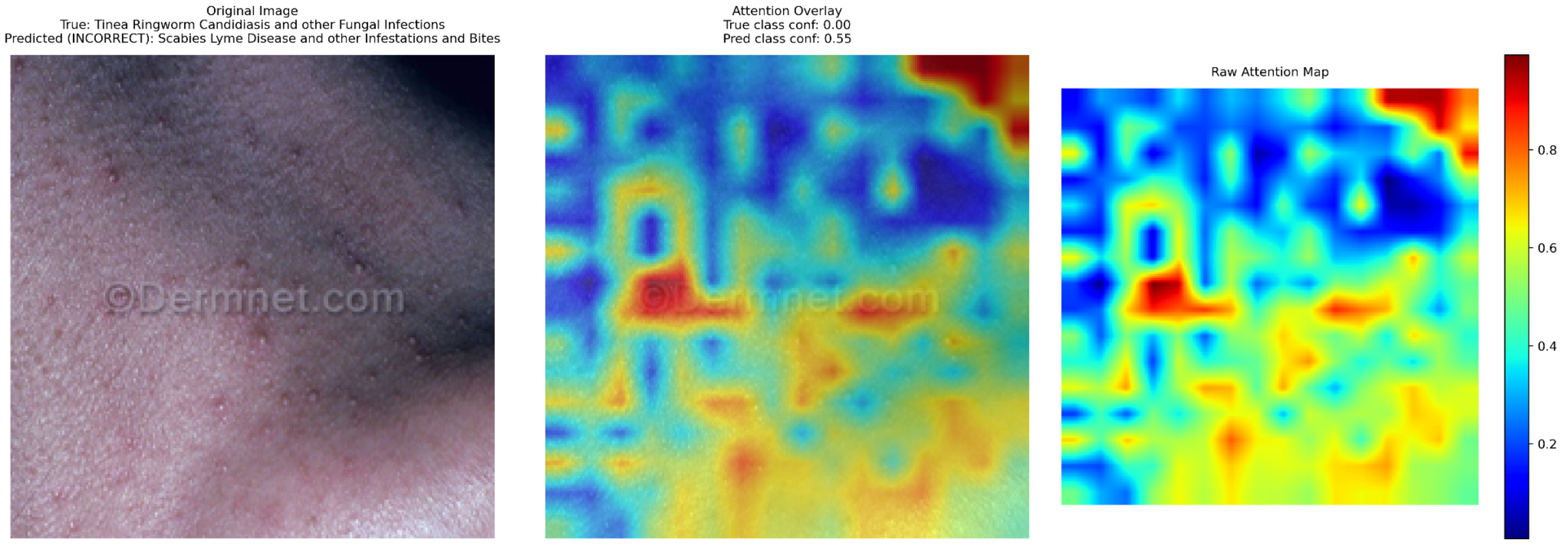

Image 15:

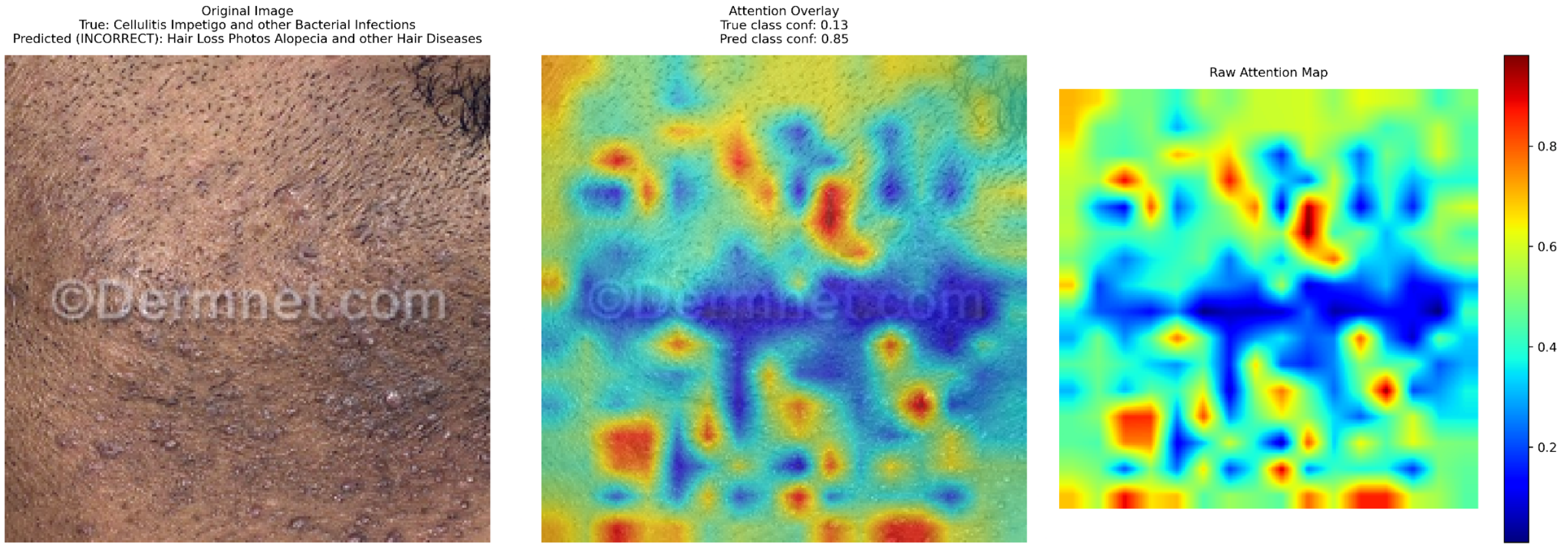

Image 16:

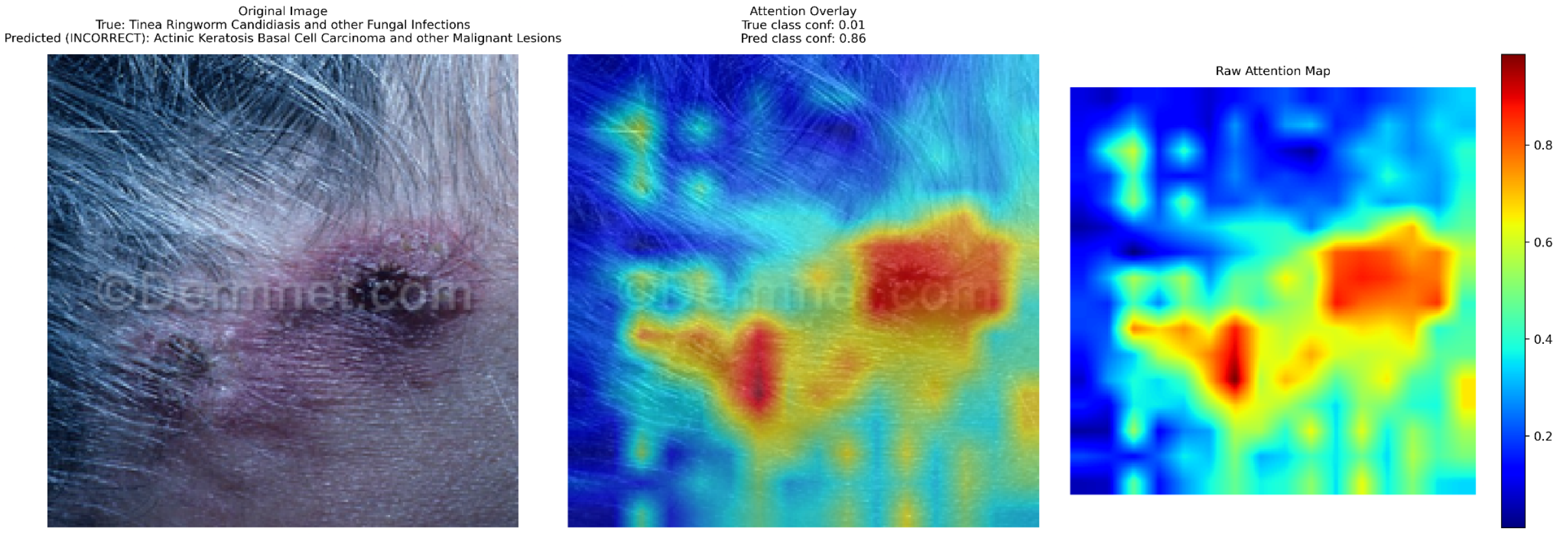

Image 17:

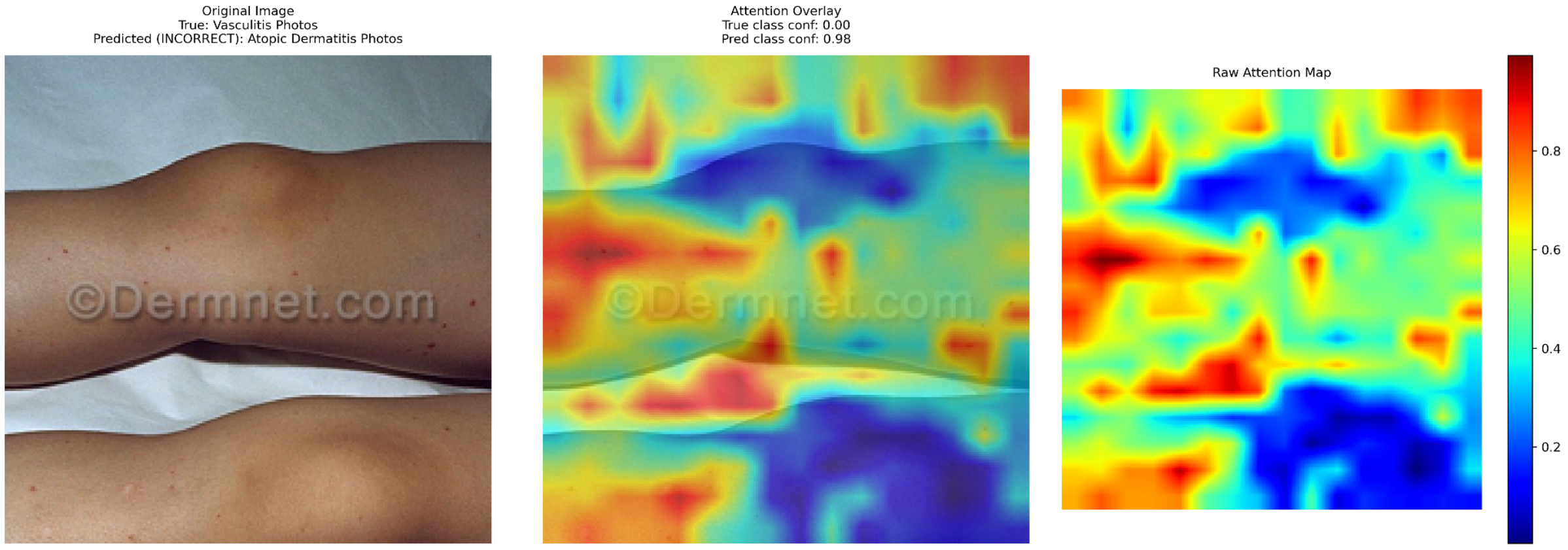

Image 18:

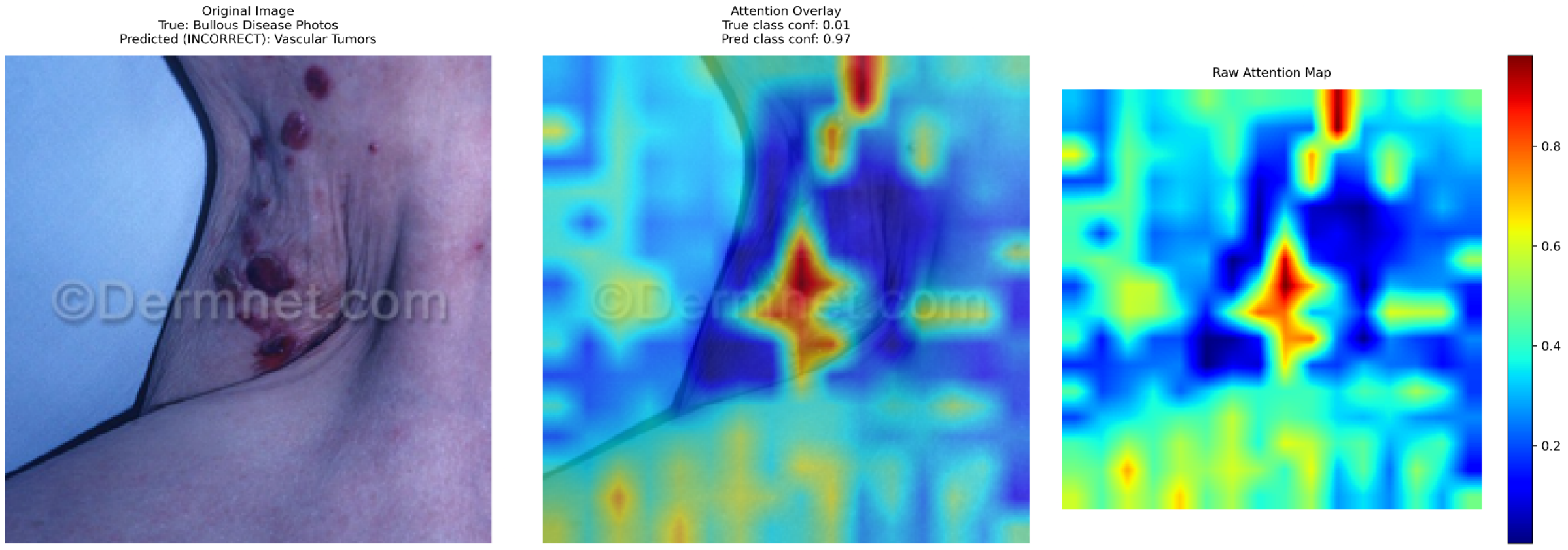

Image 19:

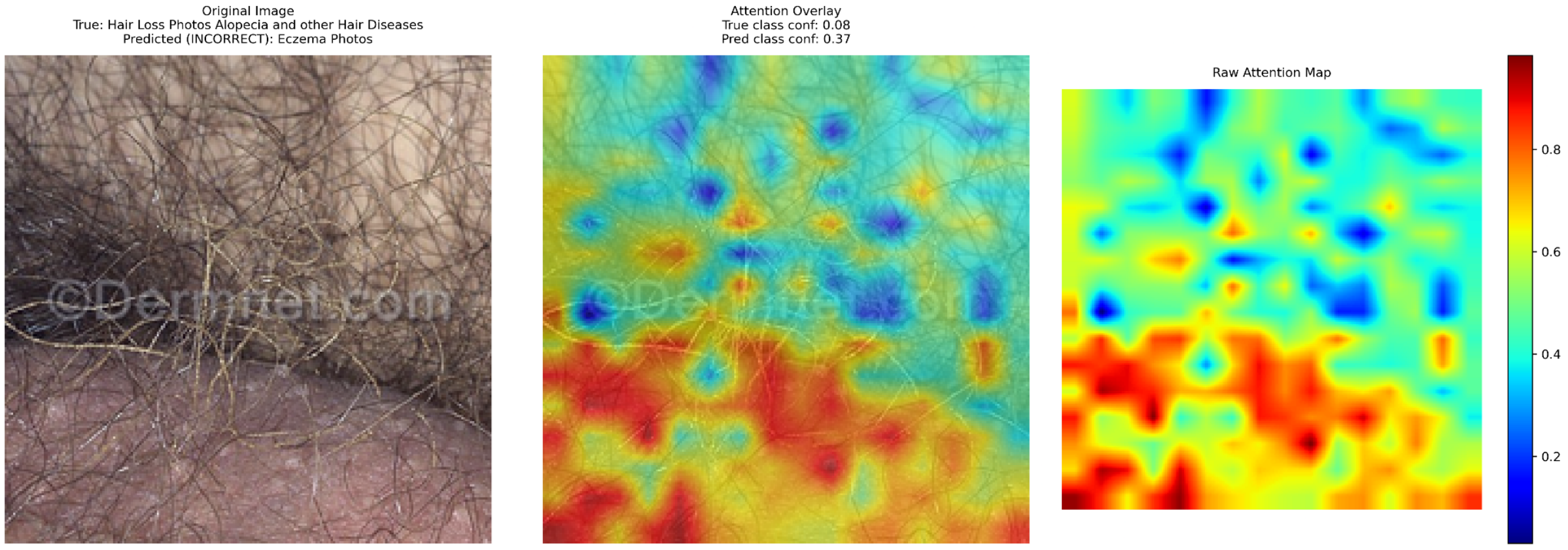

Image 20:

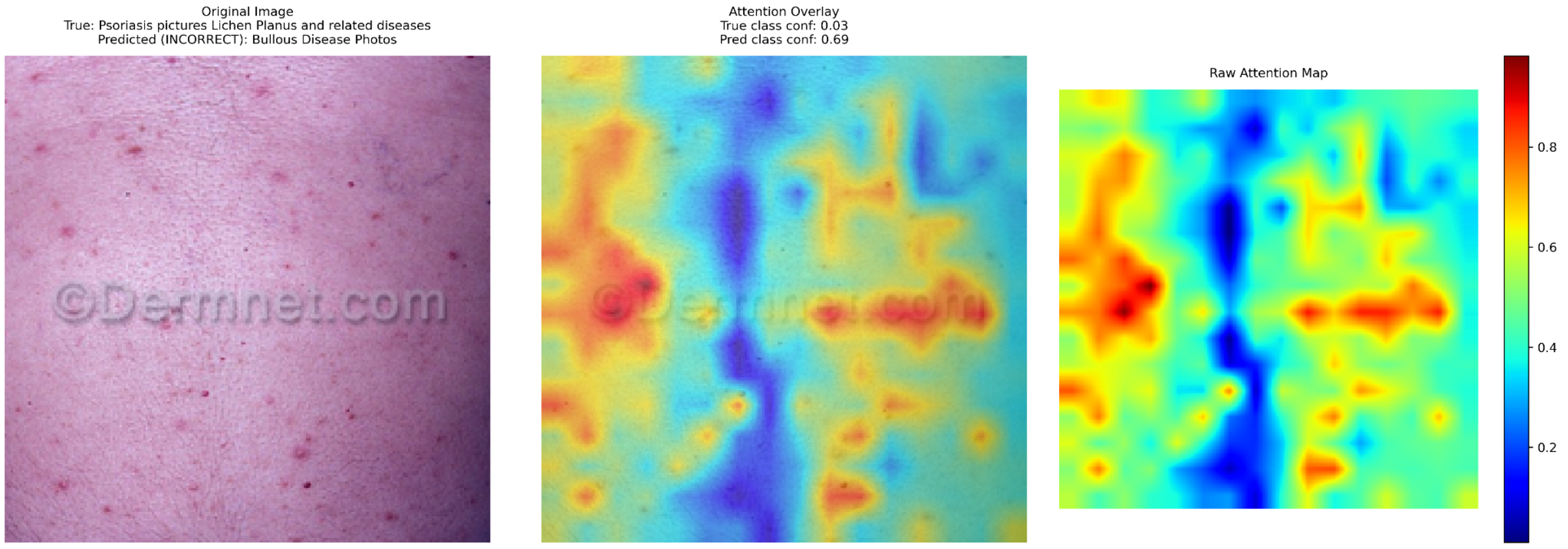

# ISIC Atlas Dataset Results

## Correctly classified images

### Image 1:

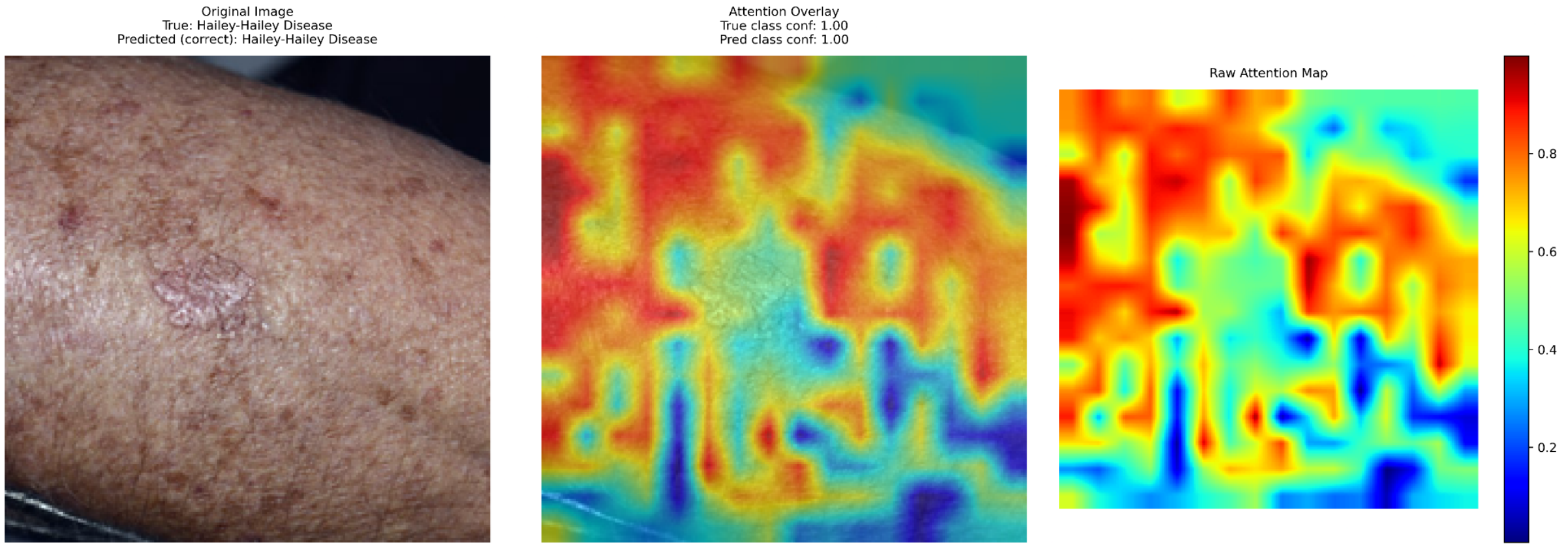

### Image 2:

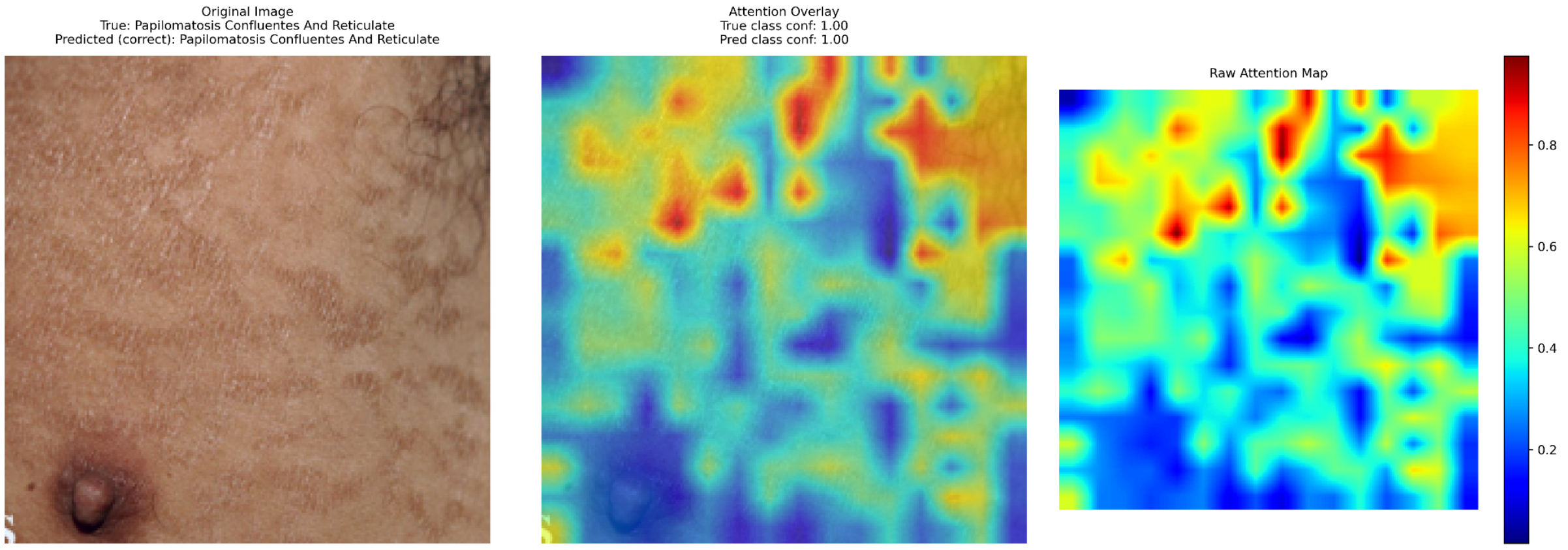

### Image 3:

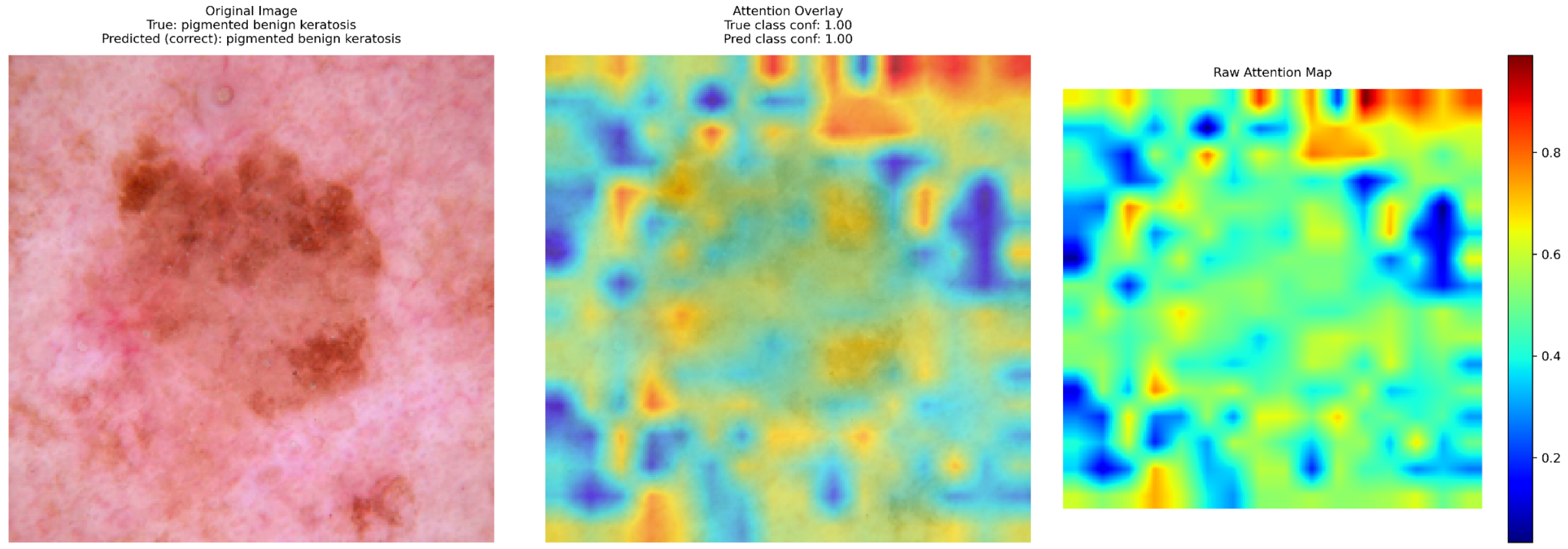

Image 4:

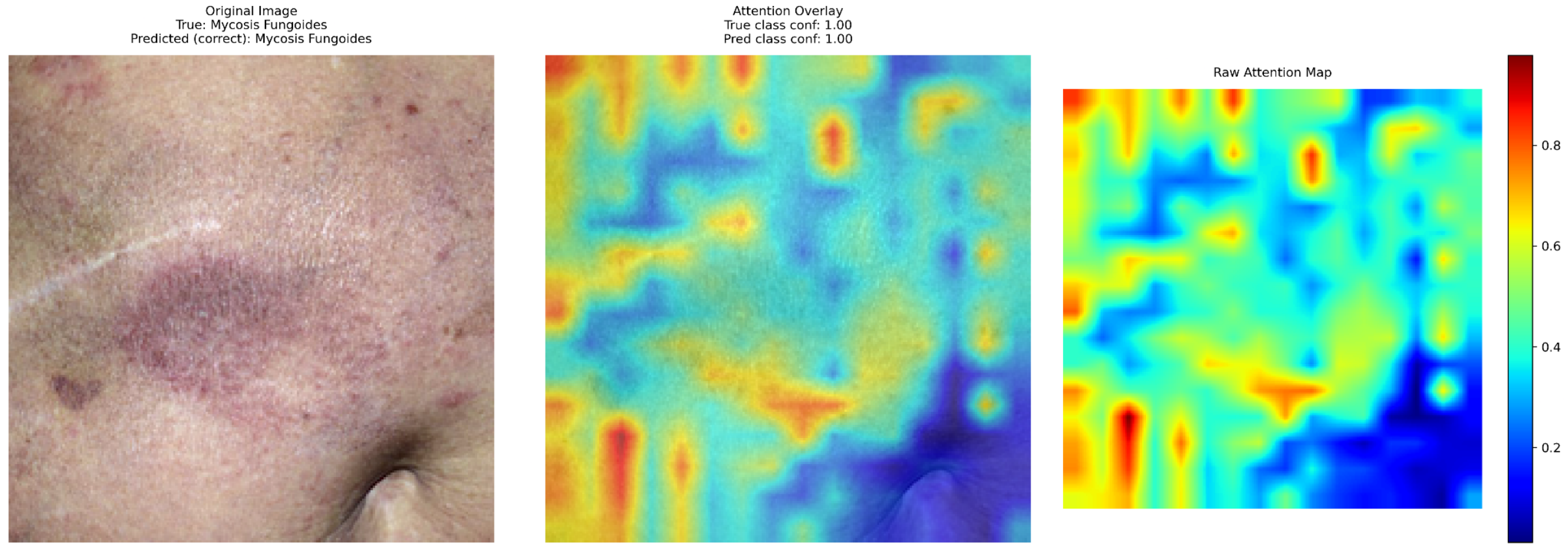

Image 5:

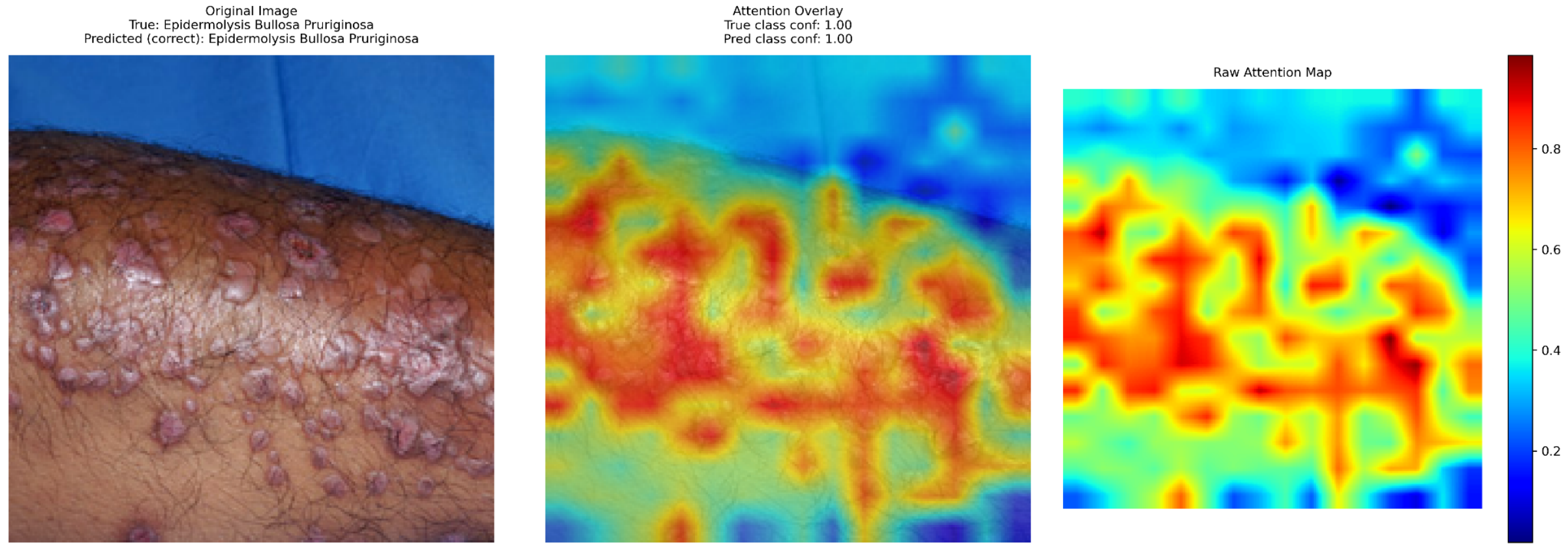

Image 6:

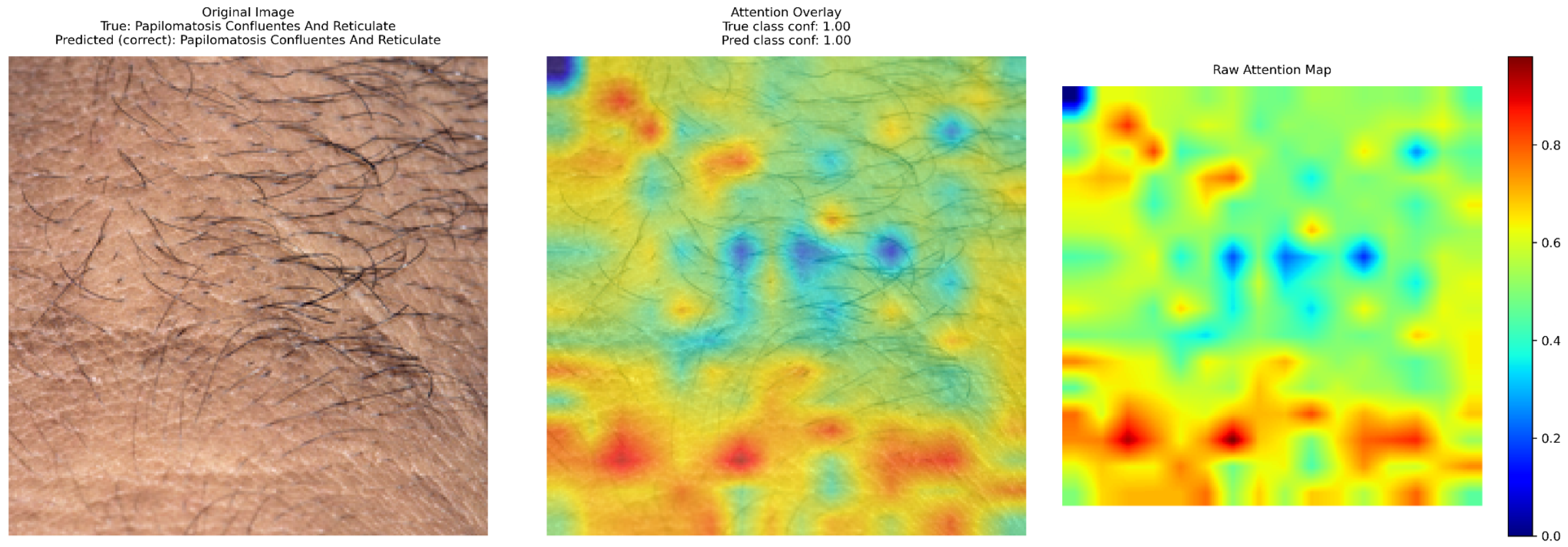

Image 7:

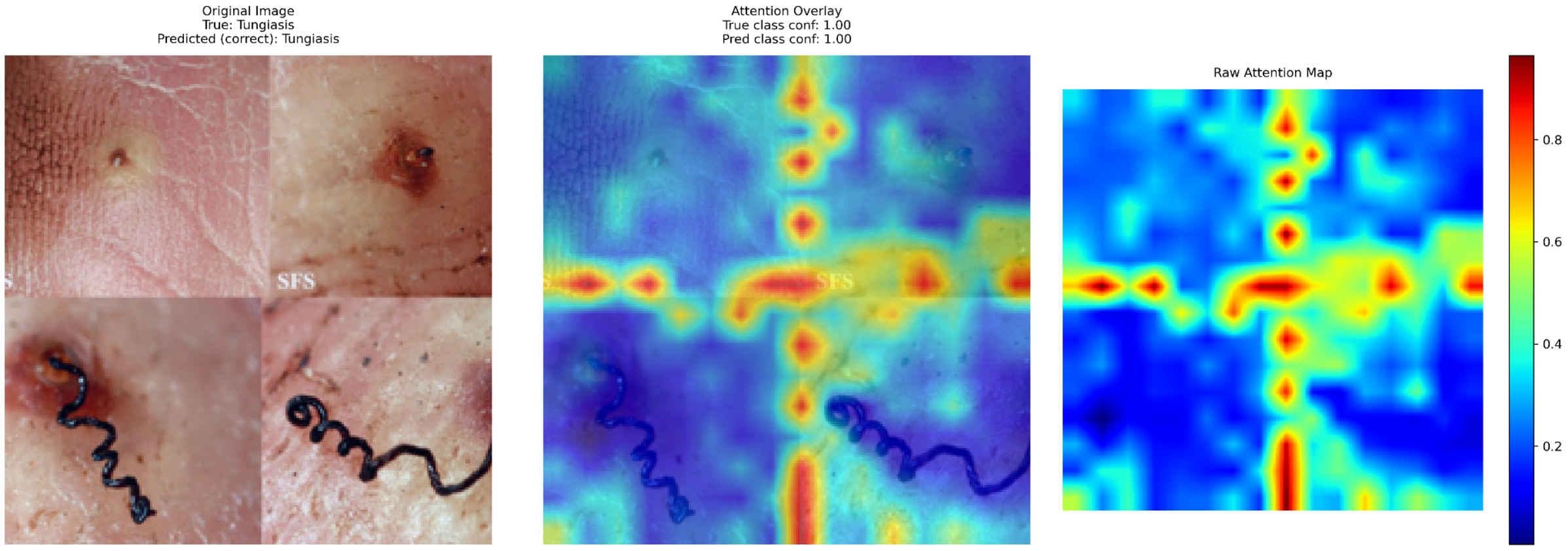

Image 8:

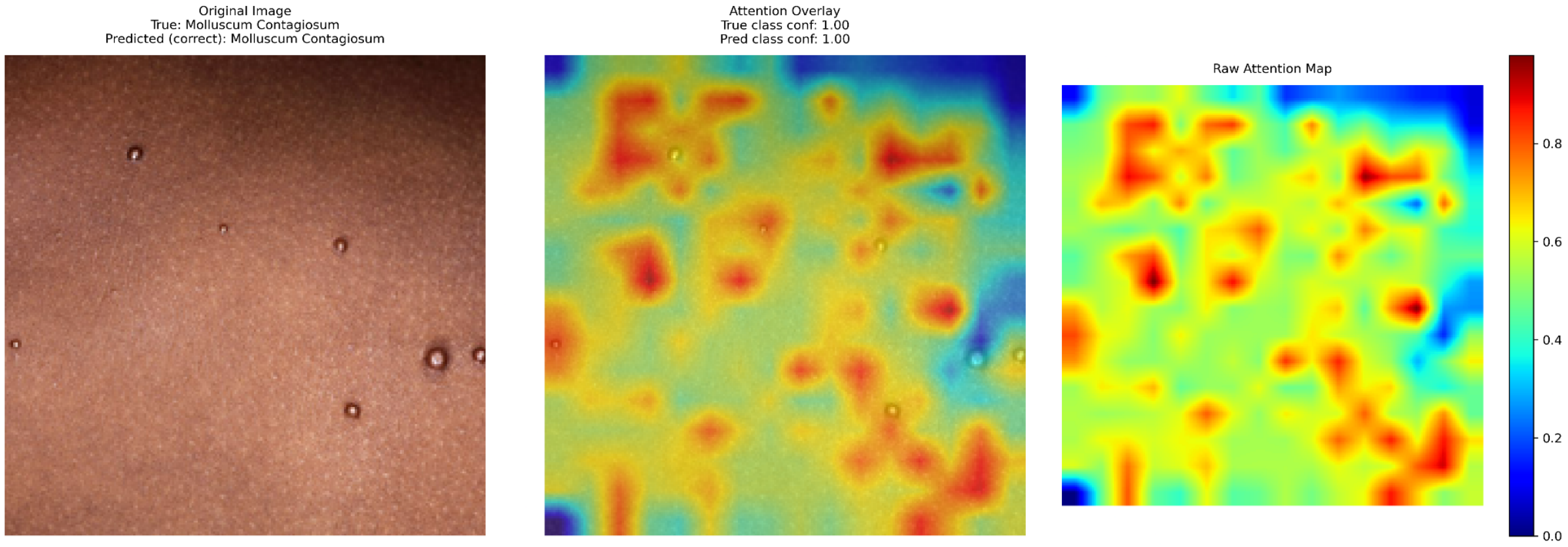

Image 9:

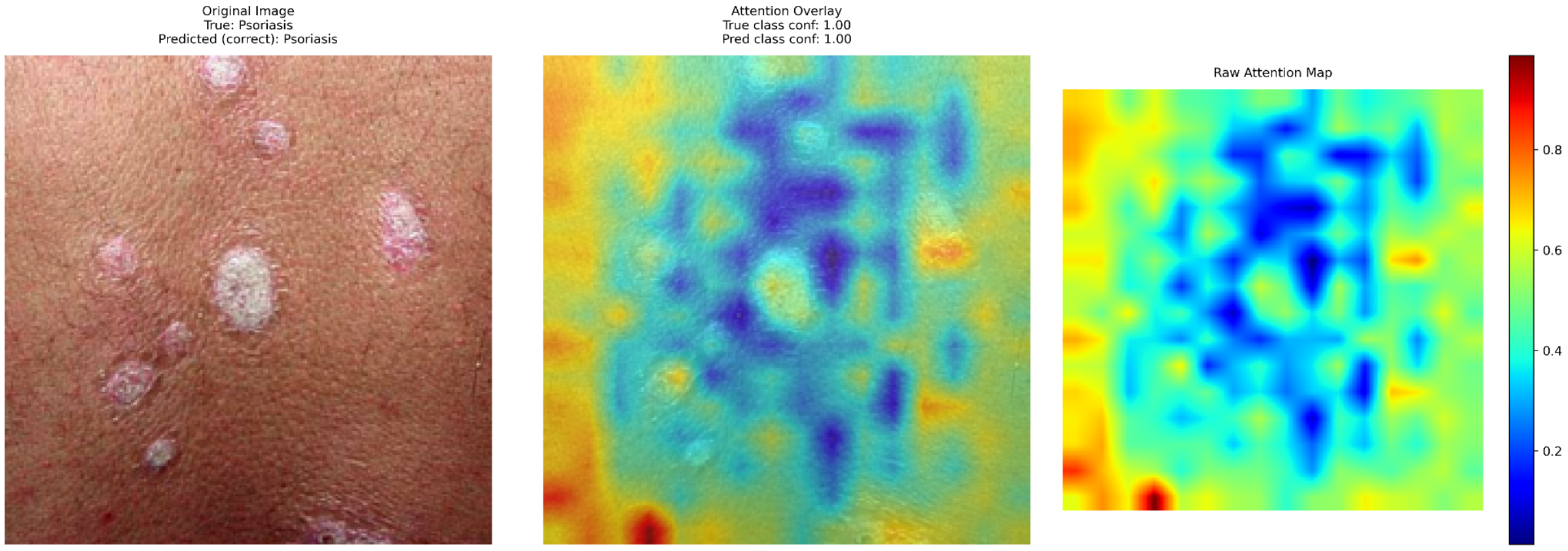

Image 10:

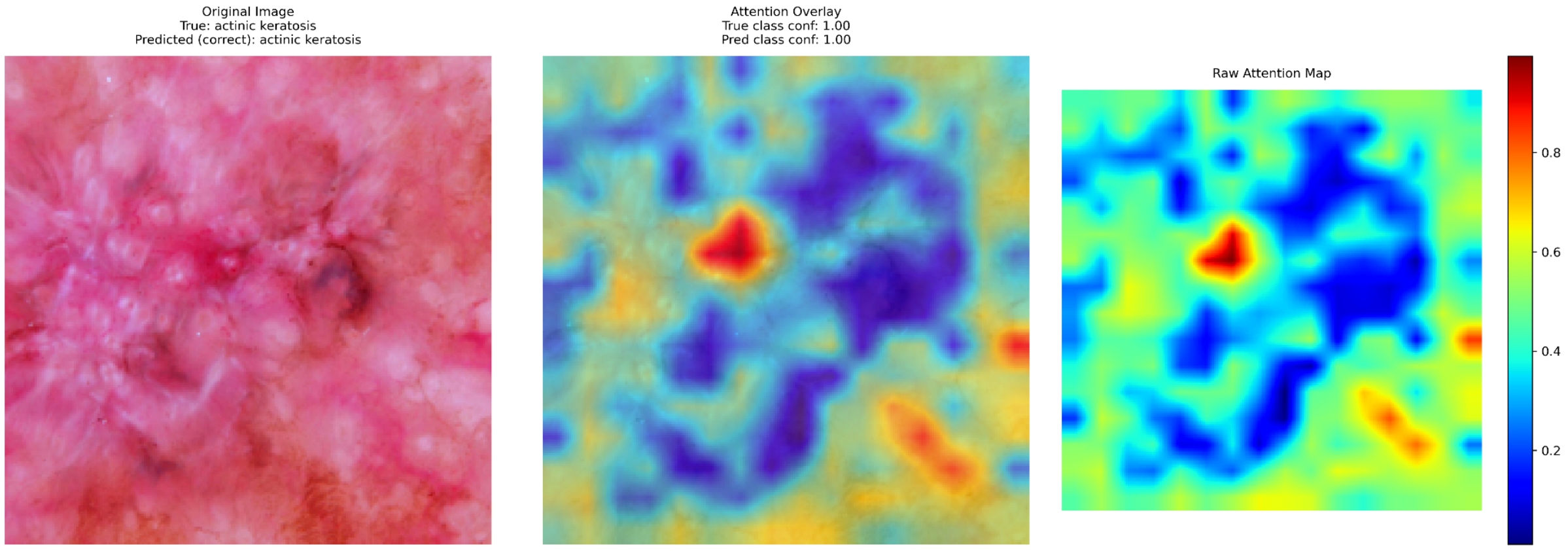

# Incorrectly classified images

## Image 11:

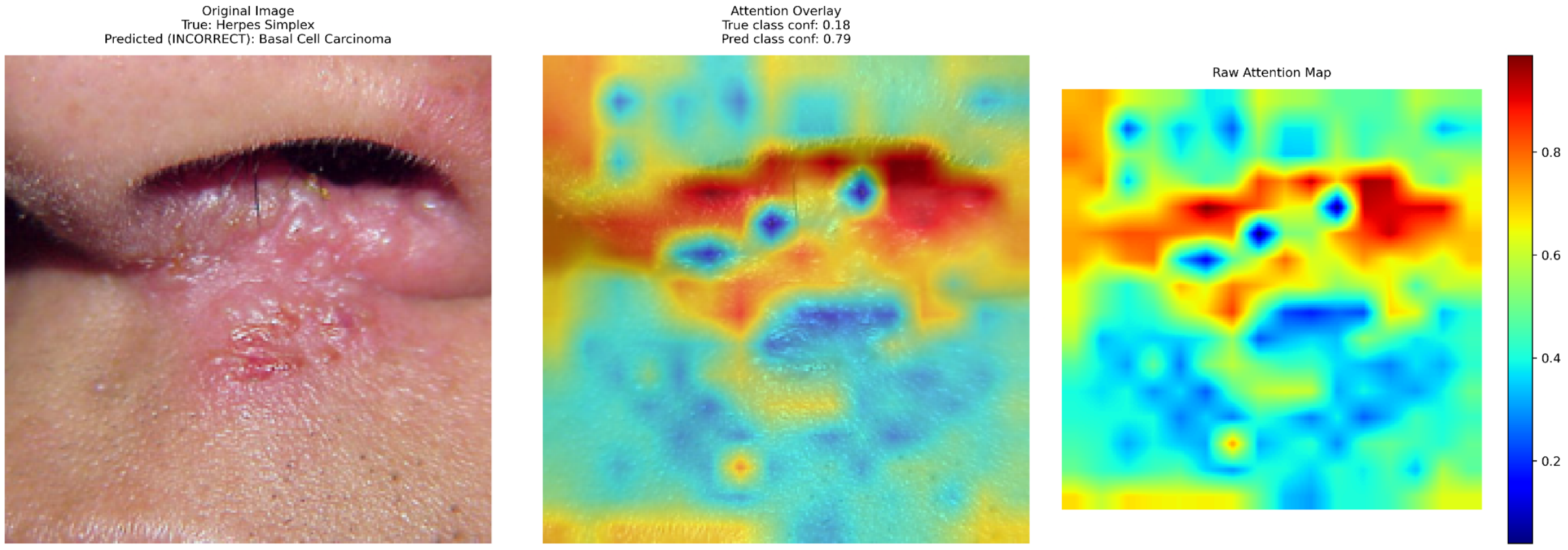

## Image 12:

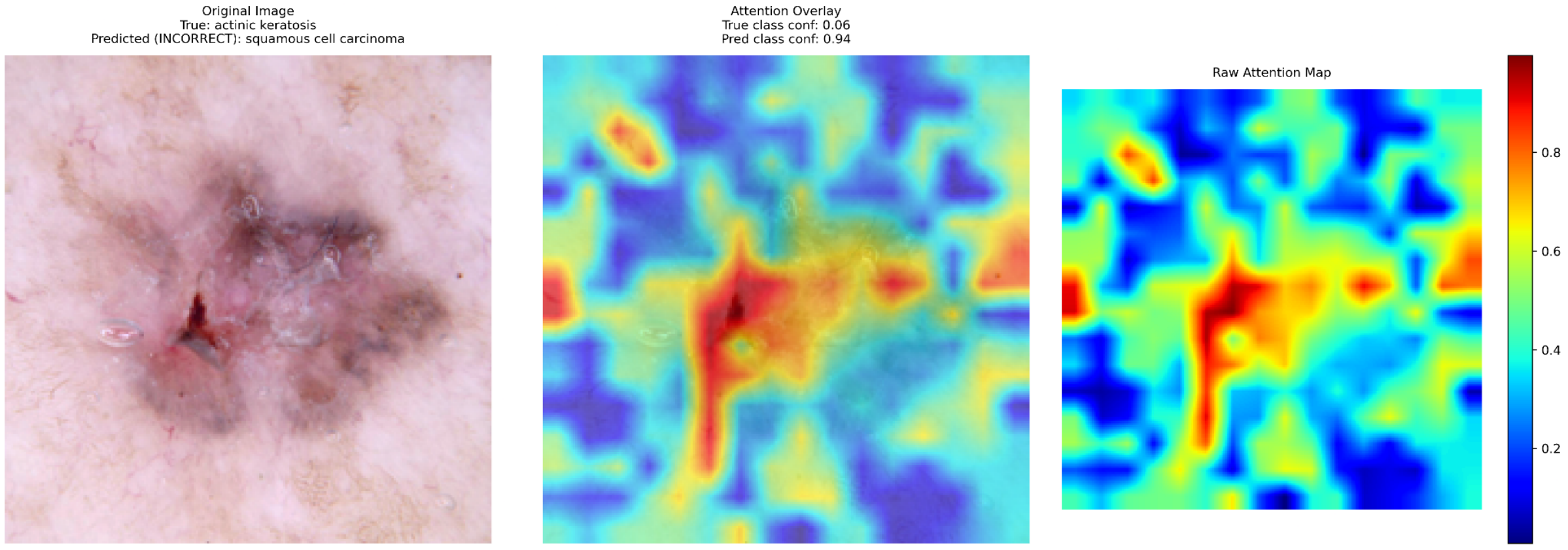

## Image 13:

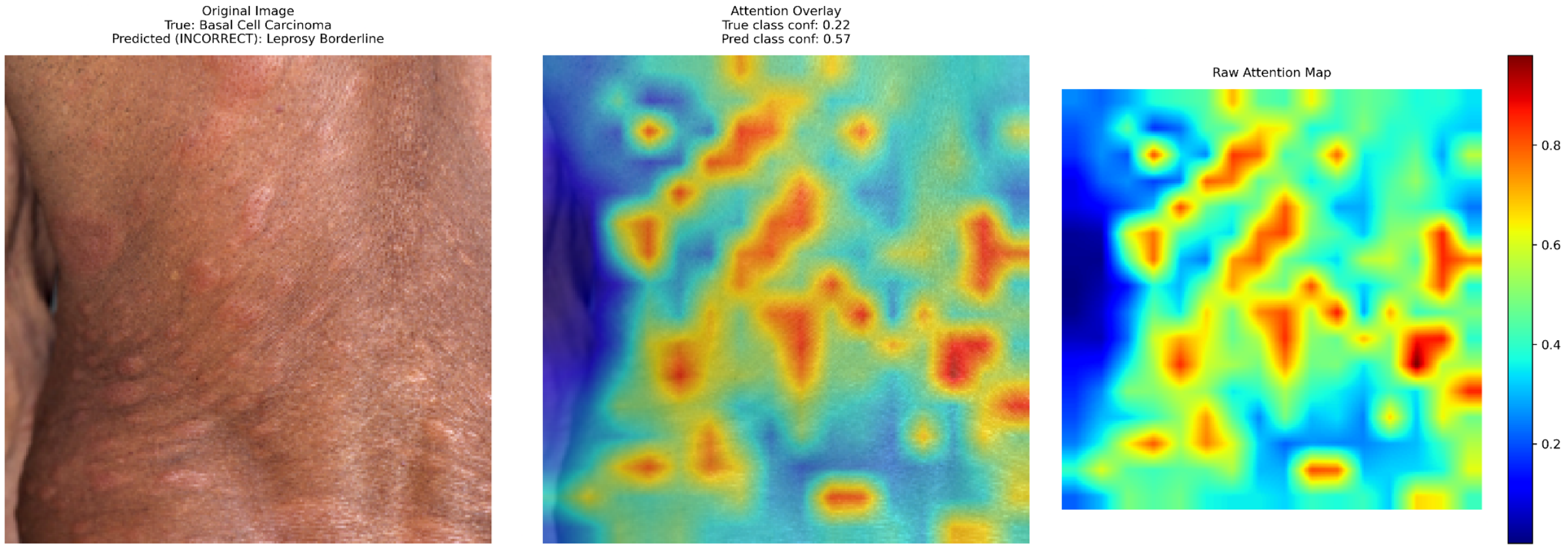

Image 14:

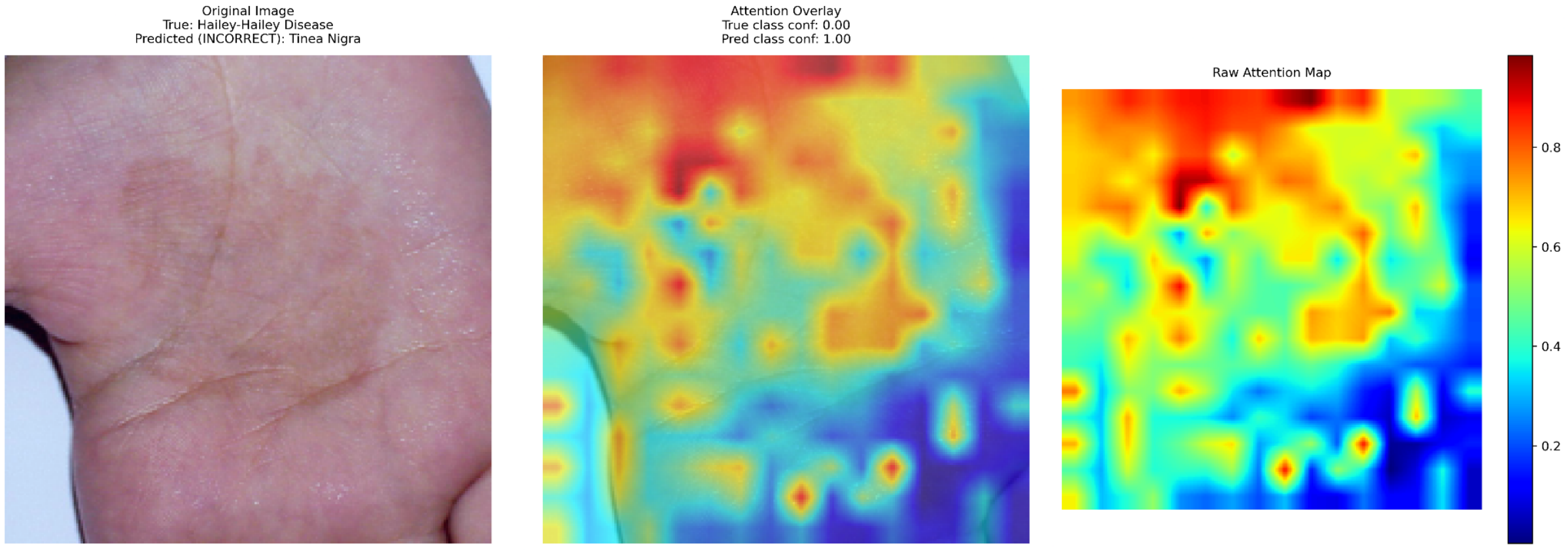

Image 15:

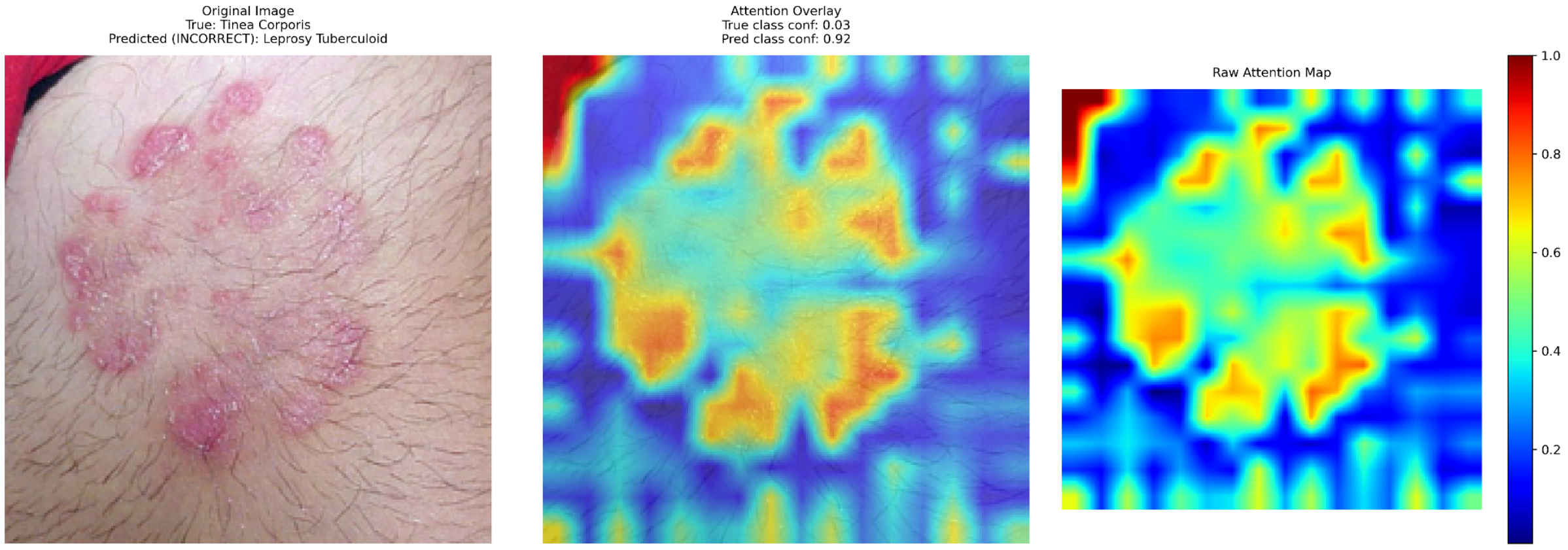

Image 16:

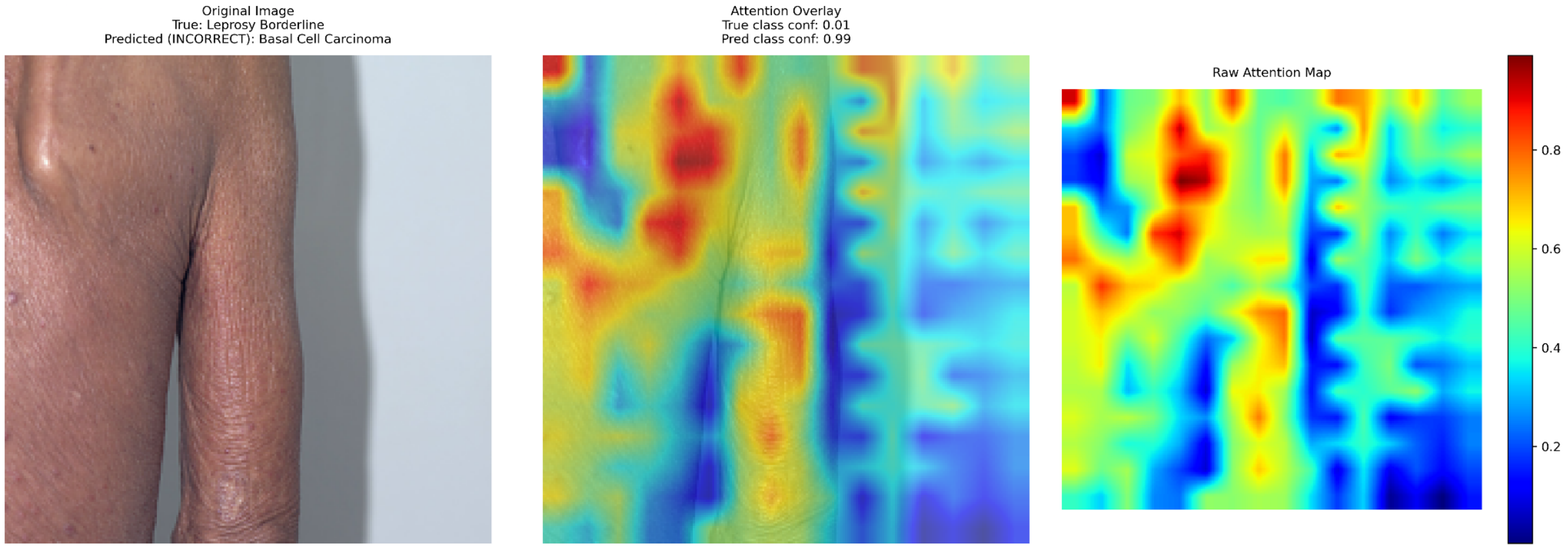

Image 17:

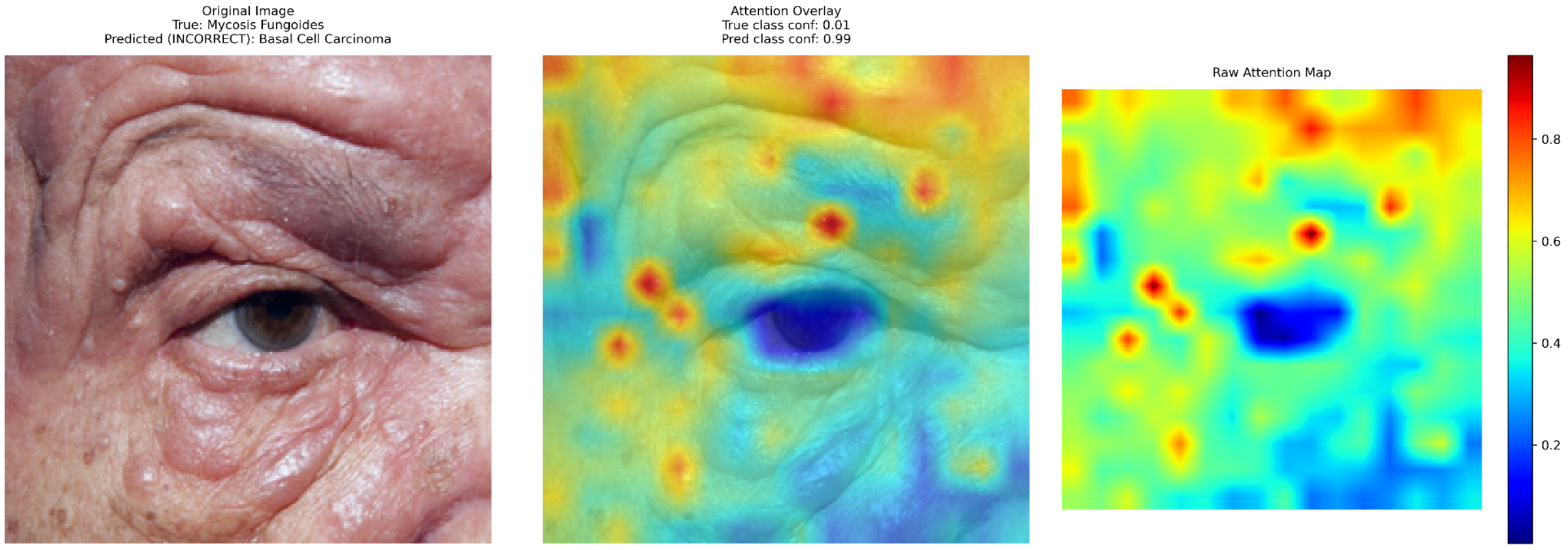

Image 18:

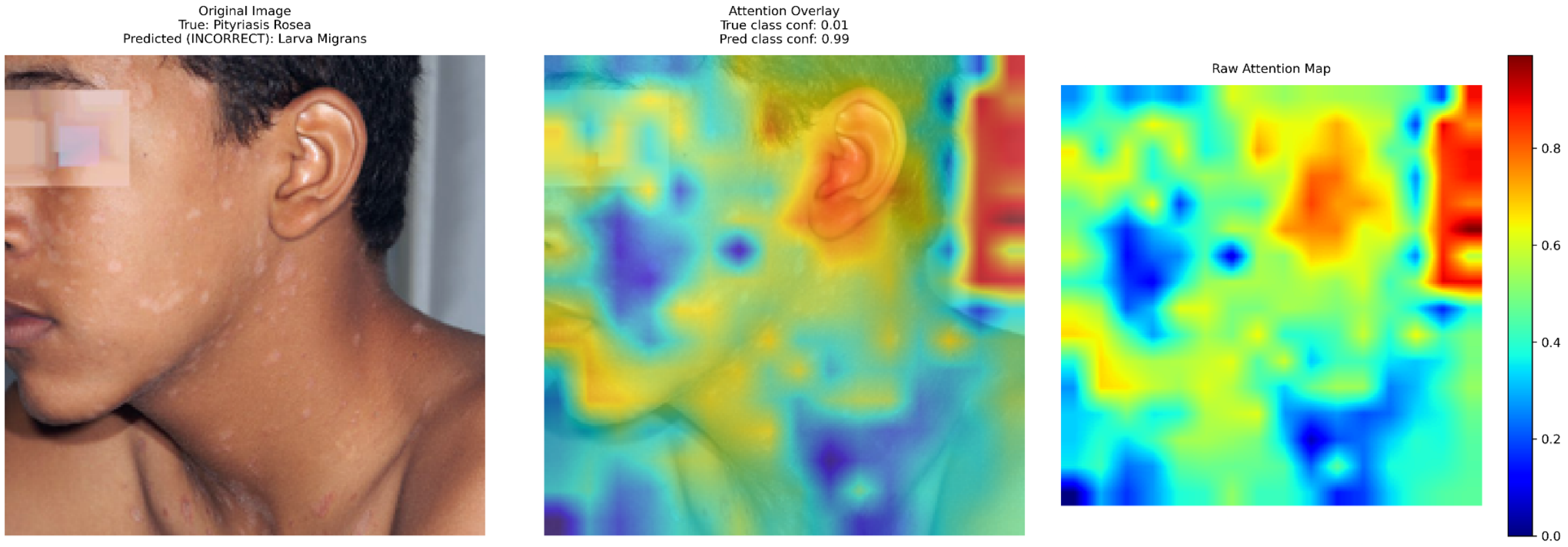

Image 19:

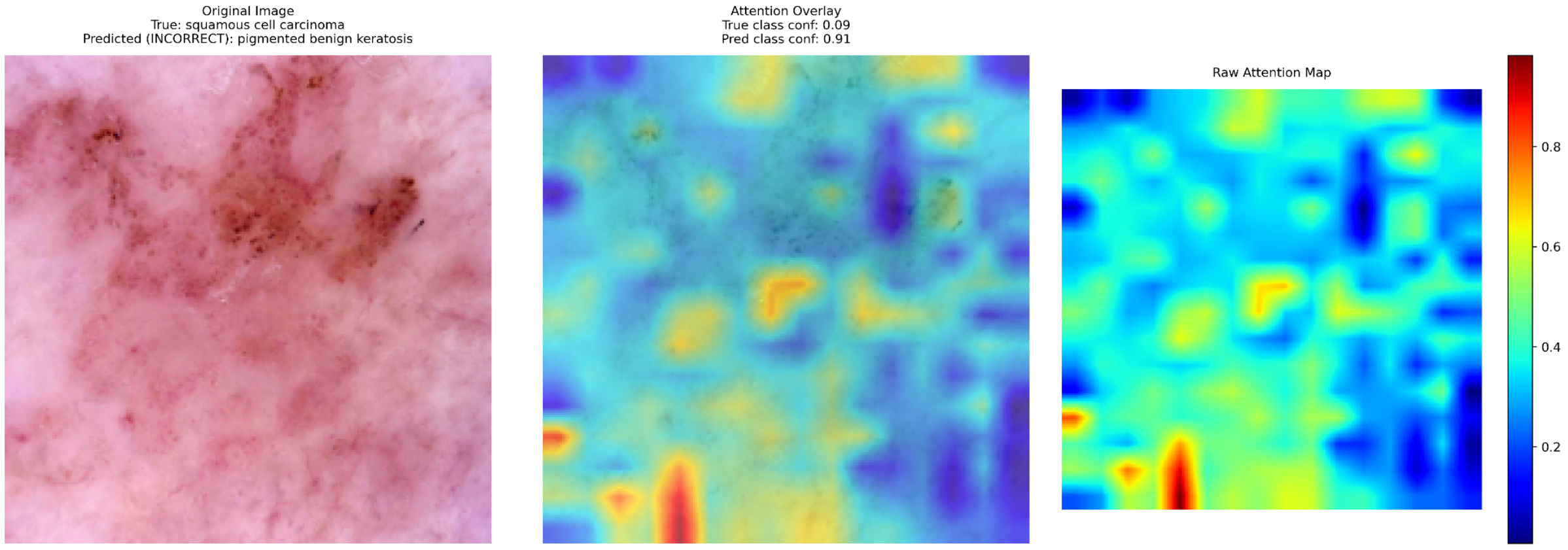

Image 20:

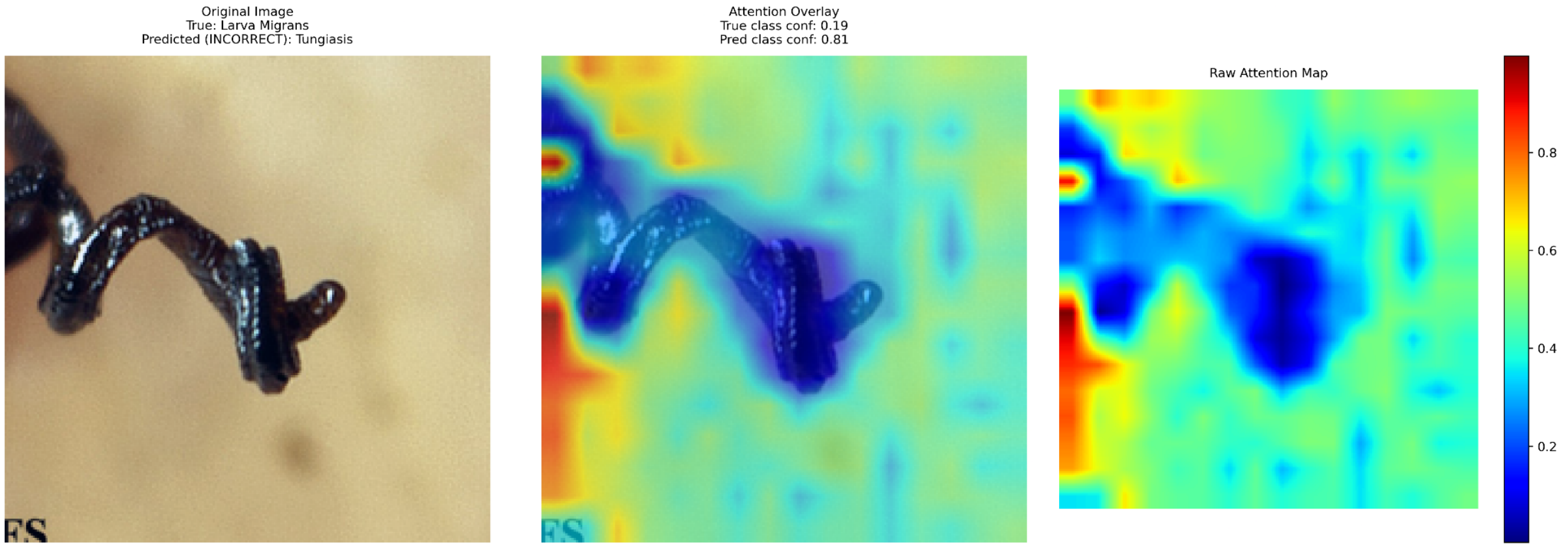